\documentclass[twocolumn,reprint,prd,preprintnumbers,amsmath,amssymb,groupedaddresst]{revtex4}
\usepackage{graphicx}
\usepackage{color}
\usepackage{dcolumn}

\newcommand {\beq}{\begin{eqnarray}}
\newcommand {\eeq}{\end{eqnarray}}
\newcommand {\non}{\nonumber\\}
\newcommand{\bi}[1]{\ensuremath{\boldsymbol{#1}}}

\def\Tr{{\rm Tr}}
\def\p{\partial}
\def\D{\mathcal{D}}

\begin{document}

\preprint{DAMTP-2011-53}
\preprint{RIKEN-MP-25}
\preprint{YGHP-11-44}

\title{Anomaly-induced charges in baryons}

\author{Minoru  {\sc Eto}}
\email{eto(at)sci.kj.yamagata-u.ac.jp}
\affiliation{{\it Department of Physics, Yamagata University, Yamagata 990-8560, Japan}}
\author{Koji {\sc Hashimoto}}
\email{koji(at)riken.jp}
\affiliation{{\it Mathematical Physics Lab., RIKEN Nishina Center, Saitama 351-0198, Japan }}
\author{Hideaki {\sc Iida}}
\email{hiida(at)riken.jp}
\affiliation{{\it Mathematical Physics Lab., RIKEN Nishina Center, Saitama 351-0198, Japan }}
\author{Takaaki {\sc Ishii}}
\email{T.Ishii(at)damtp.cam.ac.uk}
\affiliation{{\it Department of Applied Mathematics and Theoretical Physics, University of Cambridge, 
Cambridge CB3 0WA, United Kingdom}}
\author{Yu {\sc Maezawa}}
\email{maezawa(at)ribf.riken.jp}
\affiliation{{\it Mathematical Physics Lab., RIKEN Nishina Center, Saitama 351-0198, Japan }}

\begin{abstract}

We show that quantum chiral anomaly of QCD in magnetic backgrounds
induces a novel structure of electric charge inside baryons.
To illustrate the anomaly effect, we employ the Skyrme model for baryons, with the anomaly-induced 
gauged Wess-Zumino term $\sim (\pi_0 + \mbox{multi-pion})\vec{E}\cdot\vec{B}$. Due to this term, 
the Skyrmions giving a local pion condensation 
$\langle (\pi_0 + \mbox{multi-pion})\rangle\neq 0$ necessarily become a local charge source, in the background magnetic
field $\vec{B}\neq 0$. We present detailed evaluation of the anomaly effects, and calculate
the total induced charge, for various baryons in the magnetic field.

\end{abstract}

\maketitle


\section{Introduction}

The chiral anomaly is one of the central concepts in QCD, and it manifests 
nature of quantum field theories in an explicit way in our hadronic world.
As the chiral anomaly is essentially coupled to electromagnetic sector 
since the electromagnetism is a part of the chiral symmetry, the introduction
of nontrivial electromagnetic backgrounds should add a good flavor of
physics onto the chiral anomaly. In this paper, we report an interesting 
new effect induced by the chiral anomaly, for baryons in a background magnetic
field.

\begin{figure}[t]
\includegraphics[width=0.28\textwidth]{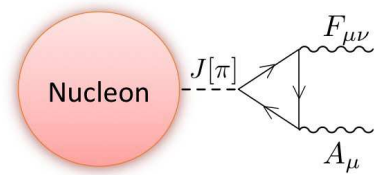}
\caption{A schematic figure for electric charge generation of a nucleon. 
In electromagnetic backgrounds, i.e., $F_{\mu\nu}\neq 0$, 
the chiral anomaly
generates an additional 
coupling to the gauge fields $A_\mu$. }
\label{chargegeneration}
\end{figure}

Our finding is that baryons in a constant magnetic background acquire
additional electric charge distribution due to the chiral anomaly. 
The result that this would generate even a total net charge is quite surprising,
but the mechanism is quite simple.
It is well-known that Wess-Zumino-Witten (WZW) term  \cite{Wess1971,Witten1983} 
actually captures the
chiral anomaly in terms of the hadronic degrees of freedom. In particular, 
this term serves as a manifestation of the famous 
$\pi_0\rightarrow 2 \gamma$ decay. Now, any baryon carries a cloud of pions
around it, and so it is a source of the pions. Once we replace one of the
two $\gamma$'s in the Wess-Zumino-Witten term by the background magnetic field,
we immediately see that the baryon can be a source of the electromagnetism
(another $\gamma$), {\it i.e.} the baryon can have an additional charge structure
due to the chiral anomaly and the pion cloud. The schematic picture of this 
mechanism is illustrated in Fig.~\ref{chargegeneration}.

In this paper, we explicitly demonstrate this mechanism in detail, with
a help of a concrete model of the pion-cloud picture of the baryons, the Skyrme model \cite{Skyrme1961} . 
In the Skyrme model, baryons are given as a solitonic object made of a local
pion condensate $\langle \pi(x) \rangle \neq 0$. Plugging the Skyrme solution
to the Wess-Zumino-Witten term, it can be shown that the magnetic field
background can induce a novel charge structure inside the baryon (Skyrmion).

In particular, we give an argument that the total charge can also be generated,
and resultantly the Gell-Mann--Nishijima formula for baryon charges can be corrected
under the magnetic field due to the anomaly, 
\begin{align}
Q_e=e\left(I_3+\frac{N_B}{2}\right) + \frac{Q_{\rm anm}}{2} \, .
\end{align}
Here in the modified formula, 
$Q_e$ is the electric charge of the baryon, $I_3$ is 
the third component of isospin, $N_B$ is the baryon number, and
the new term $Q_{\rm anm}$ is the charge generated by the anomaly and
the background magnetic field.

One may be suspicious on this generation of electric charges. However, 
for example in the renowned Witten effect \cite{Witten1979b,Wilczek1987},
monopoles are accompanied with electric charges, in the presence of
the $\theta$ term. We may regard our WZW term as an analogue of 
the $\theta$ term for the Witten effect. In addition, the chiral magnetic effect
\cite{Kharzeev2008,Fukushima2008,Eto2010,Voloshin2009b} 
in heavy ion collisions 
shares the same property too. So, it is fare to say that the generation of
the electric charge is not a unique feature of our investigation, but is a
common feature among parity-violating effects.

Quantum anomaly is literally quantum-mechanical, and thus is a tiny
effect. However, when the coupled magnetic field is strong, this effect may be
enhanced. So our physical motivation for this work is primarily oriented to
the situation in which strong magnetic field is present with a finite density 
of baryonic matter. For this, one can come up with two important physical cases:
one is a neutron star, at which neutrons are very dense and with a strong magnetic
field, and the other is a heavy ion collision at which nuclei are smashed and 
a strong electromagnetic field is expected to be created instantly.
In this paper, we do not go into these concrete cases. We concentrate on
providing a basis for that, and in particular evaluate in detail the anomaly
WZW term with the quantized Skyrmions, under a constant magnetic field.

The organization of this paper is as follows.
In section \ref{sec:2}, we provide a review of the Skyrme model and the WZW term,
with a brief introduction to the Skyrmion solution. 
In section \ref{sec:3},
we shall see explicitly that
the background magnetic field generates an additional charge structure in
the Skyrmions (baryons).
We quantize the Skyrmion and evaluate the anomaly-induced
electric current for an arbitrary baryon state.
In section \ref{sec:4}, we evaluate the multipole moments of the anomaly-induced electric
current and found a quadrupole, with a pion-mass dependence.
In section \ref{sec:5}, we discuss possible
other effect due to the background magnetic field on the baryon.
In section \ref{sec:6}, we evaluate classically the anomaly-induced charge for 
higher-charge (=multiple) Skyrmions.
The final section is for our conclusion and discussions.
Appendix \ref{app:A} is a study of the generated charge in a non-constant magnetic field. 
In Appendix \ref{app:B}, we show that the induced charge 
is due to a multi-pion effect ({\it i.e.} a pion cloud), 
and we compare our result 
with a point particle description of baryons.
The letter version of this paper is \cite{Eto:2011id}.

\section{The Skyrmions and the anomaly-induced charges}
\label{sec:2}

As briefly described in the introduction, it is indeed almost straightforward to
calculate the effect of the anomaly term for baryons in the presence of the magnetic field
background, once we adopt a concrete model of the pion cloud. Here,
we first review the Skyrme model which realizes baryons as a condensation of the pions,
and also review the gauged WZW term which manifests the chiral anomaly in QCD.

\subsection{The model}

\subsubsection{The Skyrme model}

The chiral symmetry $SU(N)_L \times SU(N)_R$ acts on left-handed and right-handed quarks as
\beq
q_L \to U_L q_L,\,
q_R \to U_R q_R,\quad \text{with}\quad U_{L,R} \in SU(N)_{L,R}.
\eeq
When the chiral condensate $\bar q_R q_L$ develops a non-zero 
vacuum expectation value by some non-perturbative effects
\beq
\left< \bar q_R q_L \right> =  - v^3 {\bf 1}_N,\qquad \text{with}\quad v = {\cal O}(\Lambda_{\rm QCD}),
\eeq
the axial-part of the chiral symmetry is spontaneously broken as
\beq
SU(N)_L \times SU(N)_R \to SU(N)_{L+R}.
\eeq
This gives rise to Nambu-Goldstone (NG) bosons, namely the pions, which takes value in the coset space 
$\frac{SU(N)_L\times SU(N)_R}{SU(N)_{L+R}}$,
\beq
U(x) = \exp\left(\frac{4i \pi^a(x)}{F_\pi} T^a\right),\quad (a = 1,2,\cdots,N^2-1).
\eeq
Here $F_\pi = 108$[MeV] is the pion decay constant and
$T^a$ is a generator of $SU(N)$ and we use the following standard normalization
\beq
\Tr[T^aT^b] = \frac{1}{2} \delta^{ab}.
\eeq
The chiral symmetry acts on the NG modes as
\beq
U \to U_L U U_R^\dag.
\eeq

For later convenience, let us define left- and right-invariant Maurer-Cartan one-forms by
\beq
L_\mu \equiv U^\dag \p_\mu U,\qquad R_\mu \equiv \p_\mu U U^\dag.
\eeq
These take their values in the algebra of $SU(N)_R$ and $SU(N)_L$, respectively. The chiral symmetry acts on them as
\beq
L_\mu \to U_R L_\mu U_R^\dag,\qquad
R_\mu \to U_L R_\mu U_L^\dag.
\eeq

We can think of $U$ as an effective low-energy field. 
Its effective Lagrangian of the leading order to ${\cal O}(\p^2)$ can be uniquely determined as
\beq
{\cal L}^{(2)}& =&  \frac{F_\pi^2}{16} \Tr\left[\p_\mu U \p^\mu U^\dag +M_\pi^2 \left(U + U^\dag - 2\right)\right]  
\nonumber \\
& =& \frac{F_\pi^2}{16}\Tr\left[- R_\mu R^\mu + M_\pi^2 \left(U + U^\dag - 2\right)\right].
\eeq
Here $M_\pi$ stands for the pion mass $M_\pi=137$[MeV] and our metric is $\eta_{\mu\nu} = {\rm diag}(+1,-1,-1,-1)$.
By expanding $L_{\mu}$ and $R_\mu$ with respect to $1/F_\pi$, one gets 
\beq
L_\mu = 4i \frac{\p_\mu \pi^a}{F_\pi}T^a + 8i \epsilon^{abc} \frac{\pi^a\p_\mu \pi^b}{F_\pi^2} T^c + \cdots,
\label{eq:exp_L}\\
R_\mu = 4i \frac{\p_\mu \pi^a}{F_\pi}T^a - 8i \epsilon^{abc} \frac{\pi^a\p_\mu \pi^b}{F_\pi^2} T^c + \cdots.
\label{eq:exp_R}
\eeq
Plugging this into ${\cal L}^{(2)}$, one obtain a standard kinetic term of the pions and corrections,
\beq
{\cal L}^{(2)} &= & \frac{1}{2}\p_\mu \pi^a \p^\mu \pi^a - \frac{M_\pi^2}{2}\pi^a\pi^a \nonumber\\
&-&  \frac{2}{3F_\pi^2}\left(\pi^a\pi^a\p_\mu \pi^b \p^\mu \pi^b - \pi^a\pi^b\p_\mu \pi^a \p^\mu \pi^b\right) \nonumber \\
&+& \frac{2M_\pi^2}{3F_\pi^2}(\pi^a\pi^a)^2 + \cdots.
\eeq

We are interested in a topological soliton made by the pions in this work.
The topological winding number is given by
\beq
\pi_3\left(SU(N)\right) = N_B \in \mathbb{Z}.
\eeq
As will be shown, $N_B$ is identified with the baryon number via the WZW term.
However, it is easy from a simple scaling argument that no topological solitons can survive from collapsing 
in the theory with ${\cal L}^{(2)}$. 
So one needs higher derivative corrections to ${\cal L}^{(2)}$.
Therefore, we take a term of order ${\cal O}(\p^4)$ which is so-called the Skyrme term
\beq
{\cal L}^{(4)} = \frac{1}{32e_s^2} \Tr\left([R_\mu,R_\nu][R^\mu,R^\nu]\right),
\eeq
with $e_s$ being a dimensionless coupling constant.
We will choose the parameter $e_s = 4.84$ by following Ref.~\cite{Adkins1984}.
Now we are ready to write down the Skyrme model with the right-invariant one form as
\beq
{\cal L} &=& \frac{F_\pi^2}{16}\Tr\left[-R_\mu R^\mu + M_\pi^2\left(U+U^\dag - 2\right)\right] \nonumber\\
&+& \frac{1}{32e_s^2} \Tr\left([R_\mu,R_\nu][R^\mu,R^\nu]\right).
\eeq

A Noether current of $SU(N)_L$ can be obtained by performing a local and infinitesimal $SU(N)_L$ rotation
\beq
\delta R_\mu =  i\p_\mu \phi_L.
\eeq
Variation of the Skyrme Lagrangian is given by
\beq
\delta {\cal L} = \Tr\left\{\frac{i}{8}\left(- F_\pi^2 R_\mu  + \frac{1}{e_s^2} [ R_\nu, [R^\mu,R^\nu]] \right)\p_\mu \phi_L\right\}.
\eeq
Then the conserved current is given by
\beq
j_L^\mu = \frac{i}{8}\left(F_\pi^2 R^\mu  - \frac{1}{e_s^2} [ R^\nu, [R_\mu,R_\nu]] \right).
\eeq
Similarly, the $SU(N)_R$ current takes the form
\beq
j_R^\mu = \frac{i}{8}\left( - F_\pi^2 L^\mu  +  \frac{1}{e_s^2} [ L^\nu, [L_\mu,L_\nu]] \right).
\eeq
These currents are related by
\beq
j_L^\mu = - U j_R^\mu U^\dag
\eeq
where we have used
\beq
U L^\mu U^\dag = U ( U^\dag \p^\mu U ) U^\dag = R^\mu.
\eeq
The equation of motion of the Skyrme model is identical to the current conservation law
if the pion mass is zero
\beq
\p_\mu j_L^\mu = 0,\quad \text{or} \quad
\p_\mu j_R^\mu = 0.
\eeq
When the pion mass is non-zero, the equation of motion becomes
\beq
\p_\mu j_L^\mu = - \frac{iF_\pi^2 m_\pi^2}{16}  \Tr\left[U-U^\dag\right].
\eeq

The vector and axial conserved currents are defined by
\beq
j_V^\mu = \frac{j_L^\mu + j_R^\mu}{2},\quad
j_A^\mu = \frac{j_L^\mu - j_R^\mu}{2}.
\eeq
The vector $SU(N)_{L+R}$ is nothing but the isospin, so we write its conserved charge as
\beq
I^a = \int d^3x\ j_V^{0a} = \int d^3x\ \Tr[(j_L^0+j_R^0)T^a].
\eeq

\subsubsection{Electromagnetic interaction}

Let us next take the electromagnetic interaction into account.
For simplicity, hereafter, we concentrate on the minimal case with two flavors $N=2$.
Since the electric charges of $u$ and $d$ quarks are $2/3$ and $-1/3$ respectively, the NG modes
are rotated under the electromagnetic $U(1)$ as
\beq
U \to e^{-i e Q} U e^{ie Q} = e^{-ieT^3}U e^{ieT^3},\quad
Q = \frac{1}{6}{\bf 1} + T^3,
\eeq
where $T^3 = \tau^3/2$ with the Pauli matrix $\tau^a$.
Thus the electromagnetic $U(1)_{\rm em}$ is a subgroup of $SU(2)_{L+R}$.

Interactions of the NG modes and the electromagnetic fields are introduced by gauging
the $U(1)_{\rm em} \subset SU(2)_{L+R}$ and replacing the partial derivative $\p_\mu$ by
a covariant derivative
\beq
\D_\mu U = \p_\mu U + i e A_\mu [T^3,U].
\eeq
The left- and right-invariant one-forms are then replaced as
\beq
R_\mu \to \tilde R_\mu \equiv \D_\mu U U^\dag,\quad
L_\mu \to \tilde L_\mu \equiv  U^\dag \D_\mu U.
\eeq
Then the total Lagrangian can be read as
\beq
{\cal L} &=& - \frac{1}{4}F_{\mu\nu}F^{\mu\nu} 
+ \frac{F_\pi^2}{16}\Tr\left[ - \tilde R_\mu \tilde R^\mu + \left(U+U^\dag - 2\right)\right] \nonumber\\
&+&  \frac{1}{32e_s^2} \Tr\left([\tilde R_\mu,\tilde R_\nu][\tilde R^\mu,\tilde R^\nu]\right).
\eeq

The classical equation of motion is derived by variational method as before.
One can easily obtain
\beq
\D_\mu \tilde{j}_L^\mu &=&  - \frac{iF_\pi^2M_\pi^2}{16}  \Tr\left[U-U^\dag\right], \\
\tilde{j}_L^\mu &\equiv& 
\frac{i}{8}\left( F_\pi^2 \tilde R^\mu  - \frac{1}{e_s^2} [ \tilde R_\nu, [\tilde R^\mu,\tilde R^\nu]] \right).
\eeq
One can also express the E.O.M. in terms of the right-invariant one-form $\tilde L_\mu$ by just replacing $\tilde R_\mu$ with $\tilde L_\mu$.
Note that since the electromagnetic charge $Q$ breaks the chiral symmetry $SU(2)_L \times SU(2)_R \to U(1)_{\rm em}$,
$\tilde{j}_L^\mu$ and $\tilde{j}_R^\mu$ are not conserved currents.
The Maxwell equation is given by
\beq
\p_\mu F^{\nu\mu} = e \tilde{j}_V^{3\nu}, \quad
 \tilde{j}_V^{3\nu} = \Tr\left[T_3(\tilde{j}_L^\nu + \tilde{j}_R^\nu)\right].
\eeq
Thus the electromagnetic charge is given by
\beq
Q_e =  e \int d^3x \ \tilde{j}_V^{3\; \nu=0} = e I^3.
\label{eq:q_em_isospin}
\eeq

\subsubsection{WZW term and Chiral anomaly}

In order to describe the baryons in the Skyrme model whose fundamental degrees of freedom are
mesons, we have to consider not only elemental particles but also topological excitations.
Since the NG modes do not carry the $U(1)_B$ charges, we need to add an extra terms to the above Lagrangian.
It is the so-called Wess--Zumino--Witten (WZW) term. With an electromagnetic field $A_\mu$, the WZW term 
in the $N=2$ flavor model is 
given by \cite{Brihaye:1984hp}
\beq
S_{\rm WZW}[A_\mu] =   \int d^4x\ \frac{e}{2}j^\mu_B A_\mu,
\label{eq:WZW}
\eeq
with the baryonic current
\beq
j^\mu_B = \frac{1}{24\pi^2} \epsilon^{\mu\nu\rho\sigma}\Tr[R_\nu R_\rho R_\sigma] 
- \frac{e}{8\pi^2} \epsilon^{\mu\nu\rho\sigma}\p_\nu \left( A_\rho P_\sigma \right).
\eeq
Here we use $\epsilon^{0123}=-1$ and define
\beq
P_\sigma &\equiv& \frac{i}{2} \Tr[\tau^3(L_\sigma + R_\sigma)].
\label{eq:def_P}
\eeq
This baryon current is clearly conserved due to the anti-symmetric tensor $\epsilon^{\mu\nu\rho\sigma}$.
On the other hand, $j_B^\mu$ appears to depend on gauge choices, at a glance. But this is not the case.
One can rewrite the baryonic current as
\beq
j^\mu_B & =& \frac{1}{24\pi^2}\epsilon^{\mu\nu\rho\sigma}\Tr[\tilde R_\nu \tilde R_\rho \tilde R_\sigma]
\nonumber \\
& & 
- \frac{ie}{32\pi^2}\epsilon^{\mu\nu\rho\sigma} F_{\nu\rho}\Tr[\tau^3(\tilde L_\sigma + \tilde R_\sigma)].
\eeq
This is manifestly gauge invariant.

The first term in the current $j^\mu_B$ gives a topological number associated with $\pi_3(SU(2)_{L-R})$. 
Indeed, the integration of it
over the space gives the topological winding number
\beq
N_B = \int d^3x\ \frac{1}{24\pi^2} \epsilon^{ijk}\Tr[R_i R_j R_k] .
\eeq
The second term in the current $j_{\rm anm}^\mu$ is a manifestation of the chiral anomaly.
Actually, plugging Eqs.~(\ref{eq:exp_L}) and (\ref{eq:exp_R}) into Eq.~(\ref{eq:WZW}), one finds the famous
$\pi^0 \to 2\gamma$ term by the anomaly
\beq
& & -\frac{N_c e^2}{48\pi^2 F_\pi} \epsilon^{\mu\nu\rho\sigma}A_\mu F_{\nu\rho} \p_\sigma \pi^0 + {\cal O}(F_\pi^{-3})
\nonumber \\
& & 
\hspace*{5mm}
= -\frac{N_c e^2}{12\pi^2 F_\pi} \pi^0 \vec E \cdot \vec B + {\cal O}(F_\pi^{-3}),
\eeq
where the equality holds up to a total derivative.

Let us next obtain the electric charge coupled to a photon fluctuation $a_\mu$ 
under a background electromagnetic field $\bar A_\mu$.
To this end, we expand the gauge field as
\beq
A_\mu = \bar A_\mu + a_\mu.
\eeq
Then the WZW action linear in $a_\mu$ gives
\beq
&& \hspace{-5mm}
S_{\rm WZW}[a_\mu] 
\nonumber \\
&=& \int\!\!  d^4x \left(\frac{e}{2} j_B^\mu(\bar A) a_\mu 
- \frac{e^2}{16\pi^2} \epsilon^{\mu\nu\rho\sigma} \bar A_\mu \p_\nu (a_\rho P_\sigma )
  \right) \non
 &=& 
 \int\!\!  d^4x \epsilon^{\mu\nu\rho\sigma}\! \left[
\left(\! \frac{e}{48\pi^2} \Tr[R_\nu R_\rho R_\sigma]
\!-\!  \frac{e^2}{16\pi^2} (\p_\nu \bar A_\rho) P_\sigma \!\! \right)a_\mu  
\right.
\non
& & 
\left.
\hspace{10mm}
- \frac{e^2}{16\pi^2}  \bar A_\mu (\p_\nu a_\rho) P_\sigma \right].
\eeq
From this, we can read the electromagnetic current $j^\mu_{\rm em, WZW} = \delta S_{\rm WZW}/\delta a_\mu$ as
\beq
&& \hspace{-5mm}
j^\mu_{\rm em,WZW} 
\non
&=&\!\! j^\mu_{\rm anm}
 \!+\!
\frac{\epsilon^{\mu\nu\rho\sigma} e}{48\pi^2}\!  \left[
 \Tr[R_\nu R_\rho R_\sigma]
\!+\! 3e  \partial_\sigma (\bar A_\rho  P_\nu) \!\right],
\label{eq:wzwcurrent}
\\
&& \hspace{-5mm} j^\mu_{\rm anm}
\equiv  -\ \epsilon^{\mu\nu\rho\sigma}\frac{e^2}{16\pi^2} (\p_\nu \bar A_\rho) P_\sigma.
\label{eq:current_anm}
\eeq
The current $j^\mu_{\rm em,WZW}$ is gauge-invariant \cite{Brihaye:1984hp}. 
The total electromagnetic current is
\beq
j^\mu_{\rm em} = e\, \tilde j_V^{3\mu} + j^\mu_{\rm em, WZW}.
\label{eq:emcurrentsum}
\eeq
This is gauge-invariant, and always conserved at on-shell, from the gauge invariance.

The total electric charge is given by the spatial integral 
\footnote{Note that, when we make the spatial integration, the last term in \eqref{eq:wzwcurrent} drops off
as it is a total derivative term. For massive pions, the pion profile of the Skyrmion always decay exponentially
asymptotically, so the surface integral derived from the integration of this total derivative term
always vanish.}
of the zero-th component of
this current \eqref{eq:emcurrentsum},
\beq
Q_e = e I_3 + \frac{e}{2}N_B 
+ \frac{e^2}{16\pi^2} \int d^3x\  B_i P_i.
\label{eq:full_charge}
\eeq
Here $B_i$ stands for a background magnetic field and
we have used $\epsilon^{0ijk} = - \epsilon^{ijk}$.
This is the expression of  gauge invariant and conserved electric charge which includes
an extra term to the well-known Gell-Mann--Nishijima formula.
The last term is a contribution from the chiral anomaly.

Note that, as we will see below shortly,
the surface term in Eq.~(\ref{eq:wzwcurrent}) does not contribute to the net electric charge
if the pion mass is non-zero.
So, we focus on the new last term of Eq.~\eqref{eq:full_charge} coming from Eq.~(\ref{eq:current_anm}), in the following of this paper.

\subsection{The Skyrmion: A nucleon as a topological soliton}
\label{sec:2B}

Let us find a solution of the classical equations of motion derived previously,
\beq
\D_\mu \tilde j_L^\mu = 0,\quad
\p_\mu F^{\nu \mu} = e \tilde j_V^\nu.
\eeq
We solve these by dealing with the electromagnetic interaction as a perturbation.
Then we expand the chiral field with respect to the electromagnetic coupling constant $e$ as
\beq
&& U = \exp\left(i \vec\tau \cdot (\vec f_0 + \vec f_1 + \cdots)\right),
\non
&& A_\mu = A^{(0)}_\mu + A^{(1)}_\mu + \cdots.
\eeq

At the zeroth order the chiral fields and electromagnetic fields are decoupled, so that the 
equations of motion are those of the Skyrme model without the electromagnetic interaction and
the Maxwell equation without a source
\beq
\p_\mu j_L^{(0)}{}^\mu =  - \frac{iF_\pi^2M_\pi^2}{16}  \Tr\left[U-U^\dag\right],\quad
\p_\mu F^{(0)\nu\mu} = 0,
\label{eq:eom_0th}
\eeq
with
\beq
j^{(0)}_L{}^\mu = \frac{i}{8}\left(F_\pi^2 R^{(0)\mu}  - \frac{1}{e_s^2} [ R^{(0)}_\nu, [R^{(0)\mu},R^{(0)\nu}]] \right).
\eeq
The second equation in Eq.~(\ref{eq:eom_0th}) is solved by
considering a constant background magnetic field, say along the $x_i$-axis
\beq
\frac{1}{2}\epsilon^{ijk}F_{jk} = B^i
\label{eq:A_0th}
\eeq
with $B^i$ being a constant.

In order to solve the first equation in Eq.~(\ref{eq:eom_0th}), it is useful to introduce a dimensionless 
coordinate
\beq
x_\mu \to \frac{1}{e_s F_\pi}x_\mu,\qquad \p_\mu \to {e_s F_\pi} \p_\mu.
\eeq 
In terms of this new coordinate, the Skyrme equation is written as
\beq
\p_\mu\left(R^{(0)\mu}  -  [ R^{(0)}_\nu, [R^{(0)\mu},R^{(0)\nu}]] \right) 
= \frac{m_\pi^2}{2}\Tr\left[U^\dag-U\right],
\label{eq:eom_0th_2}
\eeq
with a dimensionless mass in unit of $1/(e_sF_\pi)$
\beq
m_\pi \equiv \frac{M_\pi}{eF_\pi}.
\eeq
Here and after, we will use this notation.

\begin{figure}[ht]
\begin{center}
\includegraphics[width=8cm]{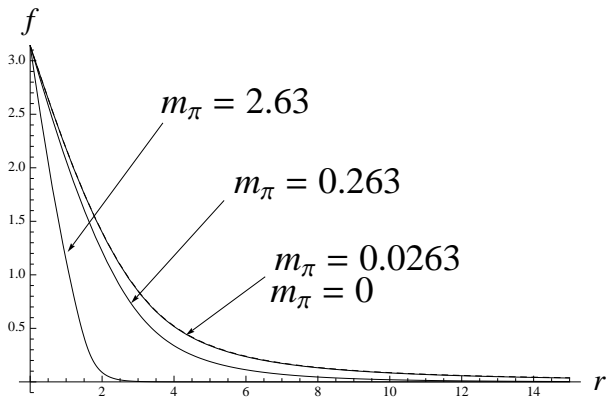}
\caption{{\footnotesize Profile functions for the hedgehog solution.}}
\label{fig:profile_hedgehog}
\end{center}
\end{figure}

Let us make a standard hedgehog (radial) ansatz,
for a static and topologically non-trivial solution with $N_B=1$,
\beq
U_0(\vec x) 
&=& \exp\left( i \vec f_0 \cdot \vec\tau \right)
= \exp\left(i f(r) \vec \tau \cdot \vec{\hat x} \right),\\
\hat x_i &=& \frac{x_i}{r}.
\eeq
One can express $U$ in a different fashion as
\beq
U_0
&=& {\bf 1}_2 \cos |\vec f_0| + i \frac{\vec f_0}{|\vec f_0|} \cdot \vec \tau \sin |\vec f_0|
\non
&=& {\bf 1}_2 \cos f + i \vec{\hat x} \cdot \vec \tau \sin f.
\eeq
Then we obtain  for a static configuration
\beq
&R^{(0)}_i = &  i f' \hat x_i  \vec{\hat x} \cdot \vec \tau
\\
 && + \frac{i}{2r}\left(\tau_i - \vec{\hat x} \cdot \vec \tau \hat x_i\right) \left( \sin2f
+ 2 i  \vec{\hat x} \cdot \vec \tau \sin^2 f\right).
\nonumber
\eeq
Putting this into Eq.~(\ref{eq:eom_0th_2}), we obtain the ordinary differential equation
\beq
&&\left(\frac{1}{4}+\frac{2\sin^2f}{r^2}\right)f'' + \frac{1}{2r}f' 
\non
&&
+ \frac{\sin 2f}{r^2} \left(f'{}^2 -\frac{1}{4} - \frac{\sin^2 f}{r^2}\right) 
= \frac{m_\pi^2}{4} \sin f. 
\eeq
The solution with a unit winding number corresponds to
\beq
\lim_{r\to\infty}f(r) = \pi,\qquad
\lim_{r\to 0} f(r) = 0.
\eeq
Numerical solutions with different $m_\pi$'s are given in Fig.~\ref{fig:profile_hedgehog}.
We adopt the physical pion mass $m_\pi^{\rm phys}=0.263$ which was determined from 
the mass splitting between nucleon and $\Delta$ \cite{Adkins1984}.
We also show the profile functions for $m_\pi = 2.63$ and $0.0263$ in order to
demonstrate a characteristic property of the profile function. We see that the larger (smaller) $m_\pi$
gives the thinner (fatter) Skyrmion. 
The profile function with $m_\pi = 0.0263$ almost coincides to that with $m_\pi = 0$,
see Fig.~\ref{fig:profile_hedgehog}.

Asymptotic behavior of $f$ 
can be found by solving the linearized equations of motion for large $r$
\beq
f'' + \frac{2}{r}f' - \frac{2f}{r^2} -m_\pi^2 f = 0.
\eeq
This is solved by
\beq
f \simeq  \left(\frac{C}{r^2} + \frac{A}{r} \right) e^{-m_\pi r},
\label{eq:asym_massless}
\eeq
where $C$ and $A (=C m_\pi)$ are constant.
For massless pion, we find 
\beq
f \simeq \frac{C}{r^2},
\label{eq:Asm0}
\eeq
with $C \simeq 8.638$. 
For massive pion, the asymptotic form decays exponentially, 
\beq
f \simeq \frac{A}{r}\ e^{-m_\pi r},
\label{eq:asym_massive}
\eeq
where $A$ is a constant which depends on $m_\pi$. For example, we find $A \simeq 2$ for
$m_\pi = 0.263$.

\section{Anomaly-induced charge}
\label{sec:3}

We substitute the Skyrme solution to the electromagnetic current calculated from
the WZW term, to evaluate the anomaly-induced electric charge of the baryons.
We here use classical Skyrmions for an illustration first, then 
we move onto quantized Skyrmions, to obtain a formula for the anomaly-induced charge
for baryon quantum states.
The anomaly-induced charge is the last term of Eq.~(\ref{eq:full_charge}).
Since we are considering a constant magnetic field background, it is enough to see $P_i$ defined in Eq.~(\ref{eq:def_P}).

\subsection{Classical evaluation}

\begin{figure}[t]
\begin{center}
\begin{tabular}{ccc}
\includegraphics[width=2.35cm]{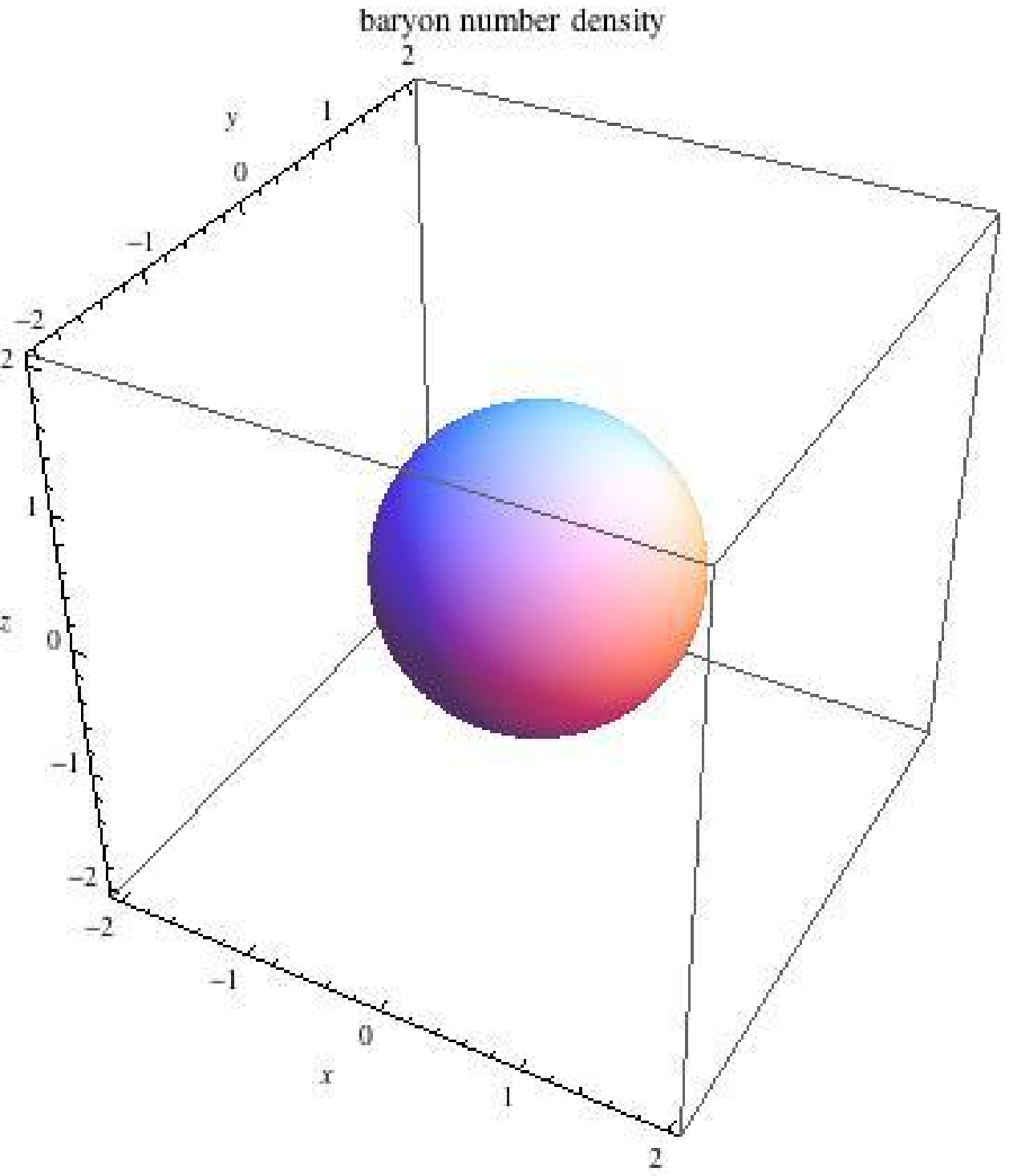} & 
\includegraphics[width=2.35cm]{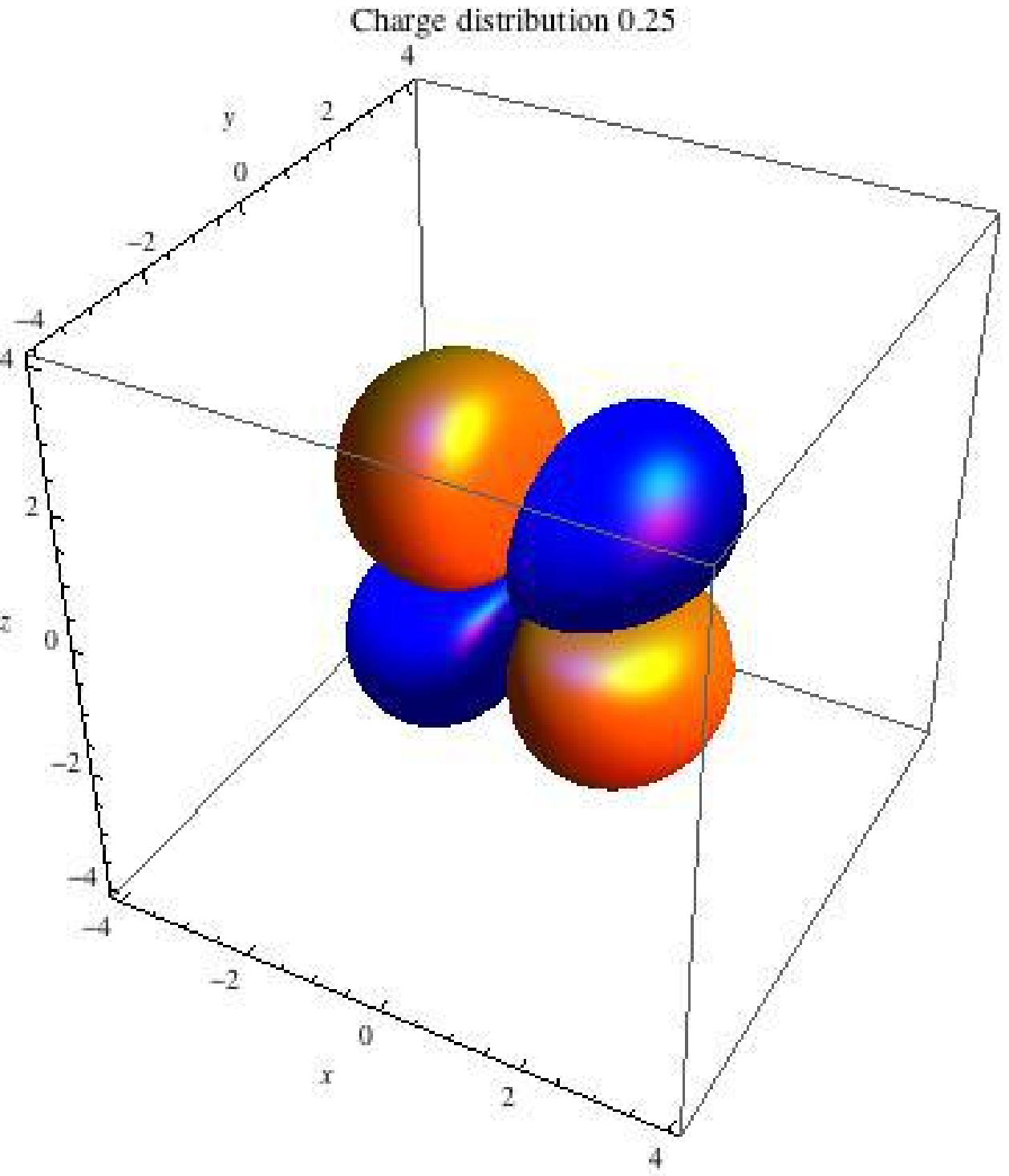} & 
\includegraphics[width=2.35cm]{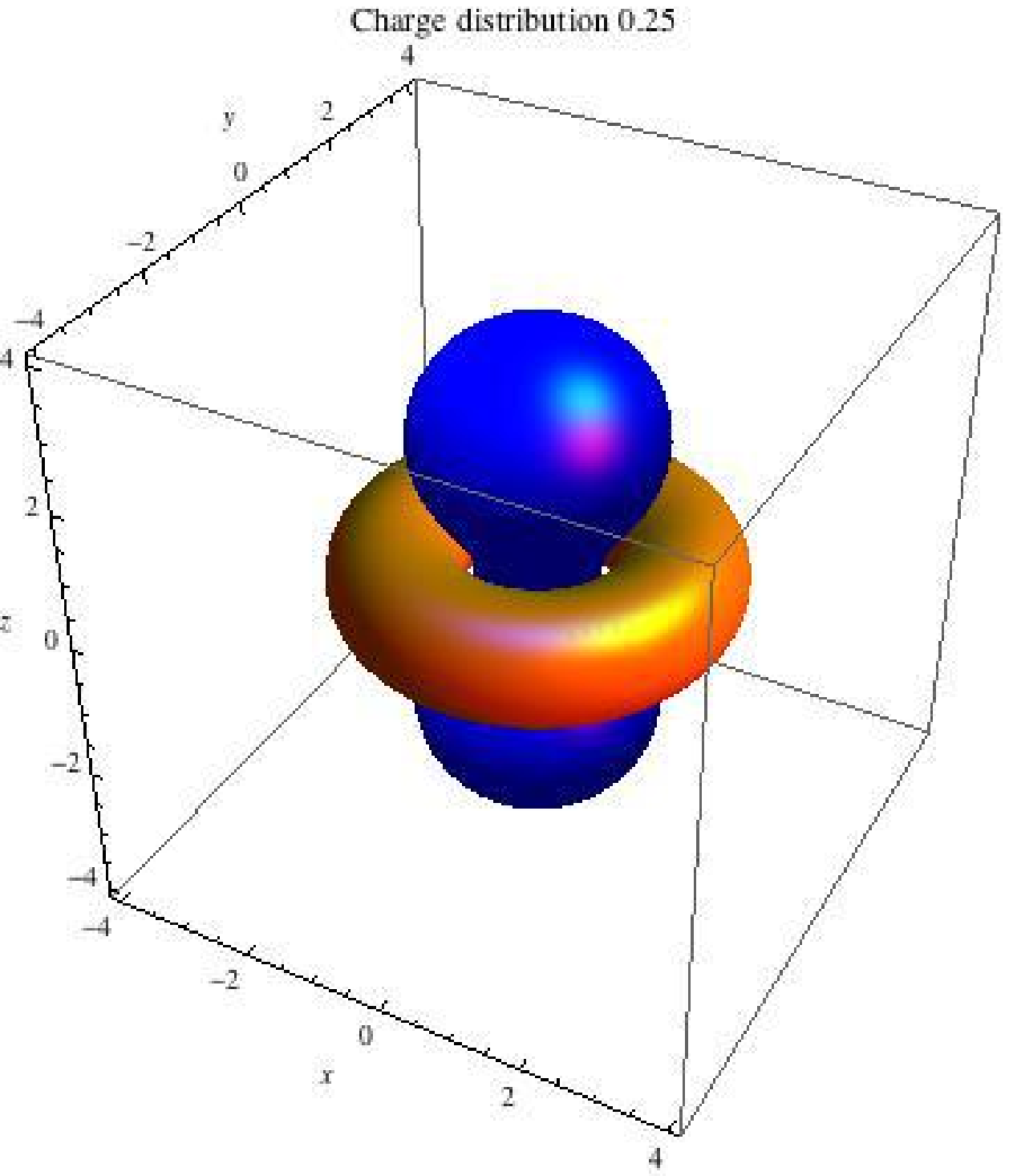}
\end{tabular}
\caption{$N_B=1$ Skyrmion solution for $m_\pi=0$. From left to right, contours plots of the baryon number density,
$P_1=\pm 0.2$ and $P_3=\pm0.2$, respectively.}
\label{fig:B=1}
\end{center}
\end{figure}

\begin{figure}[h]
\begin{center}
\includegraphics[width=6cm]{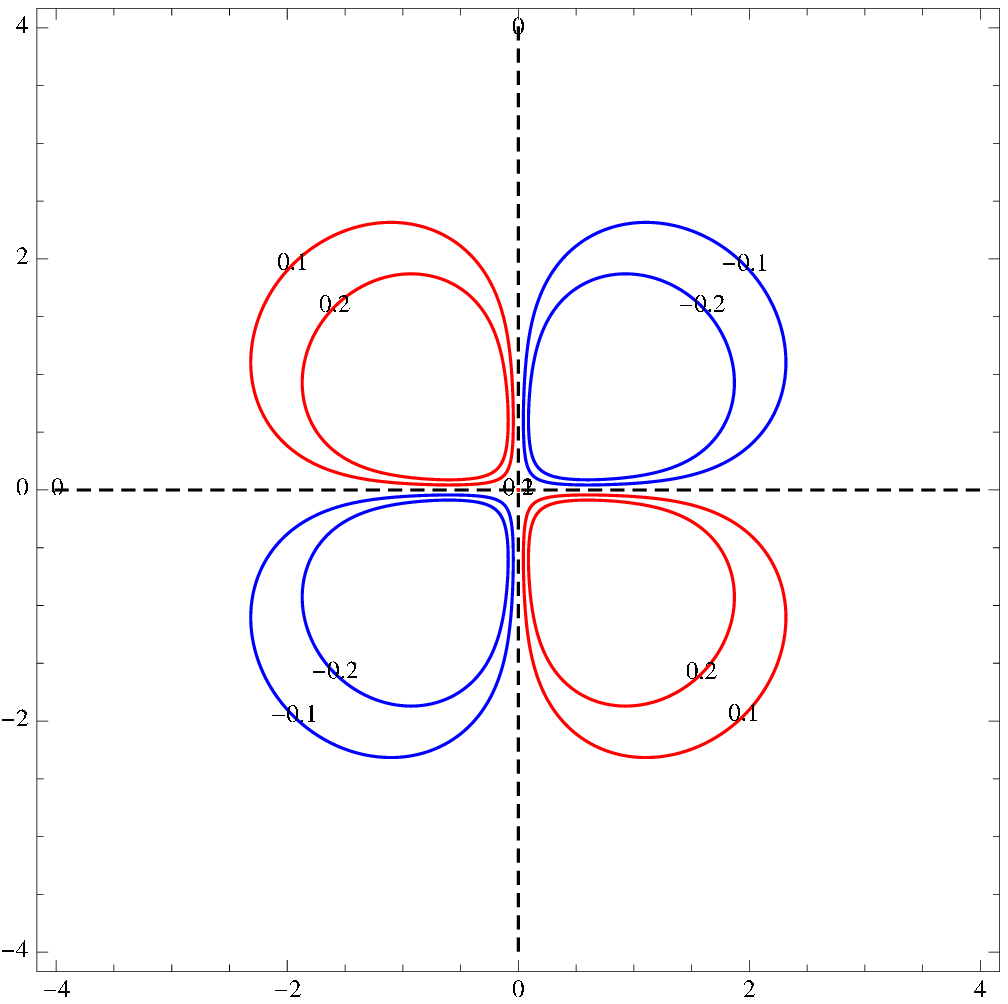} 
\includegraphics[width=6cm]{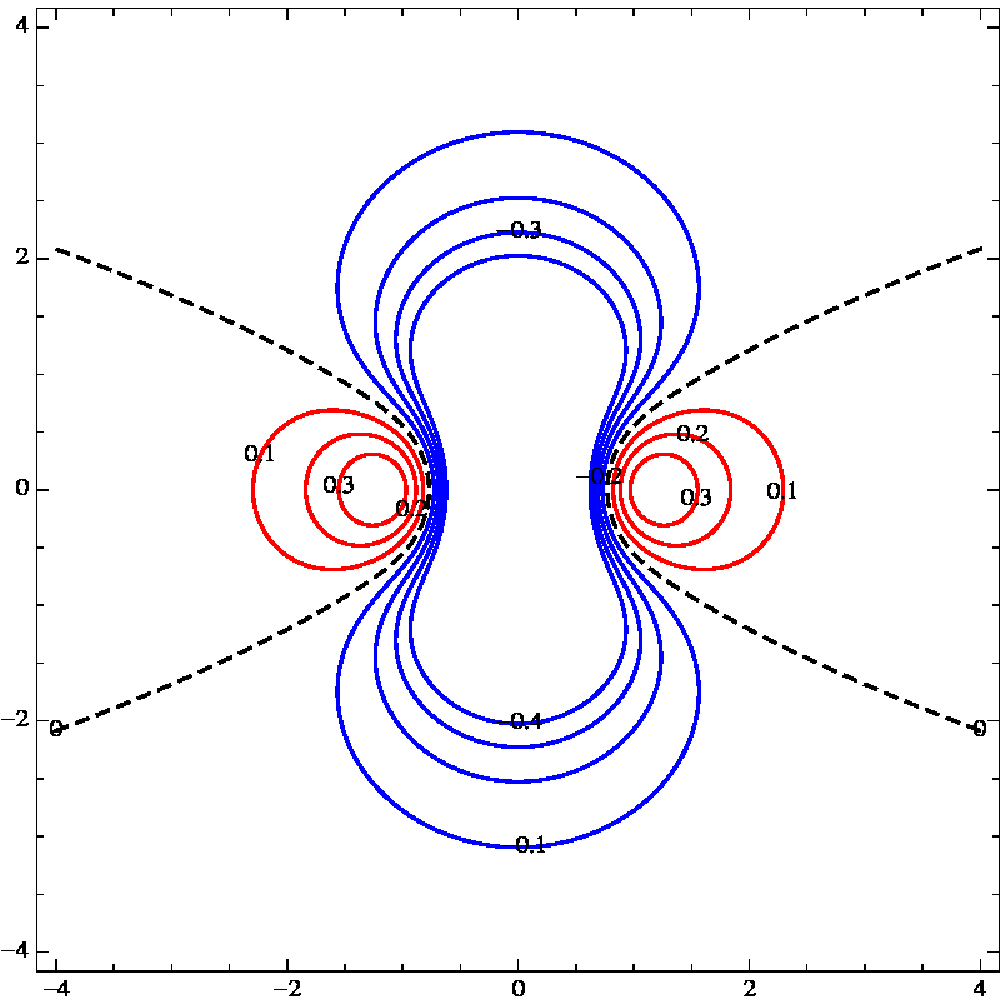}
\caption{The contour plots of the anomalous charge densities of $B=1$ Skyrmion for $\beta=0.263$
on the cross section by the $y=0$ plane. 
The top panel shows $-P_1/2$ with $B_1 \neq 0$ and the bottom panel shows $-P_3/2$ with $B_3 \neq 0$.
The blue lines have positive values and the red ones have negative values. The black broken lines correspond
to zero charge contours.}
\label{fig:B=1_cs}
\end{center}
\end{figure}

To evaluate $P_i$, let us first write down the left- and right-invariant one-forms as
\beq
R_i &=& i\left(\vec \tau \cdot \vec{\hat x}\right)f' \hat x_i + \frac{i}{2}\left(\vec\tau\cdot\p_i\vec{\hat x}\right)\sin 2f
\non
&&
+ \left[ \left(\p_i\vec{\hat x} \cdot \vec{\hat x}\right){\bf 1} + i\left(\p_i\vec{\hat x} \times \vec{\hat x}\right)\cdot\vec \tau\right] \sin^2f,\\
L_i &=&  i\left(\vec \tau \cdot \vec{\hat x}\right)f' \hat x_i + \frac{i}{2}\left(\vec\tau\cdot\p_i\vec{\hat x}\right)\sin 2f
\non
&&
+ \left[ \left(\p_i\vec{\hat x} \cdot \vec{\hat x}\right){\bf 1} - i\left(\p_i\vec{\hat x} \times \vec{\hat x}\right)\cdot\vec \tau\right] \sin^2f.
\eeq
Plugging these into Eq.~(\ref{eq:def_P}), we have
\beq
P_i = -f'\hat x_i \hat x_3 - \frac{1}{2}(\p_i \hat x_3) \sin 2f.
\label{eq:p}
\eeq
The topological charge density, $P_1$ and $P_3$ are shown in Fig.\ \ref{fig:B=1}
 at $m_\pi = 0$.
The induced electric charge densities with nonzero $m_\pi$ (see Fig.~\ref{fig:B=1_cs})
are quite similar to
those for the massless case in Fig.~\ref{fig:B=1}.
A tiny difference comes form the similar 
 but a little different behaviors in 
 profile functions $f(r)$ as shown in Fig.~\ref{fig:profile_hedgehog}.

The spatial integrations of $P_1$ and $P_3$ become
\beq
\int d^3x\ P_1 = 0,\quad
\int d^3x\ P_3 = - \frac{4\pi}{3(e_sF_\pi)^2} c_0,
\eeq
with
\beq
c_0 \equiv  \int d r \, \left( r^2f' +  r\sin 2f\right).
\eeq
Note that the integration variable $r \to  r/(e_s F_\pi)$ is the dimensionless coordinate, so that 
$c_0$ is a dimensionless number.
This means that the net induced charge is zero for $\vec B \propto (1,0,0)$, and $(0,1,0)$ whereas
it is non-zero for $\vec B \propto (0,0,1)$ where non-zero correction appears to
 the Gell-Mann--Nishijima formula.
The numerical coefficient $c_0$ can be rewritten as
\beq
c_0 = \left[ r^2 f \right]_0^\infty + \int d r \, \left(-2rf + r \sin 2f \right)
.
\eeq
The first term is surface contribution which becomes non-zero only at $m_\pi = 0$ due to
 distinct behavior at large $r$ shown in Eq.~(\ref{eq:Asm0}),
 and the second term expresses a pion-cloud effect discussed in Appendix \ref{app:B}.
Computational results of $c_0$ become
\beq
c_0 &= -14.1 + C \qquad (m_\pi=0), \label{eq:c00}\\
c_0 &= -10.2 \qquad (m_\pi = 0.263),
\eeq
where $C$ is asymptotic factor shown in Eq.~(\ref{eq:Asm0}).
Computational results for other values of the pion masses are also 
 summarized in Table \ref{tab:int}.

Finally, we evaluate contributions from the surface term in  Eq.~(\ref{eq:wzwcurrent}). 
To this end, we need to compute 
\beq
W = -\int d^3x\ \epsilon^{ijk}\p_i(\bar A_j P_k).
\eeq
Interestingly, plots of the integrand (density) are quite similar to those in Fig.~\ref{fig:B=1} if we 
choose a gauge
$\bar A_i = B_1 (0,-z/2,y/2)$ or $\bar A_i = B_3(-y/2,x/2,0)$.
The surface term can be evaluated as
\begin{eqnarray*}
W = \int d\Omega_2\left[\hat x_i \epsilon^{ijk} \bar A_j \left(f'\hat x_k \hat x_3 + \frac{(\p_k \hat x_3) \sin 2f}{2}\right)\right]_{r\to\infty}.
\end{eqnarray*}
As shown in Eq.~(\ref{eq:asym_massive}), the profile function for $m_\pi \neq 0$ is exponentially small at large $r$, so that
this integration vanishes. 

Note that in the massless case 
the asymptotic behavior given in Eq.~(\ref{eq:asym_massless}) leads to $W \neq 0$ as
\beq
W = \frac{32\pi iC}{3(e_sF_\pi)^2}B_3.
\eeq
We see that this surface term with the constant $C$ given in Eq.~(\ref{eq:asym_massless}) cannot
cancel the last term of Eq.~(\ref{eq:full_charge}).
Anyway, since the physical pion mass is not zero, we consider the new last term in Eq.~(\ref{eq:full_charge}) as
the anomaly-induced electric charge\footnote{As long as we think of the massless pion as the limit $m_\pi \to 0$,
the surface term contribution can be always ignored.}.

\subsection{Evaluation with quantized Skyrmion}

To evaluate the anomalous current and charge for each baryon state,
the Skyrmion is quantized as a slowly rotating soliton.
Quantizing the collective coordinates of soliton's moduli space $G$
is achieved by the canonical quantization of a particle on a manifold $G$.
In the case of two flavors, $G=SU(2) \simeq S^3$.
We construct the angular momentum operators acting on baryon states and the harmonic functions on $G$
corresponding to baryon wave function.
Using these, we evaluate the expectation values of the anomalous currents.

We evaluate the expectation values $\langle j_\mathrm{anm}^\mu \rangle_B$ and 
the anomalous charge $Q_\mathrm{anm}^B$
for each baryon state $B$ in the presence of a background magnetic field.
We show that the spatial components of the currents vanish: $\langle j_\mathrm{anm}^i \rangle_B =0$.
We also obtain the anomalous charge $Q_\mathrm{anm}$, which is given by integrating
$\langle j_\mathrm{anm}^0 \rangle_B$.

\subsubsection{Angular momentum operators and spherical harmonics on $S^3$}

Let $g \in G$ be a group element of a group manifold $G$.
Then there are operators $\mathcal{L}_a$ and $\mathcal{R}_a$
acting on $g$ from left and right, respectively, and satisfying the commutation relations
\begin{equation}
[\mathcal{L}_a, \mathcal{L}_b] = i f_{abc} \mathcal{L}_c, \ 
[\mathcal{R}_a, \mathcal{R}_b] = i f_{abc} \mathcal{R}_c.
\end{equation}
Here the roman indices correspond to those of the tangent space of $G$,
and $f_{abc}$ is the structure constant of the Lie algebra of $G$:
$[T_a, T_b] = i f_{abc} T_c$ and $\mathrm{Tr} (T_a T_b) = \delta_{ab}/2$.
The actions of $\mathcal{L}_a$ and $\mathcal{R}_a$ on $g$ are
\begin{align}
\mathcal{L}_a g &= - T_a g, \ \mathcal{L}_a g^{-1} = g^{-1} T_a, \nonumber \\
\mathcal{R}_a g &= g T_a, \ \mathcal{R}_a g^{-1} = - T_a g^{-1}.
\label{LR_action_g}
\end{align}
We restrict ourselves to the case $G=SU(2)$, and thus $f_{abc}=\epsilon_{abc}$.
This is the case of the two-flavor Skyrmion.

The operators $\mathcal{L}_a$ and $\mathcal{R}_a$ are precisely the angular momentum operators
with respect to the isometry $SU(2)_L \times SU(2)_R$ of $S^3$.
We introduce the scalar spherical harmonics on $S^3$, $Y_{Jm\tilde{m}}$,
where $J$ is the same magnitude spins of both $SU(2)_L$ and $SU(2)_R$,
and  $m$ and $\tilde{m}$ are the eigenvalues of their third-components, respectively.
The actions of the operators on the spherical harmonics are
\begin{align}
\mathcal{L}^2 Y_{Jm\tilde{m}} &= \mathcal{R}^2 Y_{Jm\tilde{m}} = \sqrt{J(J+1)} Y_{Jm\tilde{m}}, \nonumber \\
\mathcal{L}_\pm Y_{Jm\tilde{m}} &= \sqrt{(J \mp m)(J \pm m +1)} Y_{J(m\pm1)\tilde{m}}, \nonumber \\
\mathcal{R}_\pm Y_{Jm\tilde{m}} &= \sqrt{(J \mp \tilde{m})(J \pm \tilde{m} +1)} Y_{Jm(\tilde{m}\pm1)}, \nonumber \\
\mathcal{L}_3 Y_{Jm\tilde{m}} &= m Y_{Jm\tilde{m}}, \ \mathcal{R}_3 Y_{Jm\tilde{m}} = \tilde{m} Y_{Jm\tilde{m}},
\label{LR_action_harmonics}
\end{align}
where $\mathcal{L}_\pm = \mathcal{L}_1 \pm i \mathcal{L}_2$
and $\mathcal{R}_\pm = \mathcal{R}_1 \pm i \mathcal{R}_2$.

It is convenient to introduce a D-function $D_{ab}(g)$ so as to see
the relation of Eq.~\eqref{LR_action_g} and Eq.~\eqref{LR_action_harmonics}.
It is defined by the adjoint action of $g$,
\begin{equation}
g T_a g^{\dagger} = T_b D_{ba}(g).
\end{equation}
The action of $\mathcal{L}_a$ and $\mathcal{R}_a$ on $D_{ab}(g)$ becomes
\begin{align}
\mathcal{L}_c D_{ab}(g) &= i \epsilon_{cad} D_{db}(g), \nonumber \\
\mathcal{R}_c D_{ab}(g) &= - i \epsilon_{cdb} D_{ad}(g).
\label{LR_action_D}
\end{align}
With a little more algebra,
it can be shown that appropriate linear combinations of $D_{ab}(g)$ precisely give
the harmonic functions $Y_{Jm\tilde{m}}$.

Note that $\mathcal{L}_a$ and $\mathcal{R}_a$ are respectively the isospin and the spin operators for baryon states.
See, for instance, Ref.~\cite{Mazur:1984yf} for a discussion in three-flavor case.
The adjoint action of $g$ to the hedgehog Skyrmion gives
$U(\boldsymbol{x})=g U_0(\boldsymbol{x}) g^\dagger = U_0 ( \boldsymbol{x}^\mathrm{rot})$.
That is, $g \boldsymbol{\hat{x}\cdot\tau} g^\dagger = \boldsymbol{\hat{x}}^\mathrm{rot}\boldsymbol{\cdot\tau}$.
The transformation of the unit vector $\boldsymbol{\hat{x}}$ under the spatial rotation caused by $D(g)$ is
\begin{equation}
\hat{x}_a \to \hat{x}_a^\mathrm{rot} = D_{ab}(g) \hat{x}_b.
\label{x_rot_d_function}
\end{equation}
It is natural to identify $SU(2)_R$ as the baryon spin, where $g \to gk_R$ with $k_R \in SU(2)_R$.

\subsubsection{Absence of the spatial anomalous current}

To evaluate $\langle j_\mathrm{anm}^i \rangle_B$ in the presence of a background magnetic field,
we need to focus only on $P_0$ owing to the index structure of the WZW term.
We first write down $P_0$ and then quantize it.
Substituting a slowly rotating Skyrmion $U(\boldsymbol{x},t) = g(t) U_0(\boldsymbol{x}) g^\dagger(t)$ for $P_0$ \eqref{eq:def_P},
we obtain
\begin{equation}
P_0 = 2 \sin (2f) \epsilon_{ac3} D_{ab}(g) \hat{x}_b \mathrm{Tr} [\tau_c \dot{g} g^\dagger].
\end{equation}
In the procedure of the canonical quantization,
the time-derivative part is replaced with the angular momentum operator as follows \cite{Zahed:1986qz}:
\begin{align}
\mathcal{L}_a &= i \Lambda \mathrm{Tr} [\tau_a \dot{g} g^\dagger], \nonumber \\
\Lambda &= \frac{8 \pi}{3} \int dr r^2 \sin^2 f \left[1+4\left( f^{\prime 2} + \frac{\sin^2 f}{r^2}\right)\right].
\end{align}
Hence $P_0$ can be written in terms of raising- and lowering-operators,
\begin{align}
P_0
= -\frac{1}{\sqrt{3} \Lambda} \sin(2f) & [ \hat{x}_{+} (Y_{1--} \mathcal{L}_{+} + Y_{1+-} \mathcal{L}_{-}) \nonumber \\
& - \hat{x}_{-} (Y_{1-+} \mathcal{L}_{+} + Y_{1++} \mathcal{L}_{-}) \nonumber \\
& + \hat{x}_{3} (Y_{1-0} \mathcal{L}_{+} + Y_{1+0} \mathcal{L}_{-}) ]_\mathrm{Weyl},
\label{eq_P_0_raising_lowering_harmonics}
\end{align}
where the indices $\pm$ in $Y_{Jm\tilde{m}}$ mean $\pm1$, and $\hat{x}_\pm = \hat{x}_1 \pm i\hat{x}_2$.
The Weyl ordering for the operators is understood.

Integrating a product of three spherical harmonics over $S^3$ gives
\begin{align}
&\int \frac{d \Omega_3}{2 \pi^2} (Y_{J_1 m_1 \tilde{m}_1})^\ast Y_{J_2 m_2 \tilde{m}_2} Y_{J_3 m_3 \tilde{m}_3} \nonumber \\
&= \sqrt{\frac{(2J_2+1)(2J_3+1)}{2J_1+1}} C^{J_1 m_1}_{J_2 m_2 J_3 m_3} C^{J_1 \tilde{m}_1}_{J_2 \tilde{m}_2 J_3 \tilde{m}_3},
\label{three_harmonics_integral}
\end{align}
where $C^{J_1 m_1}_{J_2 m_2 J_3 m_3}$ is a Clebsch-Gordan coefficient of $SU(2)$.
The wave function of each baryon state is given by $Y_{Jm\tilde{m}}$.
Our primary interest is in nucleons ($I=J=1/2$) and $\Delta$ baryons ($I=J=3/2$), but here we can keep $J$ arbitrary.
We use Eq.~\eqref{three_harmonics_integral} to evaluate $P_0$ projected onto each baryon state.

In Eq.~\eqref{eq_P_0_raising_lowering_harmonics}, we need to focus only on the last line:
\begin{align}
\mathcal{O}_W
&\equiv (Y_{1-0} \mathcal{L}_{+} + Y_{1+0} \mathcal{L}_{-})_\mathrm{Weyl} \nonumber \\
&= Y_{1-0} \mathcal{L}_{+} + Y_{1+0} \mathcal{L}_{-} + \sqrt{2} Y_{100}.
\label{operator_Ow_spatial_current}
\end{align}
It is easily seen that each of the first and the second lines
in Eq.~\eqref{eq_P_0_raising_lowering_harmonics} gives no contribution.
Integrating Eq.~\eqref{operator_Ow_spatial_current} by using Eq.~\eqref{three_harmonics_integral}
along with baryon states of quantum numbers $(J,I_3,S_3)$ for (iso)spin and the third components,
we obtain
\begin{align}
&\langle \mathcal{O}_W \rangle_B
= \int \frac{d \Omega_3}{2 \pi^2} (Y_{J I_3 -S_3})^\ast \mathcal{O}_W Y_{J I_3 -S_3} \nonumber \\
&= \sqrt{3} C^{J -S_3}_{10 \, J -S_3} \left( \sqrt{(J-I_3)(J+I_3+1)} C^{JI_3}_{1-1 \, J(I_3+1)} \right. \nonumber \\
& \quad \left.+ \sqrt{(J+I_3)(J-I_3+1)} C^{JI_3}_{11 \, J(I_3-1)} + \sqrt{2} C^{JI_3}_{10 \, JI_3} \right),
\end{align}
where the minus sign appearing in front of $S_3$ is due to Eq.~\eqref{LR_action_D}.
This exactly vanishes once values of the Clebsch-Gordan coefficients are substituted \cite{Varshalovich:1988ye}:
\begin{align}
& C^{J \gamma}_{10 \, J \gamma} = \frac{\gamma}{\sqrt{J(J+1)}}, \nonumber \\
& C^{J \gamma}_{1\pm1 \, J (\gamma\mp1)} = \mp\sqrt{\frac{(J\pm\gamma)(J\mp\gamma+1)}{2J(J+1)}}.
\label{CG_coefficients}
\end{align}
Thus we see $\langle j_\mathrm{anm}^i \rangle_B =0$.

\subsubsection{Anomalous electric charge in baryons}

In $j_\mathrm{anm}^0$,
rotation of the Skyrmion is encoded in $\hat{x}_3^\mathrm{rot}$.
Hence, it is sufficient to evaluate $\langle \hat{x}_3^\mathrm{rot} \rangle_B$,
which directly leads us to $\langle j_\mathrm{anm}^0 \rangle_B$.
Since this part does not contain derivatives in time, 
we simply integrate Eq.~\eqref{x_rot_d_function} by using Eq.~\eqref{three_harmonics_integral}.
The result is
\begin{equation}
\langle \hat{x}_3^\mathrm{rot} \rangle_B = -\frac{I_3 S_3}{J(J+1)} \hat{x}_3.
\label{x3rot_for_arbitrary_spin_isospin}
\end{equation}
Below we will mainly focus on the case $J=1/2$ for simplicity.
However, thanks to Eq.~\eqref{x3rot_for_arbitrary_spin_isospin}, it is straightforward to consider higher-spin cases.
For instance,
this gives $\langle \hat{x}_3^\mathrm{rot} \rangle^{N} = -4 I_3 S_3 \hat{x}_3 /3$ for nucleons,
and $\langle \hat{x}_3^\mathrm{rot} \rangle^{\Delta} = - 4 I_3 S_3 \hat{x}_3/15$ for $\Delta$ baryons.

Let us calculate the total electric charge from the anomalous effect. 
The matrix elements of $P_\mu$ are evaluated by applying Eq.~\eqref{x3rot_for_arbitrary_spin_isospin},
\begin{align}
&\langle P_0\rangle^{N}_{I_3,S_3}=0\label{P0},\\
&\langle P_a\rangle^{N \ a=1,2}_{I_3,S_3}=-\frac{16i}{3}I_3 S_3  \left( f^\prime -\frac{\sin(2f)}{2r}\right) \hat x_a\hat x_3, 
\label{Pa} \\
&\langle P_3\rangle^{N}_{I_3,S_3}=-\frac{16i}{3}I_3 S_3 \left[ \left( f^\prime -\frac{\sin(2f)}{2r}\right) \hat x_3^2+\frac{\sin(2f)}{2r}\right]. \label{P3}
\end{align}
The anomalous charge density under a constant magnetic field $\bf B$ is indeed induced in nucleons: 
\begin{align}
\langle j^0_{\rm anm}\rangle_{I_3,S_3}^{N}=\frac{ie^2N_c}{48\pi^2}B_i\langle P_i\rangle_{I_3,S_3}^{N}.
\label{chargedensity}
\end{align}
The integration of $\langle P_i \rangle^{N}_{I_3,S_3}$ over the whole space 
yields
\begin{align}
\int d^3x \langle P_i \rangle^{N}_{I_3,S_3}&=
\left\{
\begin{array}{ll}
0 & (i=1,2), \\
-\dfrac{16\pi i}{9}(4I_3 S_3)c_0 & (i=3),
\end{array}
\right.
\nonumber
\end{align}
where $c_0=\int dr \{r^2 f^\prime+r \sin(2f)\}$.
Numerical value of $c_0$ will be shown in Table \ref{tab:int} for several pion masses. 
From Eq.~({\ref{chargedensity}), we obtain the anomalous charge for nucleons  
\begin{align}
Q^{N}_{\rm anm}
=\frac{4e^2N_c}{27\pi}I_3 S_3\frac{c_0 B_3}{(e_sF_\pi)^2}.
\label{netcharge}
\end{align} 
In this final expression we restored the rescaling factor $e_s F_\pi$ by a dimensional counting.
Equation (\ref{netcharge}) shows that an electric charge is actually induced by the anomalous effect 
{\it even for a neutron}. 

As seen from Eq.~\eqref{x3rot_for_arbitrary_spin_isospin},
dividing the result in Eq.~\eqref{netcharge} by a factor of 5 gives the anomalous charge of $\Delta$ baryons.

The plot of the charge density of $\langle j_{\rm anm}^0 \rangle$ for the quantized Skyrmion shows exactly
the same as Fig.~\ref{fig:B=1}. For a magnetic field along $x^1$ direction, the charge density plot is
symmetric, thus the total charge vanishes. However, obviously multipoles, in particular a quadrupole, may show up.
In the next section, we calculate multipoles in $\langle j_{\rm anm}^0 \rangle$.

\begin{figure}[tbp]
  \includegraphics[width=85mm]{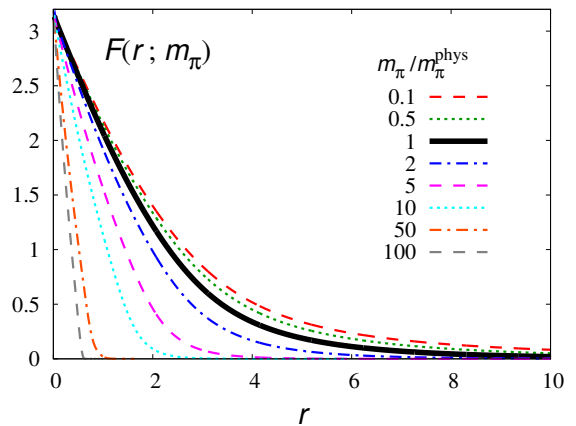}
  \caption{Behavior of the Skyrmion profile functions, $f(r)$, 
  for several pion masses.}
  \label{fig:f}
\end{figure}

\begin{figure}[tbp]
  \begin{center}
    \includegraphics[width=85mm]{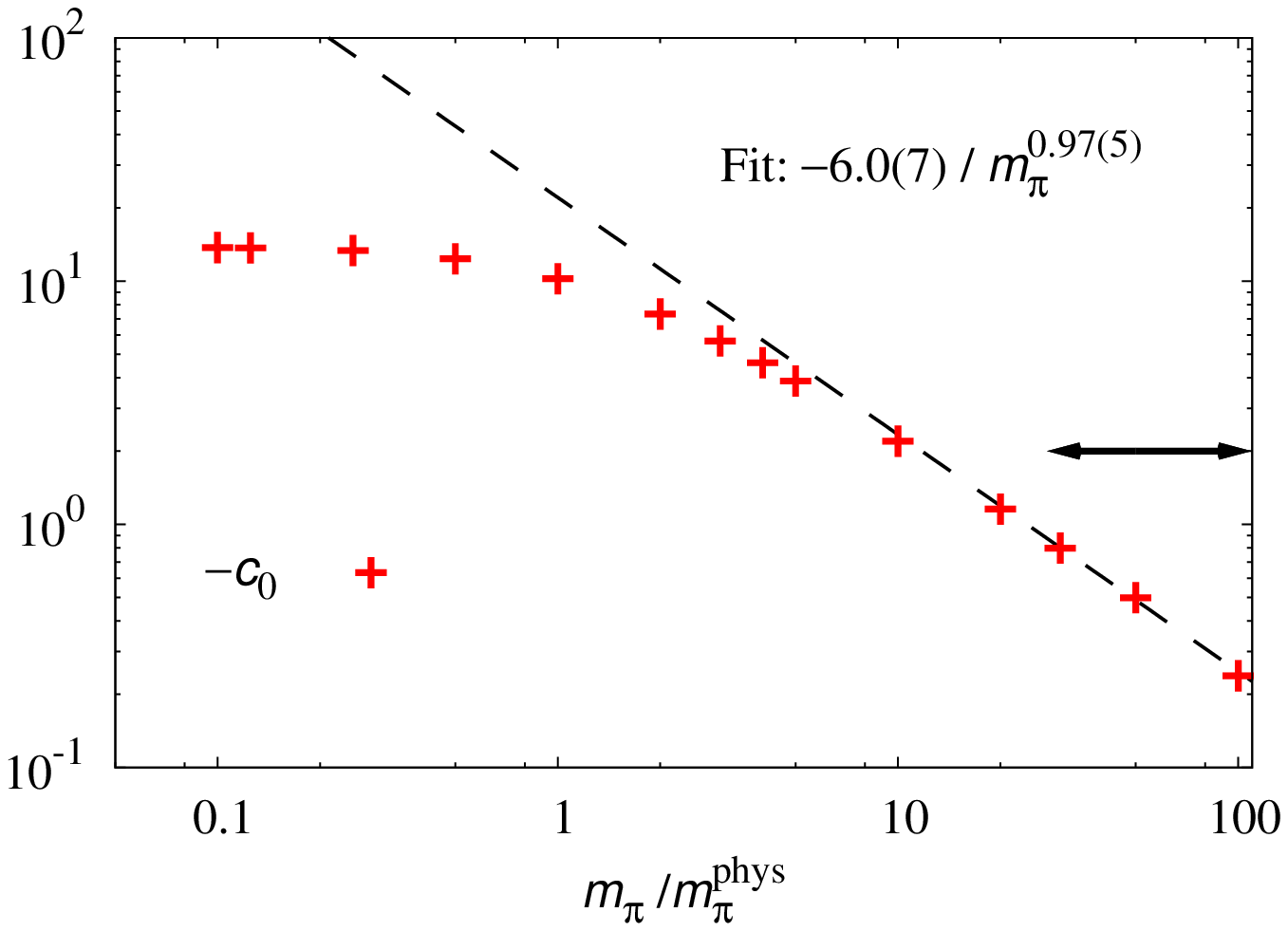}
    \includegraphics[width=85mm]{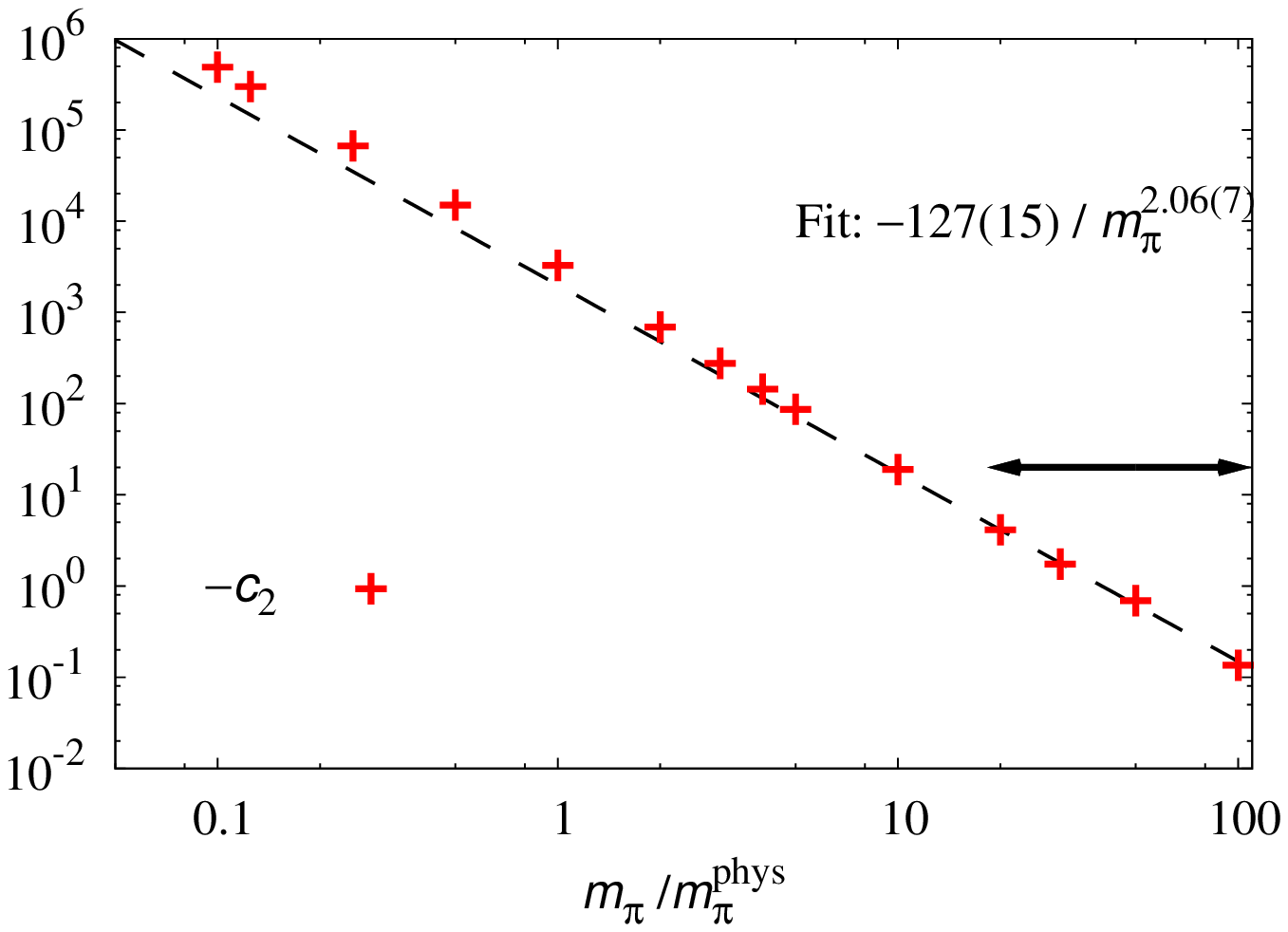}
    \caption{Results of the numerical coefficients, $-c_0$ (top) and $-c_2$ (bottom),
     as a function of $m_\pi / m_\pi^{\rm phys}$ in log-log scale.
    The dotted lines mean fit results by a function $A/m_\pi^n$ with $A$ and $n$
     being free parameters in the fit ranges shown by the arrows on the figures.
     Numerical results of the fit are also shown on the figures.}
    \label{fig:c0c2}
  \end{center}
\end{figure}

\section{Multipole moments of anomalous charges and pion-mass dependence}
\label{sec:4}

When one regards charged baryons as point-like particles,
 multipole moments are suitable physical quantities to describe
  an original charge distribution.
In this section, we extend the calculations to the higher multipole moments 
 due to the anomalous-charge distributions, 
  and estimate pion-mass dependence of the anomalous charges and the multipole moments.

First, we can easily find that the dipole moment due to the anomalous charge
 vanishes:
\begin{align}
D_i \equiv \int d^3x \, x_i \, \langle j^0_{\rm anm}\rangle^N = 0 \ \ \ \ (i=1,2,3).
\end{align}
On the other hand, the quadrupole moment:
\begin{align}
 Q_{ij} &\equiv 
 \int d^3 x \, \left( 3x_i x_j - r^2 \delta_{ij} \right) 
 \langle j^0_{\rm anm} \rangle^N
,
\end{align}
is calculated for the nucleon as
\begin{align}
Q_{ij} &=
 e \, \frac{2N_c}{135\pi} (I_3 S_3)
 \tilde{Q}_{ijk}\frac{eB_k}{(e_sF_\pi)^4} 
 \, c_2 
, \label{eq:qud} \\
\tilde{Q}_{ijk} &\equiv
\begin{pmatrix}
 -2\delta_{k3} & 0             & 3\delta_{k1} \\
 0             & -2\delta_{k3} & 3\delta_{k2} \\
 3\delta_{k1}  & 3\delta_{k2}  & 4\delta_{k3}
\end{pmatrix}_{ij}
,
\end{align}
where the numerical coefficient,
\begin{align}
c_2=\int dr \left[ 2 r^4 f^\prime - r^3 \sin(2f) \right],
\end{align}
 is shown in Tab.~\ref{tab:int} for several pion masses.
 This means that the leading multipole due to the anomalous contribution
  is the quadrupole moment.
We note that the quadrupole is induced in response to all directions of
 the external magnetic fields,
  although the anomalous charge is induced only by $B_3$
   (see Eq.~\eqref{netcharge}).

In order to extract the pion-mass dependence of the anomalous charge
 and the quadrupole moment, we calculate the Skyrmion profile function $f(r)$
  for wide pion-mass range ($ 0.1 \le m_\pi/m_\pi^{\rm phys} \le 100$).
Behavior of $f(r)$ for several pion masses is shown in Fig.~\ref{fig:f},
 where the solid line is $f(r)$ at $m_\pi = m_\pi^{\rm phys}$.
We find that the wave function shrinks with the pion mass increasing.

Since the pion mass dependence of the anomalous charge and the quadrupole moment 
 appears
 in the numerical coefficients, $c_0$ and $c_2$, via $r$ integration with $f(r;m_\pi)$,
 we focus on these coefficients.
Figure \ref{fig:c0c2} shows results of $- c_0$ (top) and $- c_2$ (bottom)
 as a function of $m_\pi / m_\pi^{\rm phys}$ in log-log scale.
Numerical values of $c_0$ and $c_2$ are also summarized in Tab.~\ref{tab:int}. 
We can see that $c_0$ becomes almost plateau at small pion-mass 
 ($m_\pi/m_\pi^{\rm phys} < 1$),
 whereas that decreases linearly at large pion-mass 
 ($m_\pi/m_\pi^{\rm phys} > 10$).
We fit the results by a function, $A/m_\pi^n$, with $A$ and $n$ being
 free parameters, and obtain $c_0 \sim - 6.0(7) / m_\pi^{0.97(5)}$ in a range
  of $30 \le m_\pi / m_\pi^{\rm phys} \le 100$,
  shown by the dashed line on Fig.~\ref{fig:c0c2} (top).
This implies that the pion-mass dependence of the anomalous charge
 is  $Q_{\rm anm} \propto 1/m_\pi$ at large pion-mass.

On the other hand, $c_2$ behaves linearly for all pion-mass region.
We also fit the results by $A/m_\pi^n$, and obtain
 $c_2 \sim - 127(15) / m_\pi^{2.06(7)}$ in a range of 
  $20 \le m_\pi / m_\pi^{\rm phys} \le 100$,
   shown by the dashed line on Fig.~\ref{fig:c0c2} (bottom).
Although the fit is performed at large pion-mass region,
 almost all results of $c_2$ is located around the dashed line.
This implies that the quadrupole moment due to the anomaly
 behaves as $Q_{ij} \propto 1/m_\pi^2$.

Note that we have evaluated $j^\mu_{\rm anm}$ which is
only a part of the total electromagnetic current.\footnote{It is possible that 
the last term in \eqref{eq:wzwcurrent} gives additional multipoles
although it is negligible for the total charge. However, the term itself is
not gauge-invariant and once combined with the baryon number
term (the second term in \eqref{eq:wzwcurrent}) it becomes gauge-invariant. 
In this paper we evaluate only the gauge-invariant $j^\mu_{\rm anm}$ 
for the quadrupole moment, and the other terms (which can be evaluated
if a back-reaction to the Skyrmion profile can be computed) 
are left for our future work.}

\begin{table}[tbp]
 \begin{center}
  {\renewcommand{\arraystretch}{1.2} \tabcolsep = 2mm
 \newcolumntype{.}{D{.}{.}{6}}
 \begin{tabular}{.ll}
 \hline\hline
 \multicolumn{1}{c}{$m_\pi / m_\pi^{\rm phys}$} &
 \multicolumn{1}{c}{$c_0$} & 
 \multicolumn{1}{c}{$c_2$} \\ 
 \hline
 \multicolumn{1}{.}{0} &
 \multicolumn{1}{l}{$-$1.41$\times 10^{ 1}$} & 
 \multicolumn{1}{c}{$-\infty$} \\ 
     0.1   &      $-$1.37$\times 10^{ 1}$ & $-$4.90$\times 10^{ 5}$    \\
     0.125 &      $-$1.37$\times 10^{ 1}$ & $-$2.99$\times 10^{ 5}$    \\
     0.25  &      $-$1.34$\times 10^{ 1}$ & $-$6.68$\times 10^{ 4}$    \\
     0.5   &      $-$1.23$\times 10^{ 1}$ & $-$1.50$\times 10^{ 4}$    \\
     1     &      $-$1.02$\times 10^{ 1}$ & $-$3.27$\times 10^{ 3}$    \\
     2     &      $-$7.32                 & $-$6.91$\times 10^{ 2}$    \\
     3     &      $-$5.67                 & $-$2.76$\times 10^{ 2}$    \\
     4     &      $-$4.62                 & $-$1.44$\times 10^{ 2}$    \\
     5     &      $-$3.88                 & $-$8.63$\times 10^{ 1}$    \\
    10     &      $-$2.20                 & $-$1.89$\times 10^{ 1}$    \\
    20     &      $-$1.15                 & $-$4.14                    \\
    30     &      $-$7.97$\times 10^{-1}$ & $-$1.74                    \\
    50     &      $-$4.99$\times 10^{-1}$ & $-$6.90$\times 10^{-1}$    \\
   100     &      $-$2.38$\times 10^{-1}$ & $-$1.35$\times 10^{-1}$    \\
 \hline\hline
 \end{tabular}}
 \end{center}
 \caption{Numerical results of the coefficients $c_0$ and $c_2$.
 We neglect the surface contribution at $m_\pi =0$ shown in Eq.~(\ref{eq:c00}) due to
  distinct behavior of the Skyrmion profile function $f(r)$ at $r \rightarrow \infty$, 
  i.e. $f(r) \sim 1/r^2$, discussed in Sec.~\ref{sec:2B}.}
 \label{tab:int}
\end{table}

\section{No contribution from other corrections}
\label{sec:5}

In this section, we study other effects of the background magnetic field to the 
the electric charge of the nucleon. 
Our aim is to show that the anomaly-induced electric charge is
not cancelled by the other electromagnetic effects. 

The total electric charge is written as a modified Gell-Mann--Nishijima formula,
\begin{align}
Q_e = e I_3 + \frac{e N_B}{2} + \frac{Q_{\rm anm}}{2}.
\label{GNmod}
\end{align}
The first term stems from the electromagnetic current in the original Skyrme
model. The second term is due to the baryon number coupling to the electromagnetic 
potential. The last one is the anomaly-induced electric charge which is nonzero 
only when we have the background electromagnetic field.

We are working with the perturbative expansion with respect to the background
magnetic field $eB$. What we found for the anomaly-induced charge is 
\begin{align}
Q_{\rm anm} = {\cal O}(e^2 B).
\end{align}
In the total charge formula \eqref{GNmod}, the second term is due to the baryon
charge which is a topological charge for the Skyrme model, thus not corrected
by the background magnetic field. On the other hand, the first term can be 
corrected in the presence of $B$. 
If a correction
of the order ${\cal O} (eB)$ appears from the $I_3$ term in the charge formula,
then it may possibly cancel our anomaly-induced charge. 
In the following, we shall present an argument 
showing that
there is no such correction of ${\cal O}(eB)$ to the $I_3$ term. 

First, let us examine if there is a correction to the  electromagnetic current itself in the Skyrme
model. One may naively think that, since the Skyrme solution itself is corrected by the 
background electromagnetic field, the current may also be corrected. However, this is not
the case for the Skyrmion. The reason is that for the Skyrmion, the electromagnetic $U(1)$ 
is identical to a part of the isospin, and the action itself has the isospin structure from the first
place. In fact, the relevant $I_3$ is indeed expressed by a part of the flavor $SU(2)$ rotation,
and thus the current $I_3$ is universally expressed as
\begin{align}
I_3 = \frac{i}{2}
\left[
a_0 \frac{\partial}{\partial a_3} - a_3 \frac{\partial}{\partial a_0} - a_1 \frac{\partial}{\partial a_2}
+a_2 \frac{\partial}{\partial a_1}
\right]
\end{align}
Here, there is no room for $eB$ to show up, thus we can safely use this expression for
$I_3$ in the electric charge formula \eqref{GNmod}.

Then, the issue is whether the expectation value of $I_3$ in the background magnetic field is
corrected or not. The background magnetic field modifies the wave function of the Skyrmion,
so in principle this corrected wave function may give a correction to $\langle I_3 \rangle$,
which is of importance for us. We shall show in the following that there is no such correction
at ${\cal O}(eB)$.

To proceed, we need to know how the Skyrme moduli wave function is corrected. 
Due to the background magnetic field, there appears a potential in the moduli space,
then two of the moduli parameters are lifted to become pseudo-moduli parameters.
The corrected quantum mechanics of the moduli and the pseudo-moduli is
written as 
\begin{align}
S = 2 \lambda \sum_{i=0}^{3}
\left[
\left(\dot{a}_i\right)^2 
\right]- e B_3 V(\vec{a})
\label{qmcor}
\end{align}
where the potential of the quantum mechanics is of the form
$V(\vec{a}) = \left((a_1)^2 + (a_2)^2\right)\tilde{V}(\vec{a})$.
The function $\tilde{V}(\vec{a})$ is a polynomial of $a_i$ (with a finite order), with
just numerical coefficients.
The potential $V(\vec{a})$ breaks the $SU(2)$ symmetry of the system down to the diagonal $U(1)$.

We briefly explain how to derive the form (\ref{qmcor}) of the induced potential. In the Skyrme model,
the electromagnetic interaction enters as
\begin{align}
\hat{R}_\mu \equiv D_\mu U U^\dagger, \quad
D_\mu U \equiv \partial_\mu + i e A_\mu [q,U].
\end{align}
For a background magnetic field $B_3$, we consider $A_1 = -B_3 x^2$, thus among $\hat{R}_\mu$
the electromagnetic contribution appears only in $\hat{R}_1$ as
\begin{align}
& \hat{R}_1 = \partial_1 U U^\dagger + \delta \hat{R}_1,
\\
& \delta \hat{R}_1 \equiv
- i e B_3 x^2 [q, G U_0 G^\dagger] G U_0^\dagger G^\dagger.
\end{align}
Here we have already substituted the Skyrme solution.
Assuming that $G$ is dependent on $t$ and 
plugging this $\hat{R}$ into the action, we obtain a correction to the Skyrme Lagrangian 
at ${\cal O}(eB)$ as
\begin{align}
\delta L = \frac{F_\pi^2}{8} 
{\rm Tr} [R_1  \delta \hat{R}_1]
+ \frac{1}{8 e_s^2} {\rm Tr}
[R_\mu, R_1][R^\mu, \delta \hat{R}_1].
\end{align}
Since we know that the Skyrme solution has the particular dimension dependence 
$x \rightarrow x/(e_s F_\pi)$, we obtain
\begin{align}
\delta S = \int \! d^4x  \; \delta L
= \int \! dt \frac{1}{e_s^3 F_\pi} e B_3 V(\vec{a}),
\end{align}
where $V(\vec{a})$ is a polynomial in $a_i$, with only dimensionless numerical coefficients.  This is 
nothing but the potential in Eq.~(\ref{qmcor}).
As the potential $V(\vec{a})$ should vanish when $G$ corresponds to the electromagnetic
direction, {\it i.e.} the $\tau_3$ direction, $V(\vec{a})$ is proportional to $(a_1)^2+(a_2)^2$.

With the potential, the Skyrme wave function $\psi(\vec{a})$ is modified. We may apply a well-known
perturbation technique for quantum mechanics, and obtain the corrected nucleon wave function as 
\begin{align}
| l=1/2\rangle  \; =\;  &  \; |l= 1/2\rangle_0 \nonumber \\
& + e B_3 \sum_{n=1}^{\infty} \frac{V_{l=1/2, \; l=n+1/2}}{E_{l=1/2}-E_{l=n+1/2}}
|l=n+1/2\rangle_0
\nonumber \\
& + {\cal O}((eB)^2).
\end{align}
Here $V_{l=1/2,\;l=n+1/2}$ is the matrix element of the operator $V$ appearing in the
quantum mechanics (\ref{qmcor}), and the state with subscript $0$ is the one without the perturbation. 
In the current case the states have degenerate energy,
but the expression above is universal.

Now, using this corrected wave function, we evaluate the expectation value of $I_3$.
Since we have 
\begin{align}
{}_0\langle l=n\!+\!1/2\; | I_3 |\; l\!=\!1/2\!\;\rangle_0 = 0
\end{align}
for $n\geq 1$, we obtain
\begin{align}
\langle I_3 \rangle = \langle I_3 \rangle_0 + {\cal O}((eB)^2),
\end{align} 
where  $\langle I_3 \rangle_0$ is the third component of the isospin of the leading
(uncorrected) order wave function.
Therefore, the electromagnetic correction to the charge formula starts at ${\cal O}((eB)^2)$, 
which is at higher order compared to the anomaly-induced charge $Q_{\rm anm}$.
This means that our anomaly-induced charge $Q_{\rm anm}$ is the leading-order correction
of ${\cal O}(eB)$, 
and cannot be cancelled by the other effect of the background magnetic field.

Here we have presented an argument that the total induced charge due to the anomaly is not cancelled by
other corrections due to the magnetic field. This argument may be reenforced and supplemented by
an explicit computation of a back-reaction to the Skyrme configuration itself, from the magnetic field.
The calculation of the back-reaction is quite complicated, so we leave it to our future work.

\section{Higher-charge Skyrmions}
\label{sec:6}

In this section, we study classical higher-charge Skyrmions and the anomaly-induced charges.
To this end, we will utilize the so-called rational map ansatz \cite{Houghton:1997kg} which is a reasonable method
giving a good approximation. 
In this section we use another notation based on a standard textbook \cite{Manton:2004tk}.
A main difference from the previous sections is the dimensionless coordinate
\beq
x_\mu \to \frac{2x_\mu}{e_s F_\pi},\qquad
\p_\mu \to \frac{e_s F_\pi\p_\mu}{2}.
\eeq
Let us first give a brief review on the rational map ansatz.
A solution of the Skyrmion $U(\bi x)$ with $U(\bi x) \to {\bf 1}$ as $|\bi x| \to \infty$
gives a map from ${\bf R}^3 + \{\infty\} \simeq S^3$ to $SU(2) \simeq S^3$.
The map is characterized by the homotopy group $\pi_3(SU(2)) = {\bf Z}$. 
More explicitly, it can be expressed as
\beq
U(\bi x) = \exp\left(
i f_B(r) \vec \tau \cdot \vec n(\theta,\phi)
\right),
\label{eq:ansatz}
\eeq
where $\{f_B,\bi n\}$ is a coordinate of $SU(2)$
under a constraint $f_B \in [0,\pi]$ and $|\vec n| = 1$:
Namely, we decompose $SU(2)$ into $I_{[0,\pi]} \times S^2$.
The parameters $(r,\theta,\phi)$ are standard spherical coordinates on
the space ${\bf R}^3 \simeq {\bf R}_{\ge 0} \times S^2$.
In order to get the map of degree $N_B \in {\bf Z}$, we assume that $f_B$ is a one-to-one
map from 
$R_{\ge 0} \to I_{[0,\pi]}$.
Then $\vec n (\theta,\phi)$ should give a map $S^2 \to S^2$ of degree $N_B$.
Let us introduce the stereographic projection which is useful to find the generic map of degree $N_B$,
\beq
z(\theta,\phi) = e^{i\phi}\tan \frac{\theta}{2}.
\eeq
For example, the $N_B=1$ hedgehog ansatz (the one-to-one map from $S^2$ to $S^2$)
can be expressed by
\beq
\vec n (z,\bar z) 
= \left( 
\frac{z+\bar z}{1+|z|^2}, 
-i\frac{z-\bar z}{1+|z|^2}, 
\frac{1-|z|^2}{1+|z|^2}
\right).
\eeq
This can be easily extended to a $N_B$-to-one map from $S^2$ to $S^2$ by replacing $z$ with
any rational maps $w(z): {\bf C}$ to ${\bf C}$,
\beq
w(z) = \frac{P(z)}{Q(z)}.
\eeq
Here $P(z)$ and $Q(z)$ are holomorphic functions in $z$ and we set $N_B = \max\{\deg P,\deg Q\}$.
Thus, we obtain the map from $S^2$ to $S^2$ of degree $N_B$,
\beq
\vec n  
= \left( 
\frac{w+\bar w}{1+|w|^2}, 
-i\frac{w-\bar w}{1+|w|^2}, 
\frac{1-|w|^2}{1+|w|^2}
\right).
\eeq
Plugging this into Eq.~(\ref{eq:ansatz}), we reach the map from $S^3$ to $SU(2)$ of
degree $N_B$. This is called the rational map ansatz.

The baryon number can be expressed by 
\beq
\!\!N_B \!=\! -\! \int\!\! \frac{f_B'}{2\pi^2} \left(\frac{\sin f_B}{r} \frac{ 1 + |z|^2}{1+|w|^2} \right)^{\!\!\!2} \left| \frac{dw}{dz}\right|^2
r^2 dr d\Omega_z,
\label{eq:b-num}
\eeq
where $d\Omega_z$ is the usual area element on a 2-sphere
\beq
d\Omega_z = \frac{2 i dz d\bar z}{(1 + |z|^2)^2} = \sin \theta d\theta dr.
\eeq
By making use of the following pull-back
\beq
\left(\frac{ 1 + |z|^2}{1+|w|^2} \right)^2 \left| \frac{dw}{dz} \right|^2  d\Omega_z
= d\Omega_w, 
\eeq
it is easy to change the integral area over $S^2$ to the target space of the rational map $S^2$ as
\beq
&& \text{r.h.s of Eq.(\ref{eq:b-num})} = - \int \frac{f_B'}{2\pi^2} \sin^2 f_B \ 
dr d\Omega_w 
\non
&& \quad = - \frac{2N_B}{\pi} \int_\pi^0 \sin^2 f_B\ df_B = N_B,
\eeq
where we have used $\int d\Omega_w = 4\pi N_B$.
The Skyrmion energy in the $F_\pi/(4e_s)$ energy unit and $2/(e_s F_\pi)$ length unit
with the rational map ansatz is given by
\beq
E = 4\pi \int ^\infty_0 dr \bigg[
r^2 f_B'{}^2  +  2 N_B (f_B'{}^2 + 1)\sin^2 f_B \non
+ I \frac{\sin^4 f_B}{r^2} + 8 m_\pi^2(1-\cos f_B)
\bigg],
\eeq
where we have introduced
\beq
I \equiv \frac{1}{4\pi} \int \left(\frac{ 1 + |z|^2}{1+|w|^2} \right)^4 \left| \frac{dw}{dz} \right|^4  d\Omega_z.
\eeq

In order to find the best approximation, we need to seek an appropriate rational map $w(z)$.
We should choose $w$ in such a way that $I$ is minimized. Though this is not easy task, by using
a numerical method, the rational maps $w(z)$ for several $N_B$ were found in Ref.~\cite{Houghton:1997kg}. 
For instance, the following rational maps for $N_B = 1,2,\cdots,8$ are known as
\beq
w_1 &=& z,\\
w_2 &=& z^2,\\
w_3 &=& \frac{z^3-\sqrt 3 i z}{\sqrt 3 i z^2 -1},\\
w_4 &=& \frac{z^4 + 2 \sqrt 3 i z^2 + 1}{z^4 - 2\sqrt 3 i z^2 + 1},\\
w_5 &=& \frac{z\left(z^4+b z^2+a\right)}{a z^4-b z^2+1},\\
w_6 &=& \frac{z^4+i c}{z^2\left(i c z^4+1\right)},\\
w_7 &=& \frac{z^7-7 z^5-7z^2-1}{z^7+7z^5-7z^2+1},\\
w_8 &=& \frac{z^6-d}{z^2\left(d z^6+1\right)},
\eeq
with $a=3.07$, $b=3.94$, $c=0.16$ and $d=0.14$.

The last task is to determine the profile function $f_B$.
Because no analytic solutions have been known, we need to solve equations of motion
numerically,
\beq
&&
\left(1+\frac{2 N_B}{r^2} \sin^2f_B\right) f_B''  + \frac{2}{r}  f_B' 
\non
&&
+ 
\frac{N_B\sin2f_B}{r^2}\left( f_B'^2 -1  - \frac{I}{N_B}\frac{ \sin^2f_B}{r^2}\right)\non
&& - 4m_\pi^2 \sin f_B =0,
\eeq
with the boundary condition $f_B(0) = \pi$ and $f_B(\infty) = 0$.
We show several numerical solutions for $N_B=2$ with different pion masses in 
Fig.~\ref{fig:profiles}. 
\begin{figure}[t]
\begin{center}
\includegraphics[height=5.5cm]{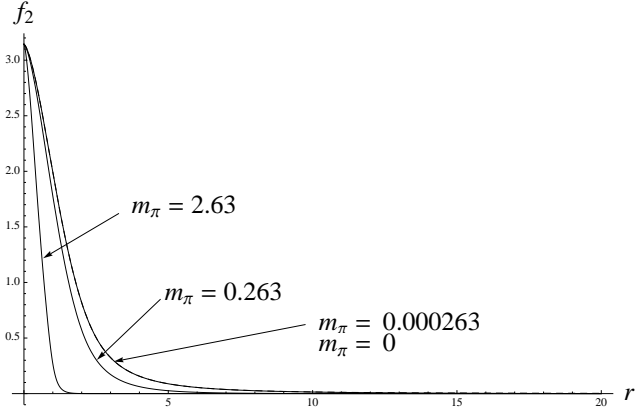}
\caption{The profile functions $f_{B=2}$ for $N_B=2$.}
\label{fig:profiles}
\end{center}
\end{figure}

Now we are ready to evaluate the anomaly induced electric charge from Eqs.~(\ref{eq:def_P}) and
(\ref{eq:full_charge}).
As before, what we need is only $P_i$ which can be obtained from
\beq
R_i &=&  i (\vec \tau\cdot\vec n) f_B'\, \hat{x}_i + \frac{i}{2} (\vec \tau \cdot \p_i \vec n) \sin 2 f_B  
\non
& & +
\left\{(\p_i \vec n \cdot \vec n) {\bf 1} + i (\p_i \vec n \times \vec n)\cdot \vec \tau\right\} \sin^2 f_B,\\
L_i &=&   i (\vec \tau\cdot\vec n) f_B'\, \hat{x}_i + \frac{i}{2} (\vec \tau \cdot \p_i \vec n) \sin 2 f_B  
\non
& &
+
\left\{(\p_i \vec n \cdot \vec n) {\bf 1} - i (\p_i \vec n \times \vec n)\cdot \vec \tau\right\} \sin^2 f_B.
\eeq
By plugging these into Eq.~(\ref{eq:def_P}), we get
\beq
P_i = - f'_Bn_3\, \hat{x}_i - \frac{1}{2}(\p_i n_3) \sin 2f_B.
\eeq
Note that, as expected, replacement $n_3$ with $\hat x_3$ gives us $N_B=1$ hedgehog solution.

The induced charge densities for the $N_B=2$ solution are shown in Fig.~\ref{fig:B=2}.
As one can see, the $N_B=1$ and $N_B=2$ charge distributions are quite similar, even though
the baryon charge distributions are totally different.
However, one can find differences if paying attention to the detail structures.
As can be seen in Figs. \ref{fig:B=1_cs} and \ref{fig:B=2_cs}, the $N_B=2$ densities are fatter than
those of $N_B=1$. Also the $N_B=2$ configuration has an internal structure.
\begin{figure}[t]
\begin{center}
\begin{tabular}{ccc}
\includegraphics[width=2.35cm]{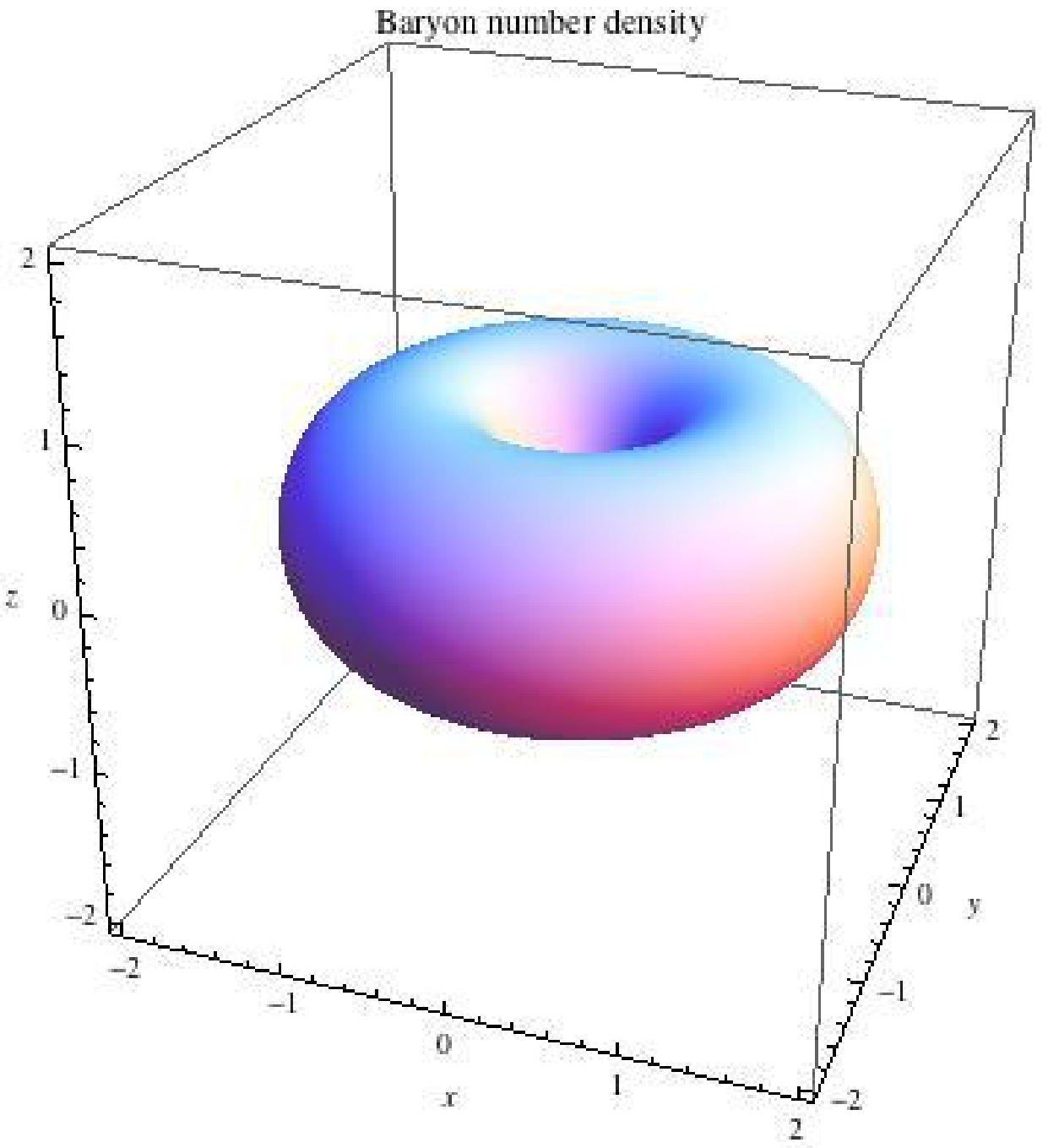} & 
\includegraphics[width=2.35cm]{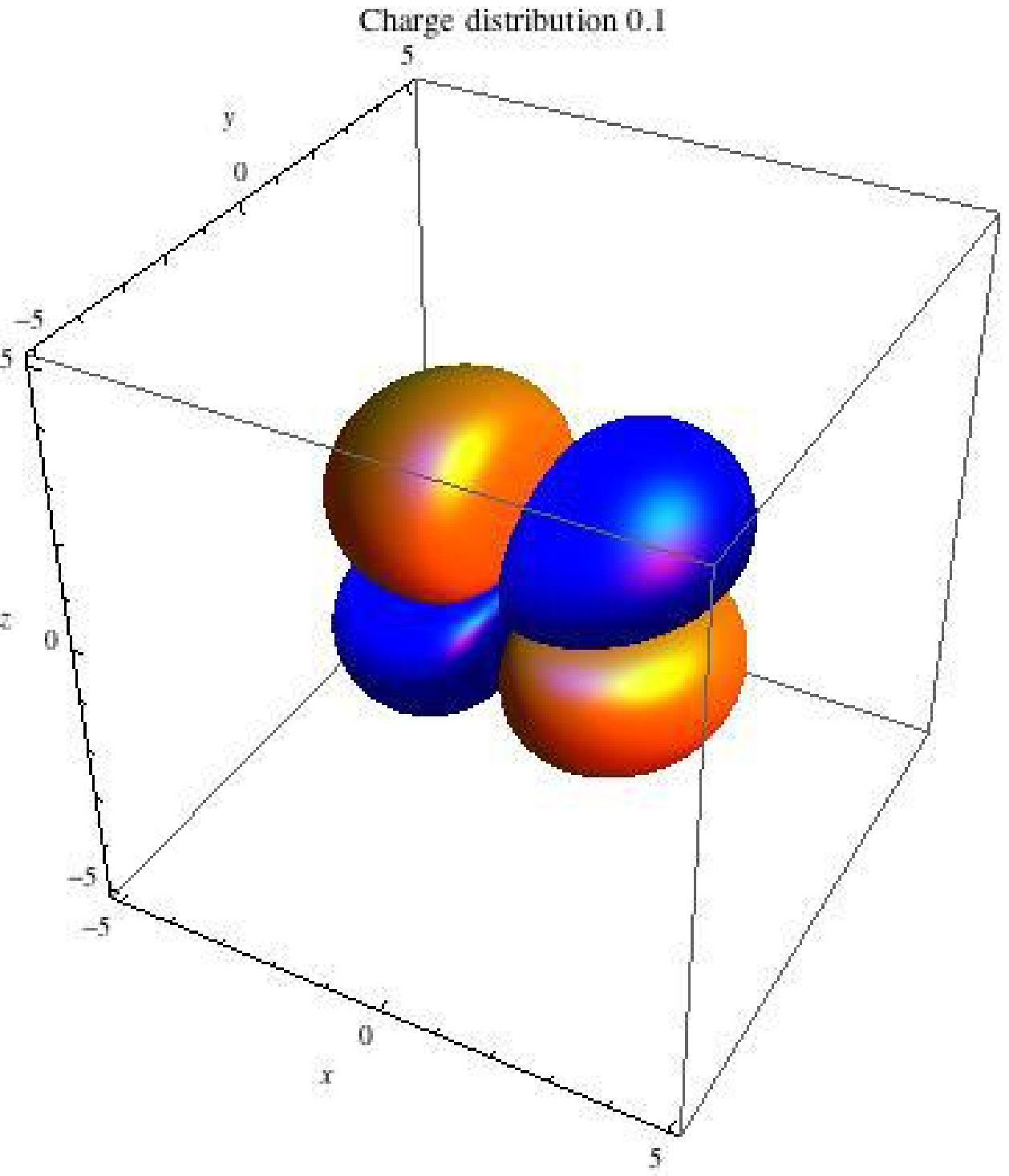} & 
\includegraphics[width=2.35cm]{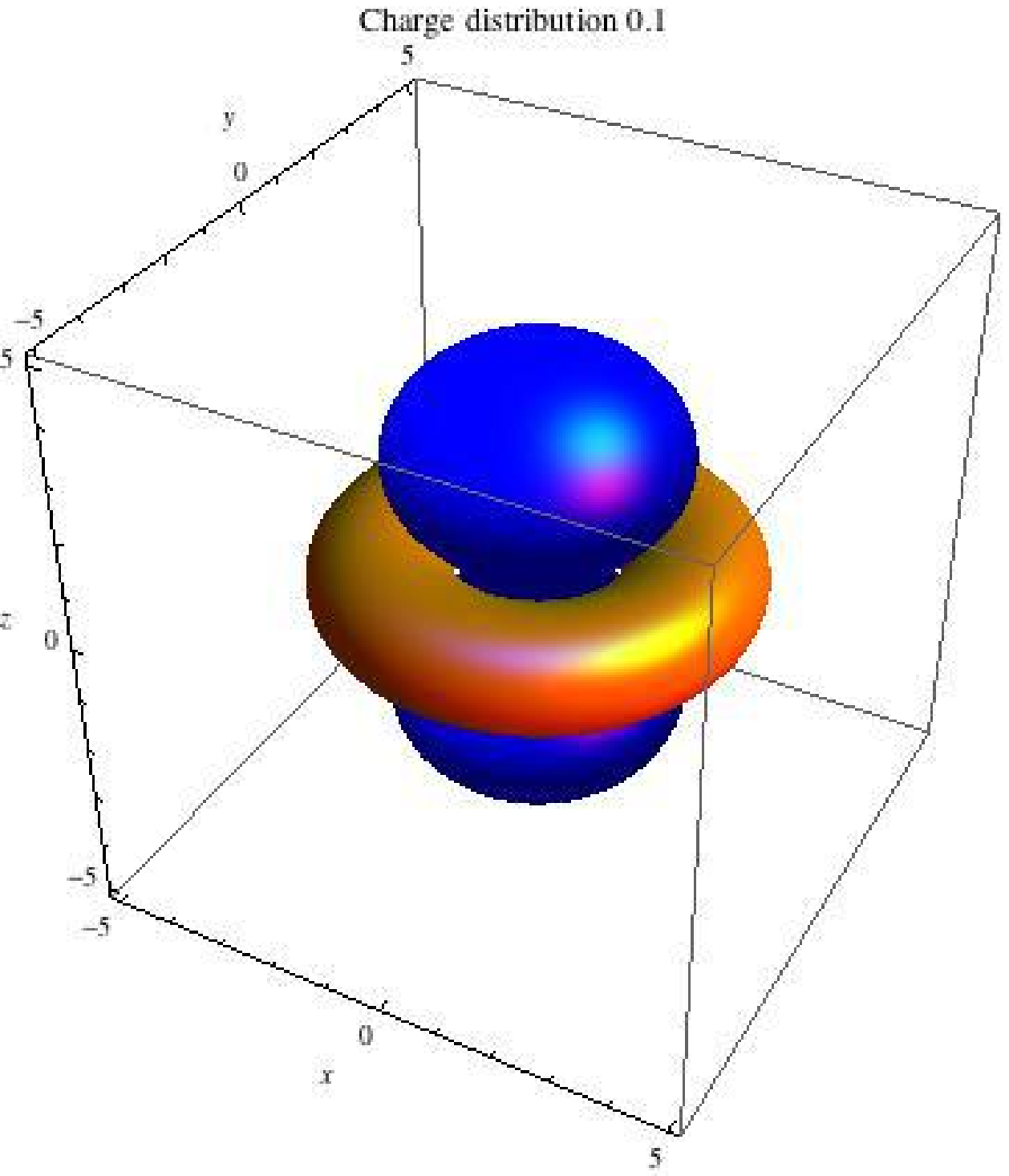}
\end{tabular}
\caption{$N_B=2$ Skyrmion solution. From left to right, the baryon number density,
$P_1$ and $P_3$, respectively.}
\label{fig:B=2}
\end{center}
\end{figure}
\begin{figure}[t]
\begin{center}
\includegraphics[width=6cm]{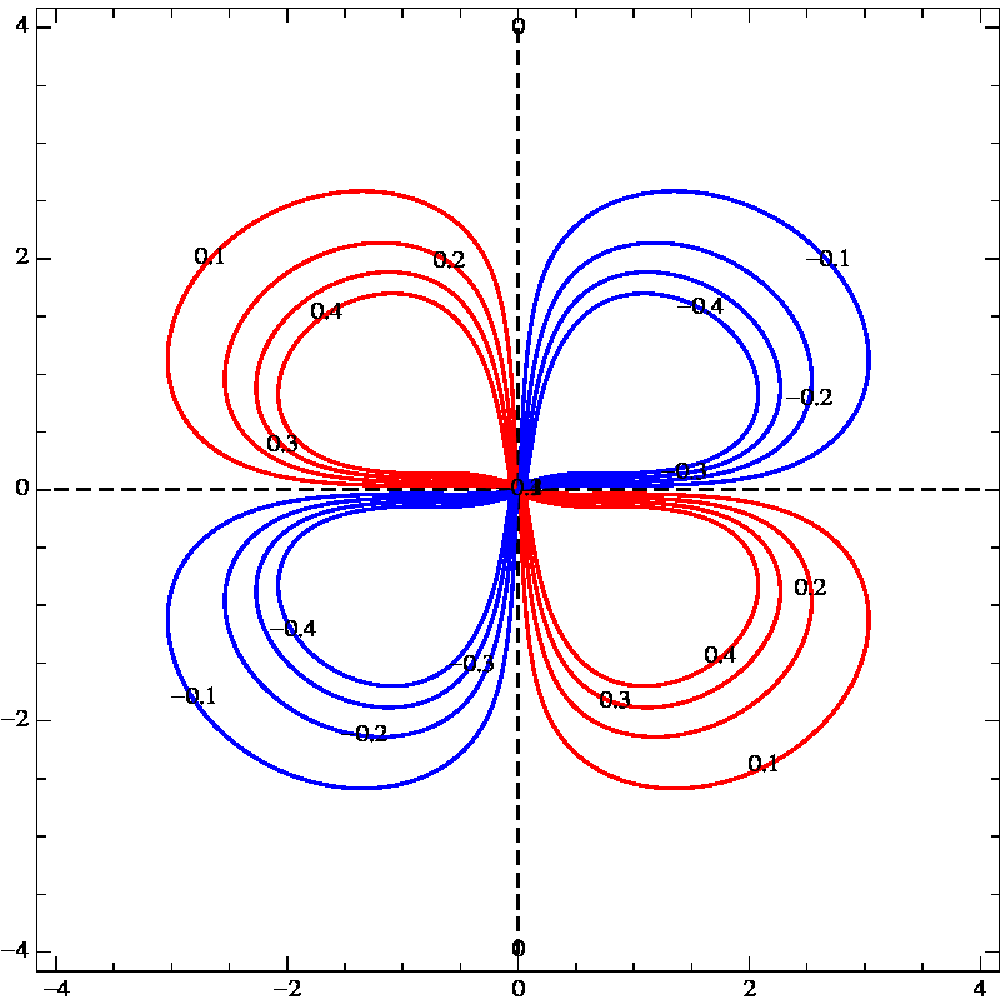} 
\includegraphics[width=6cm]{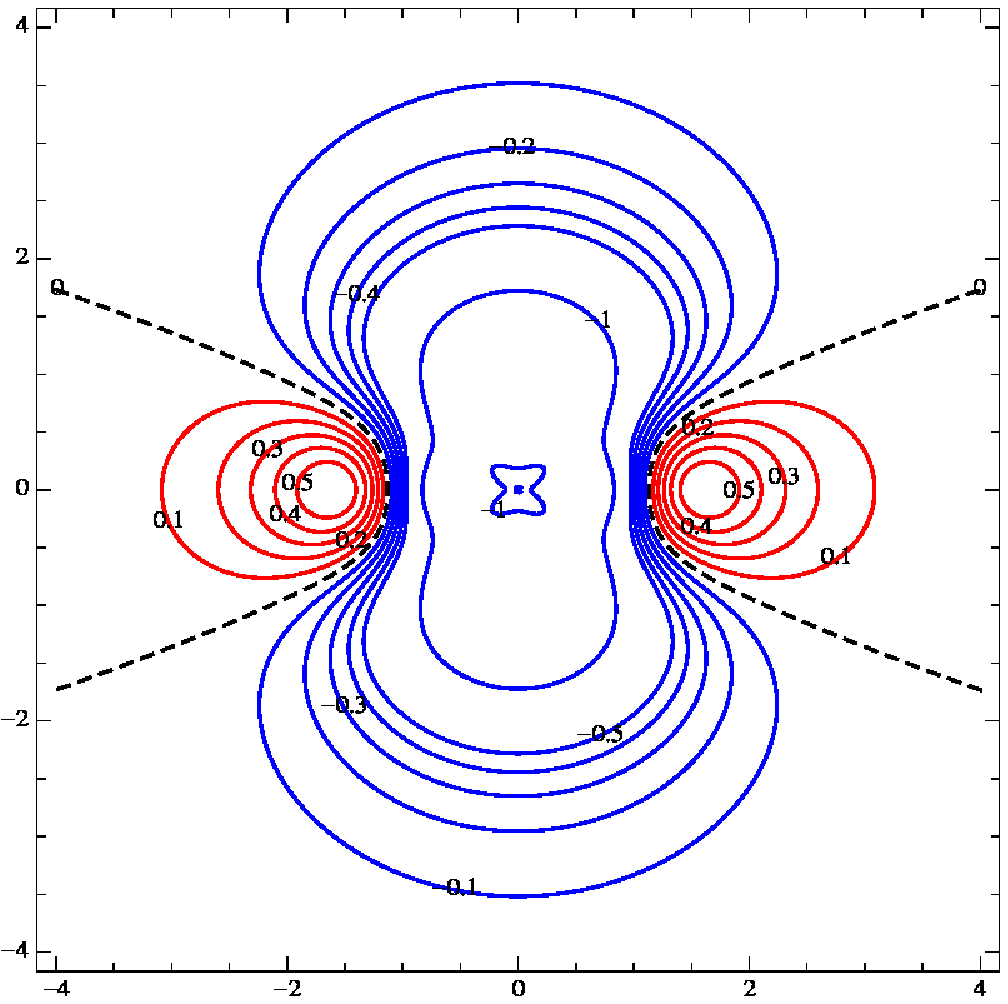}
\caption{The contour plots of the anomalous charge densities of $B=2$ Skyrmion on the cross section by the $y=0$ plane. 
The top panel shows $-P_1/2$ with $B_1 \neq 0$ and the bottom panel shows $-P_3/2$ with $B_3 \neq 0$.
The blue lines have positive values and the red ones have negative values. The black broken lines correspond
to zero charge contours.}
\label{fig:B=2_cs}
\end{center}
\end{figure}

The anomaly induced charges of the classical Skyrmions with $N_B = 1,2,\cdots,8$
under a constant background magnetic field $\vec B = (0,0,B_3)$
\beq
Q_{\rm anm}^{\rm classical} &=& \frac{e^2}{16\pi^2(e_s F_\pi)^2} B_3 \tilde c_0,\\
\tilde c_0 &=& 4 \int d^3 x\ P_3,
\eeq 
are summarized in the Table \ref{tab:sum_charge}.
The pre-factor $4$ is needed because of the dimensionless coordinate
$x_\mu \to 2x_\mu/{e_s F_\pi}$.
Note that $c_0$ and $\tilde c_0$ for $N_B=1$ are related by $\tilde c_0 = 4\pi c_0/3$.
We find that the classical anomaly-induced charge is not proportional to the baryon charge $N_B$.
It is intriguing that $N_B=4$ and $7$ Skyrmions have zero induced electric charge.
From the values given in the Table \ref{tab:sum_charge}, we observe that higher-charge Skyrmions tend to
cancel the total induced charge. A natural reason for this cancellation is as follows. Each Skyrmion has
a classical orientation in spin and isospin space, and to form a bound state of the Skyrmions the orientations
should be arranged to cancel each other. Our formula of the anomaly-induced charge depends on the 
signs of the quantum spin and isospin, so, accordingly the total anomaly-induced charge would tend to cancel
each other.
Although we have not performed 
quantization of the higher-charge Skyrmions, we expect that this cancellation 
should occur even at the quantized level.

\begin{table}[t]
\begin{center}
\begin{tabular}{c|cccccccc}
\hline
\hline
$N_B$ & 1 & 2 & 3 & 4 & 5 & 6 & 7 & 8\\
\hline
$\tilde c_0$ & $-43.2$ & $-105$ & $-60.3$ & $0.00$ & $-13.3$ & $28.7$ &  $0.00$ & $-11.6$\\
\hline
\hline
\end{tabular}
\caption{The anomaly induced charge of $N_B=1,2,\cdots,8$ Skyrmions under a constant magnetic field background.
The dimensionless pion mass is chosen to be $m_\pi = 0.263$.}
\label{tab:sum_charge}
\end{center}
\end{table}

Let us finally display the baryon number densities and anomaly induced electric
charges of the Skyrmions with $N_B=3,4,\cdots,8$ and $N_B=17$ \cite{Eto:2011id},
see Fig.~\ref{fig:multi}.
The anomaly-induced charge densities exhibit amusing shapes.
Possible interpretation of the shape is an open question. 

\begin{figure}[t]
\begin{center}
\begin{tabular}{ccc}
\includegraphics[width=2.35cm]{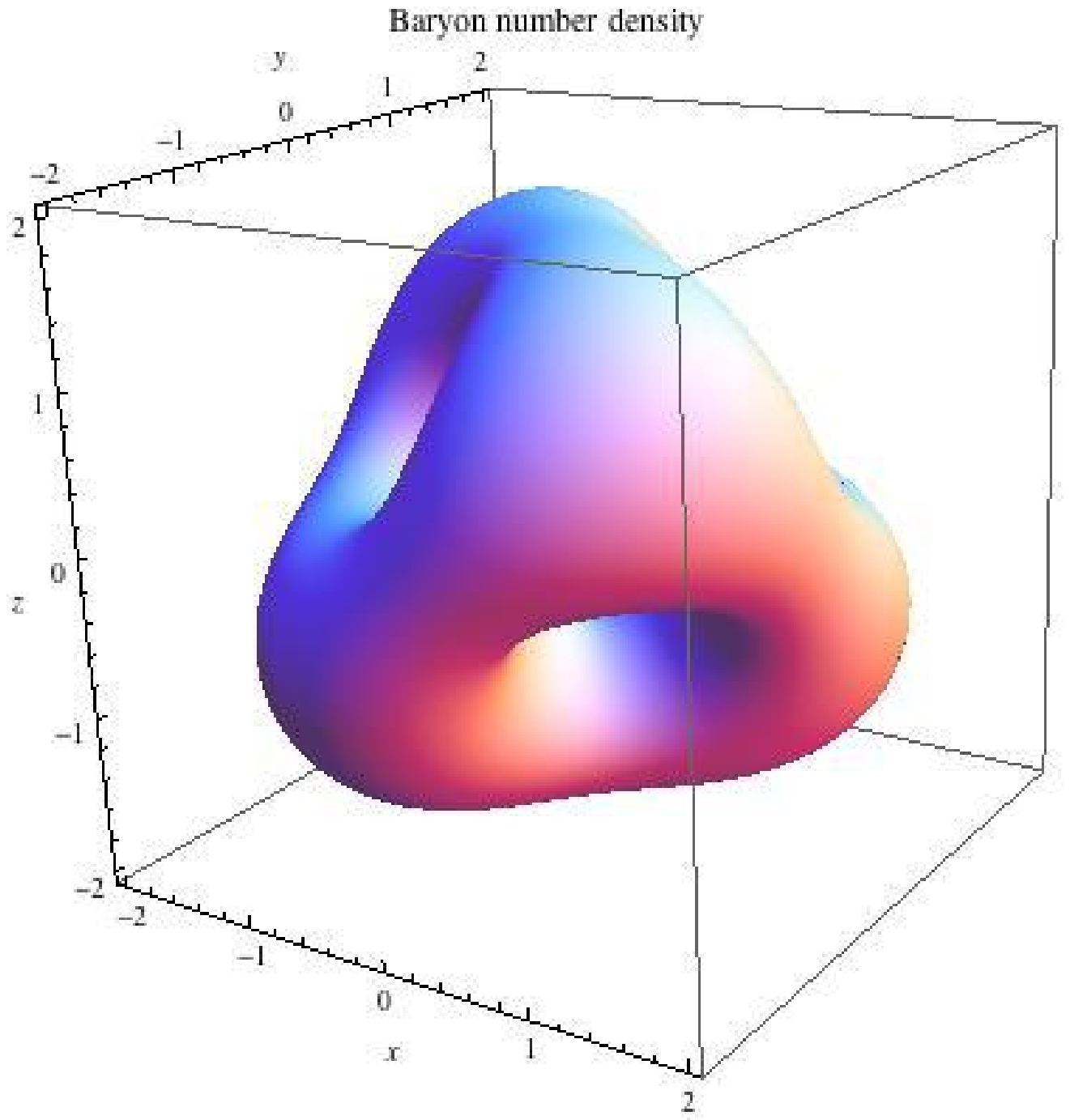} & 
\includegraphics[width=2.35cm]{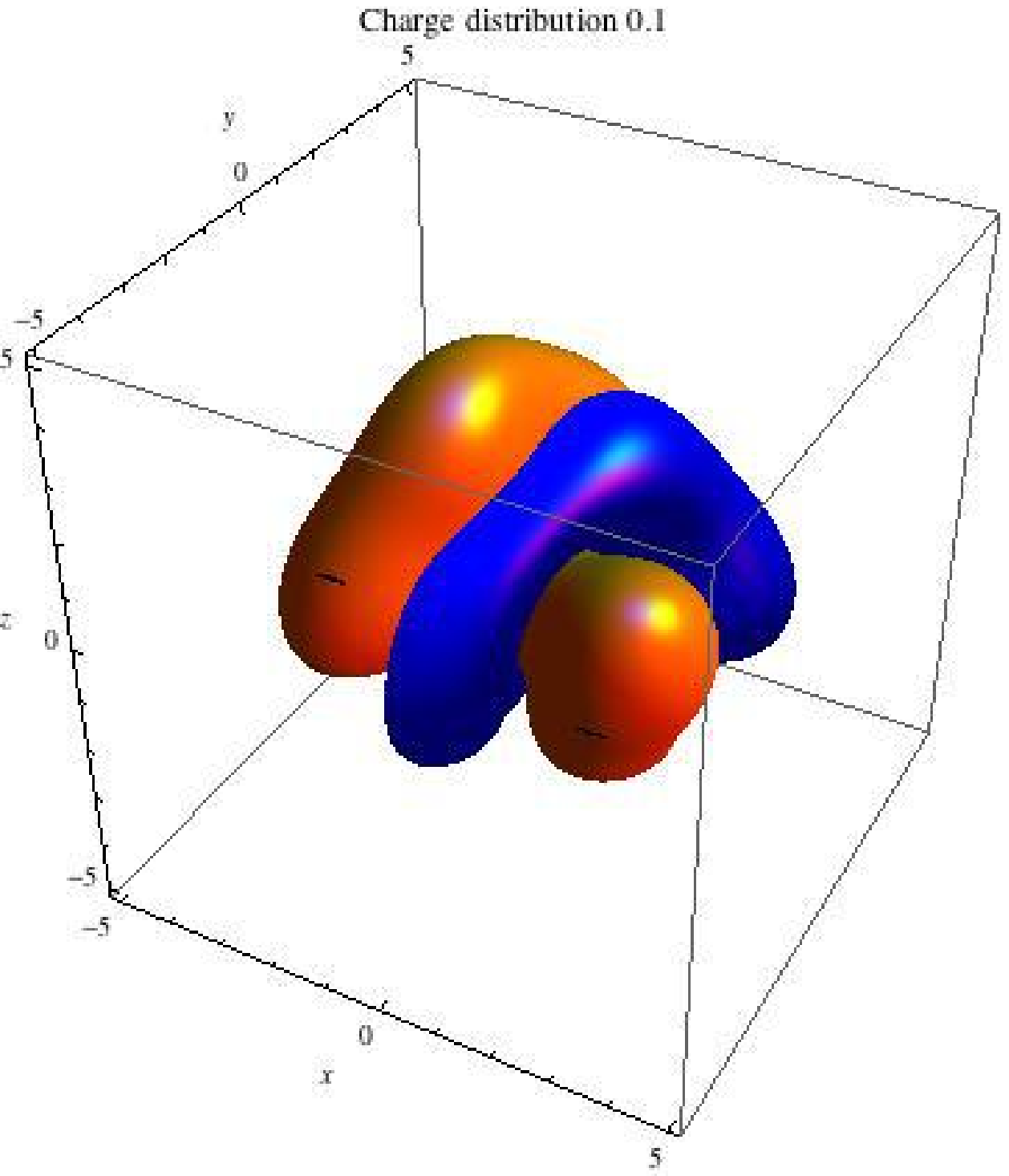} & 
\includegraphics[width=2.35cm]{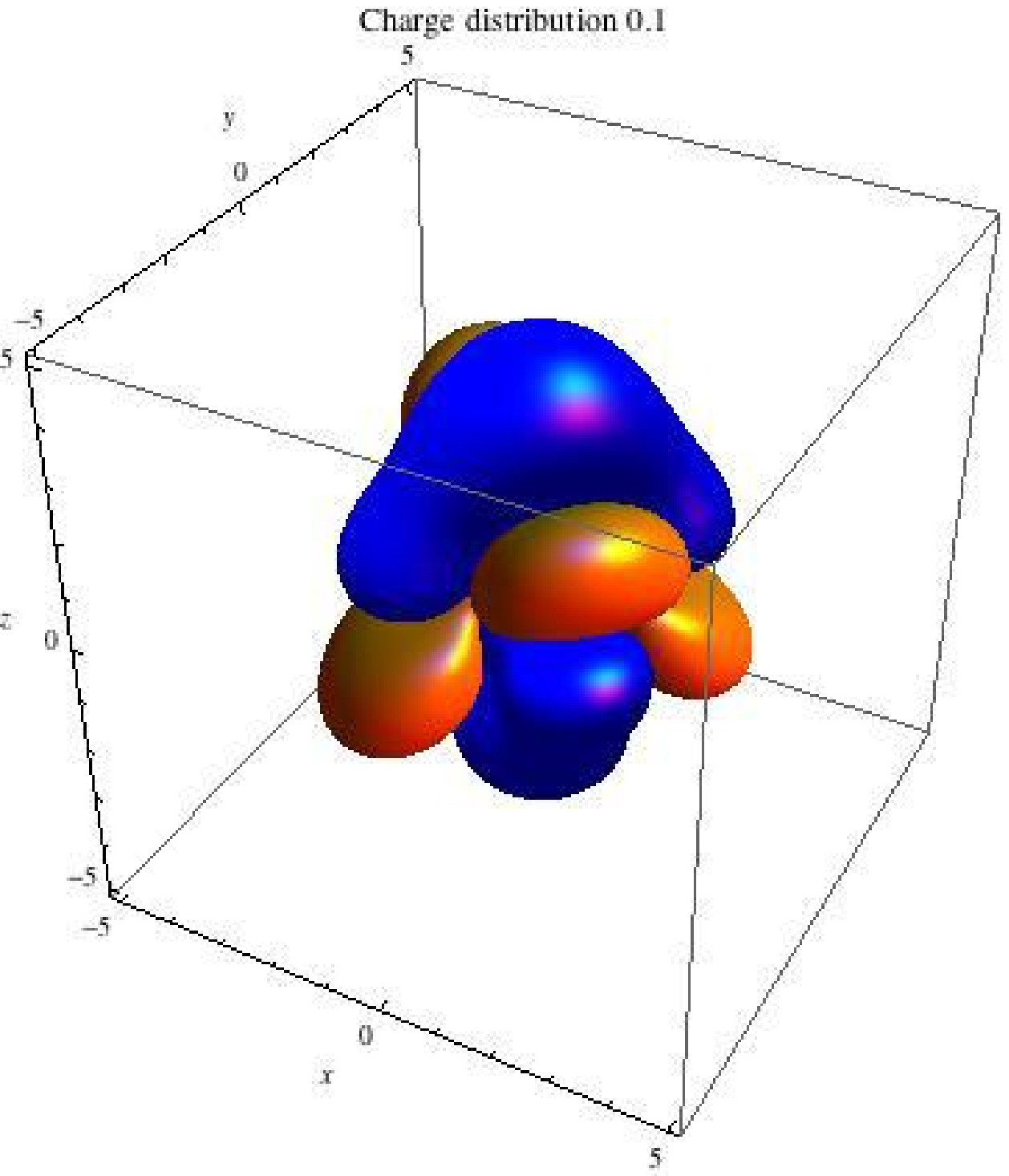}\\
\includegraphics[width=2.35cm]{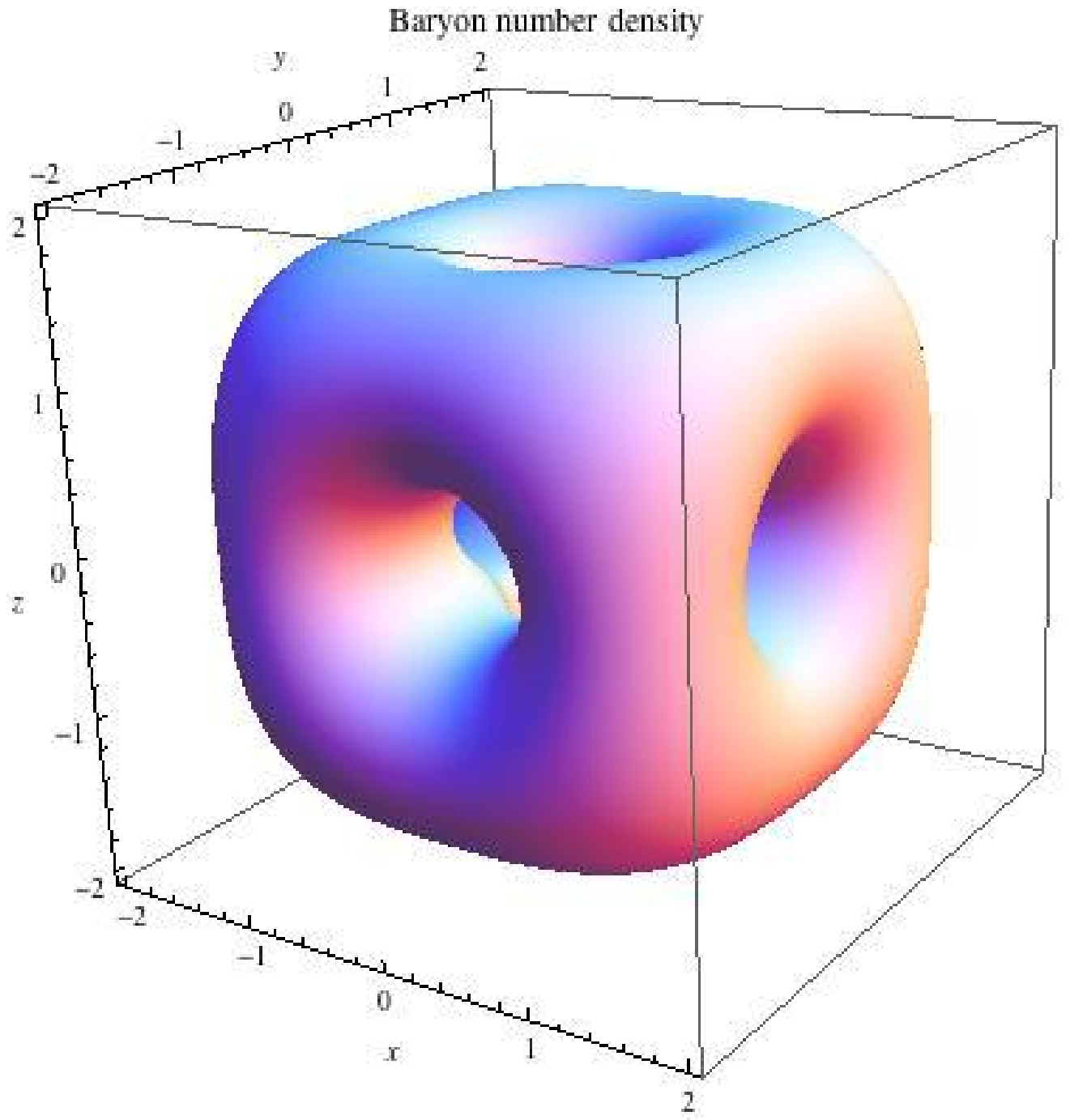} & 
\includegraphics[width=2.35cm]{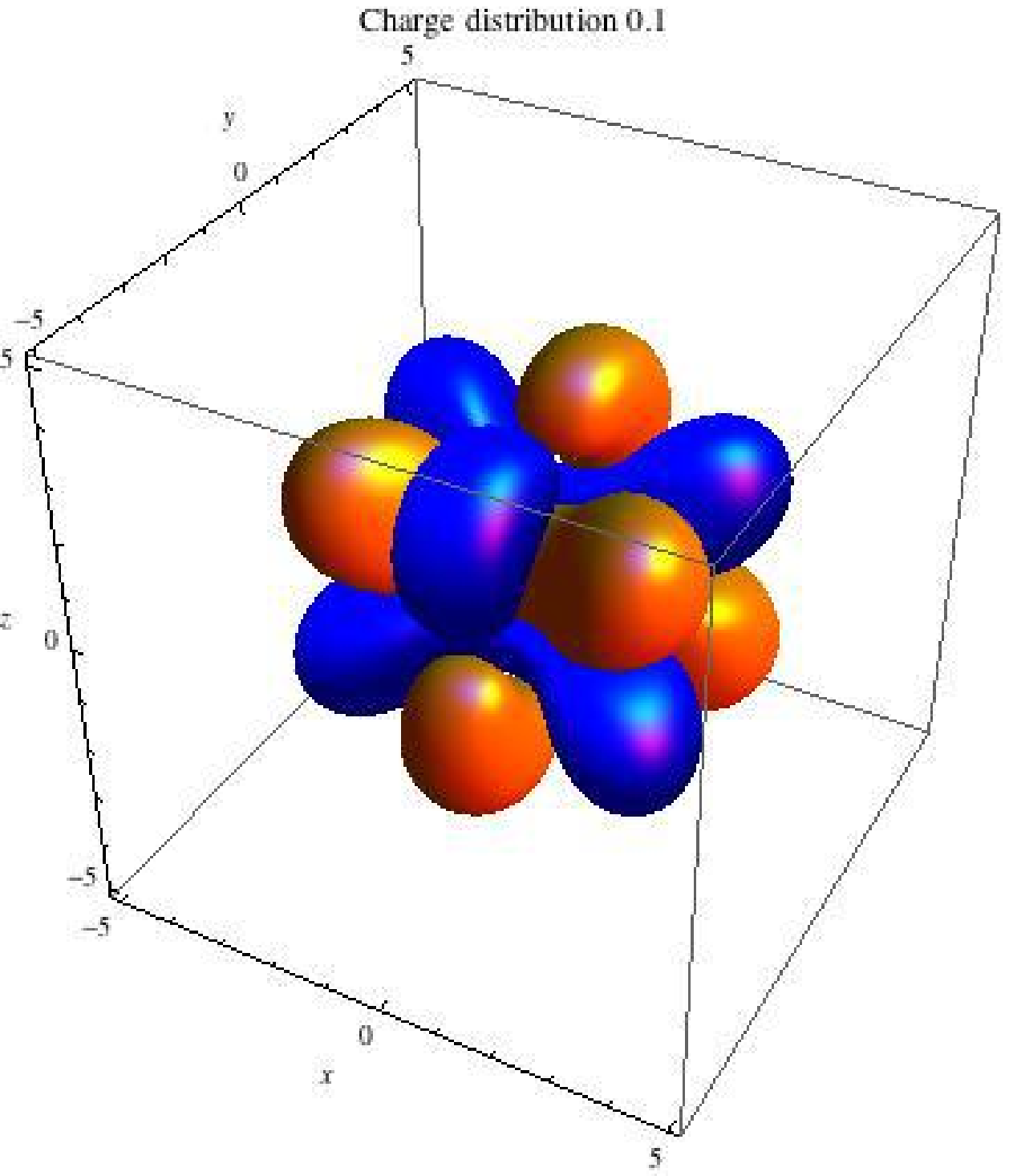} & 
\includegraphics[width=2.35cm]{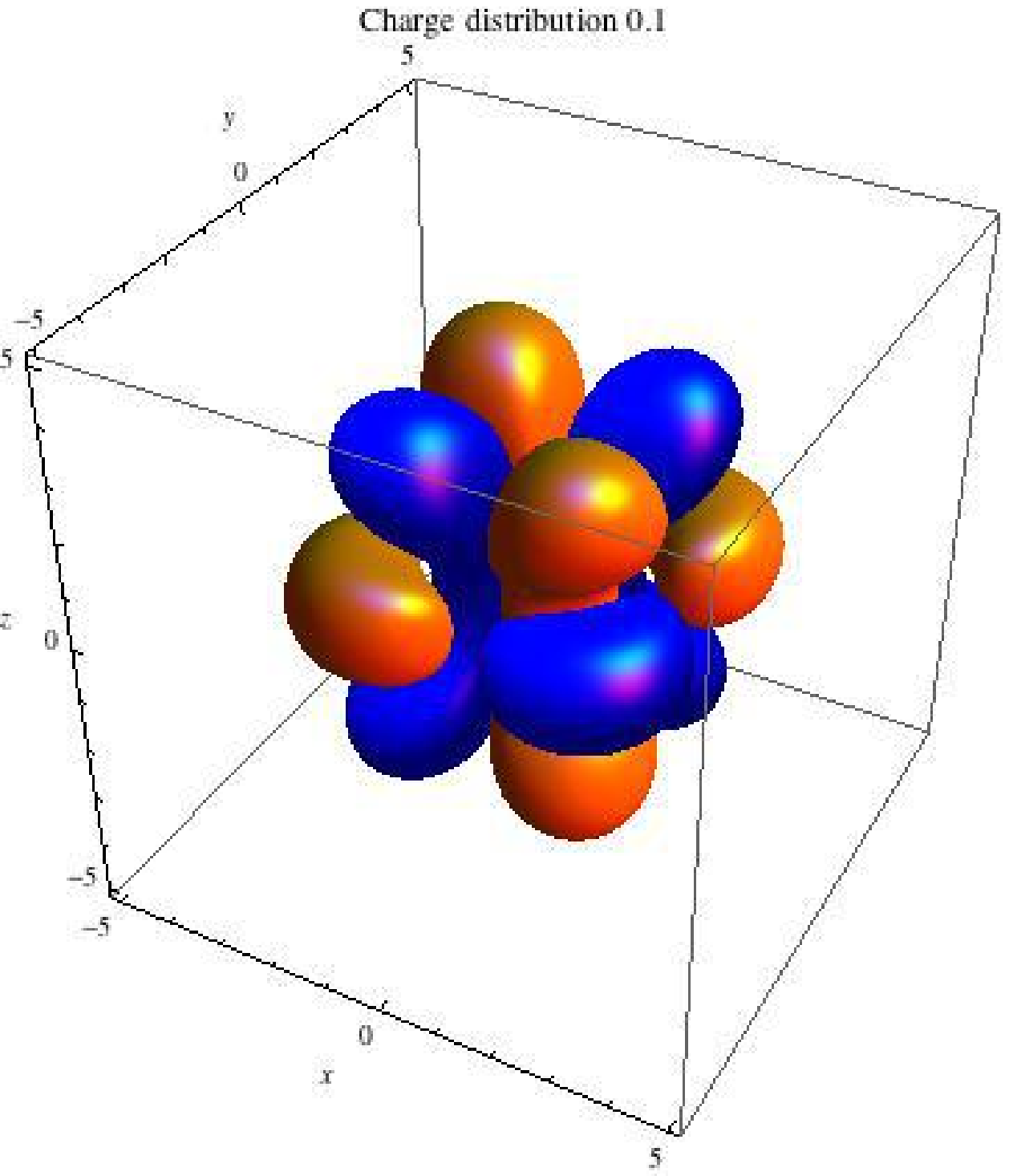}\\
\includegraphics[width=2.35cm]{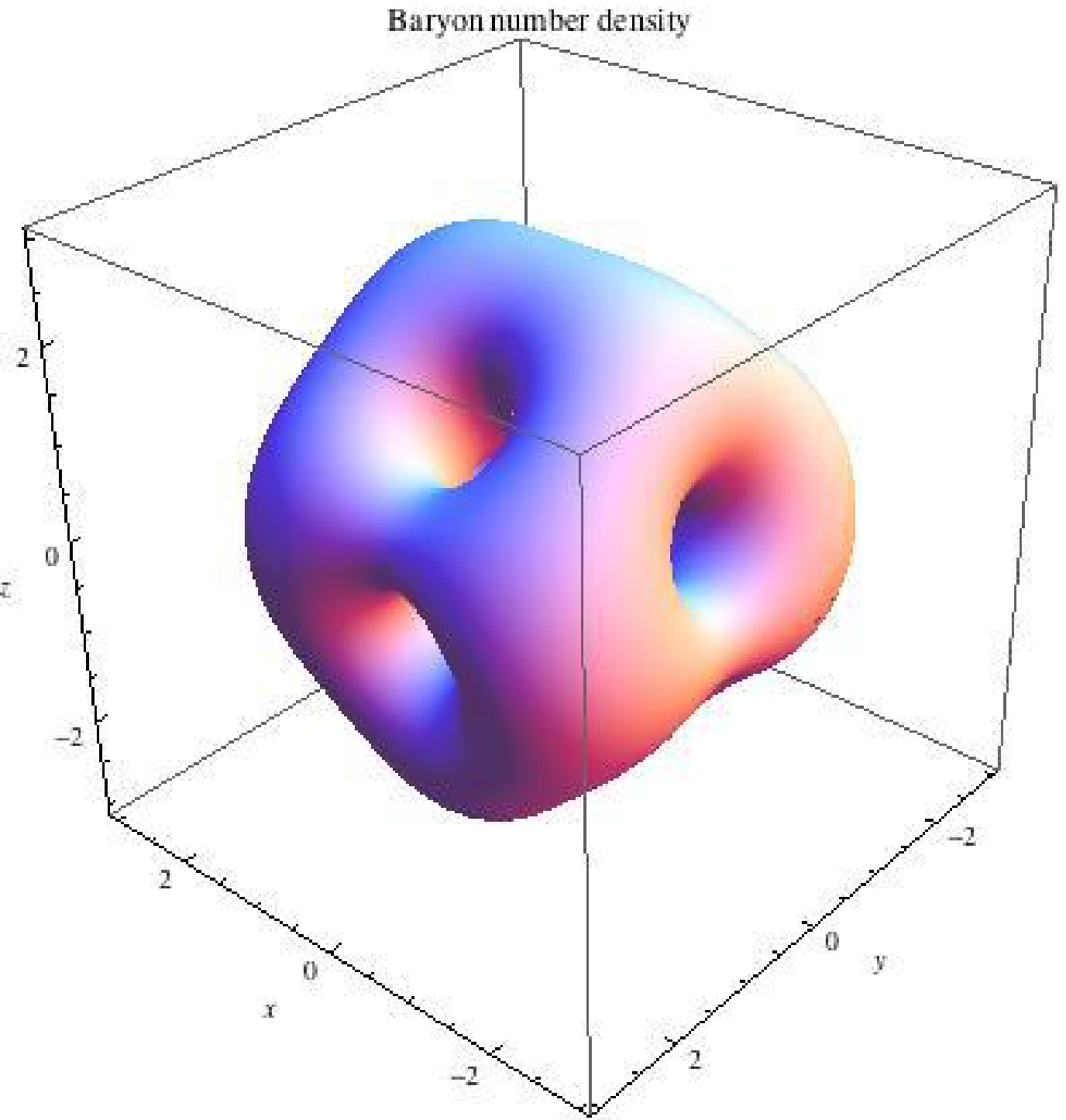} & 
\includegraphics[width=2.35cm]{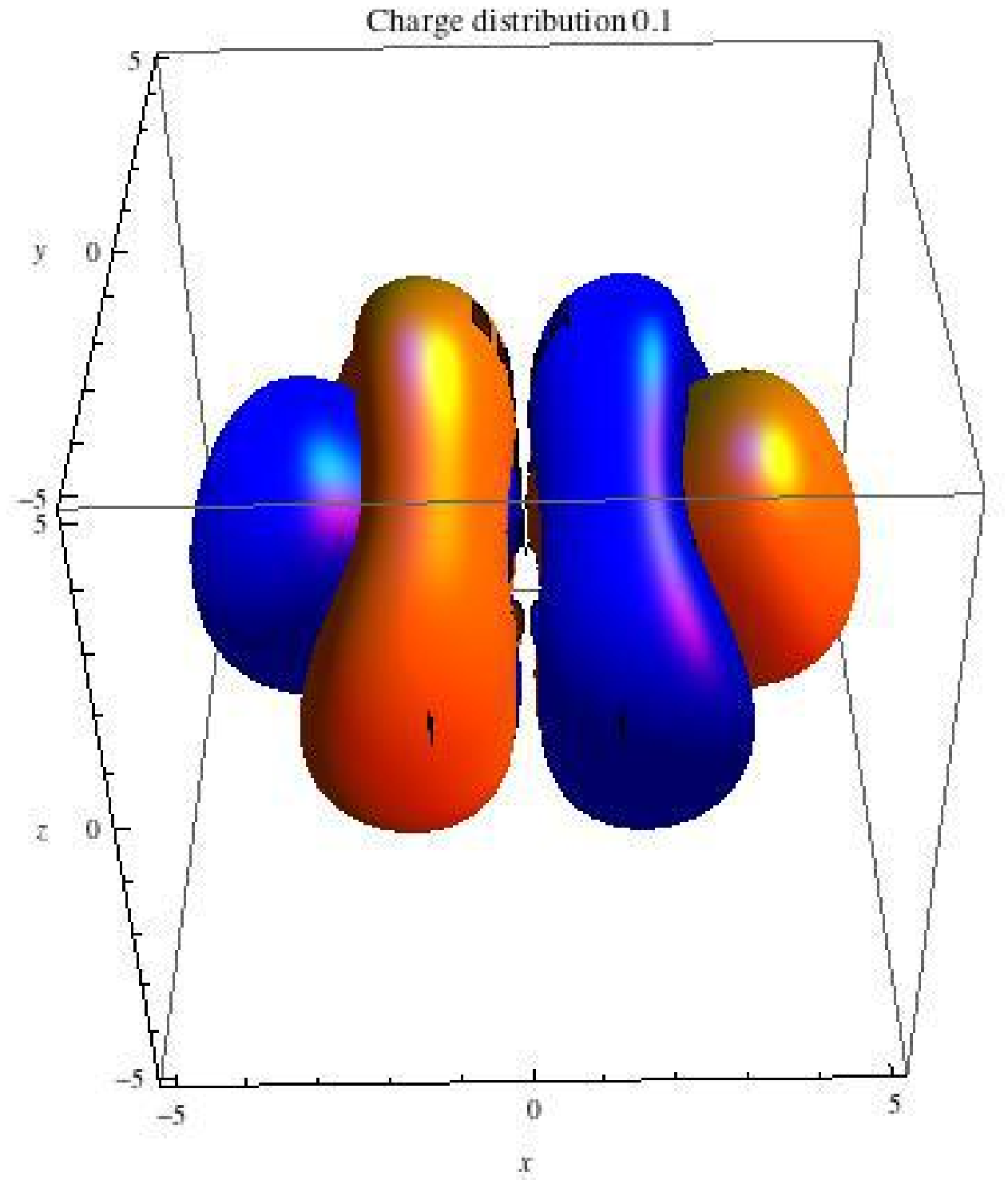} & 
\includegraphics[width=2.35cm]{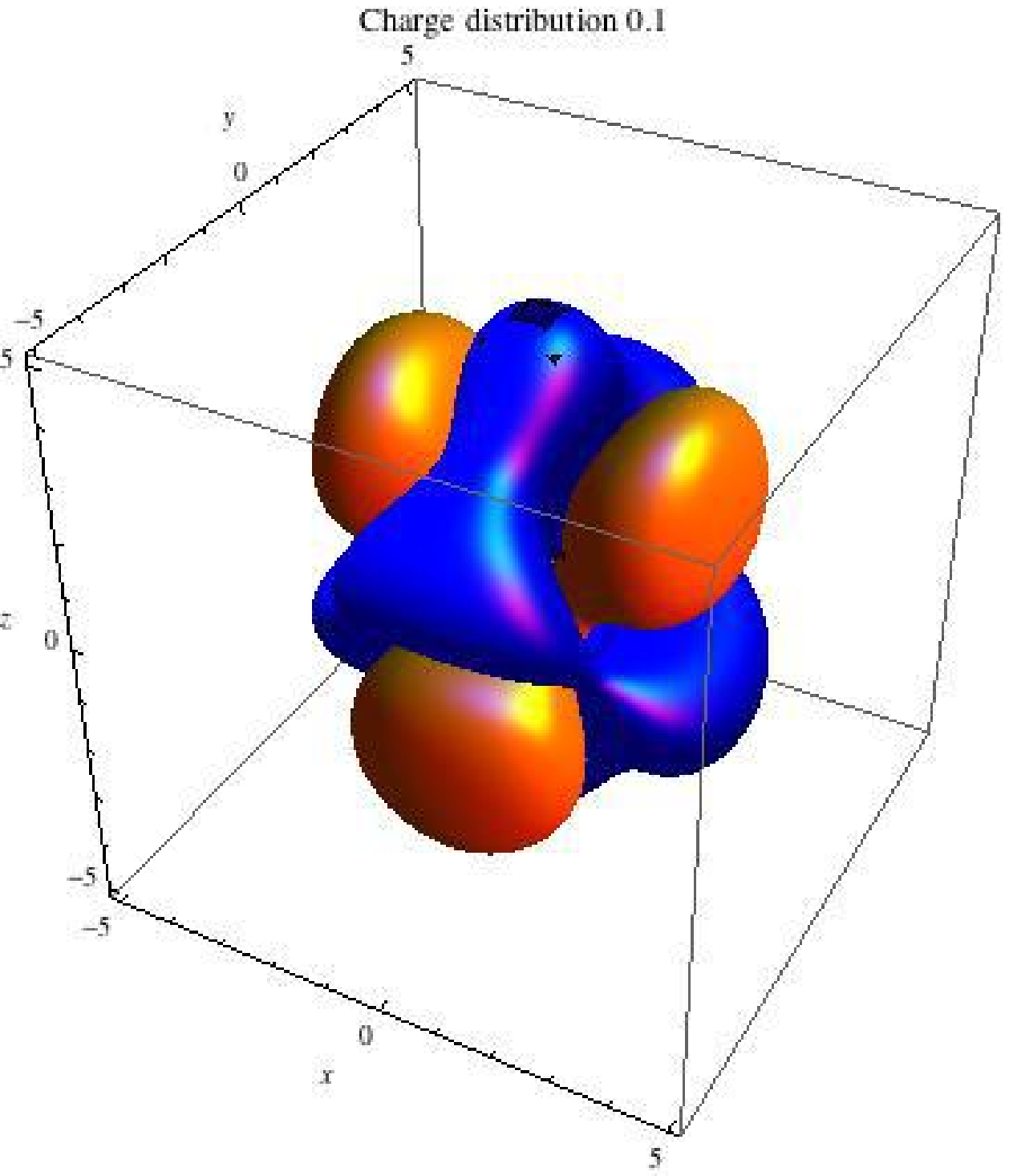}\\
\includegraphics[width=2.35cm]{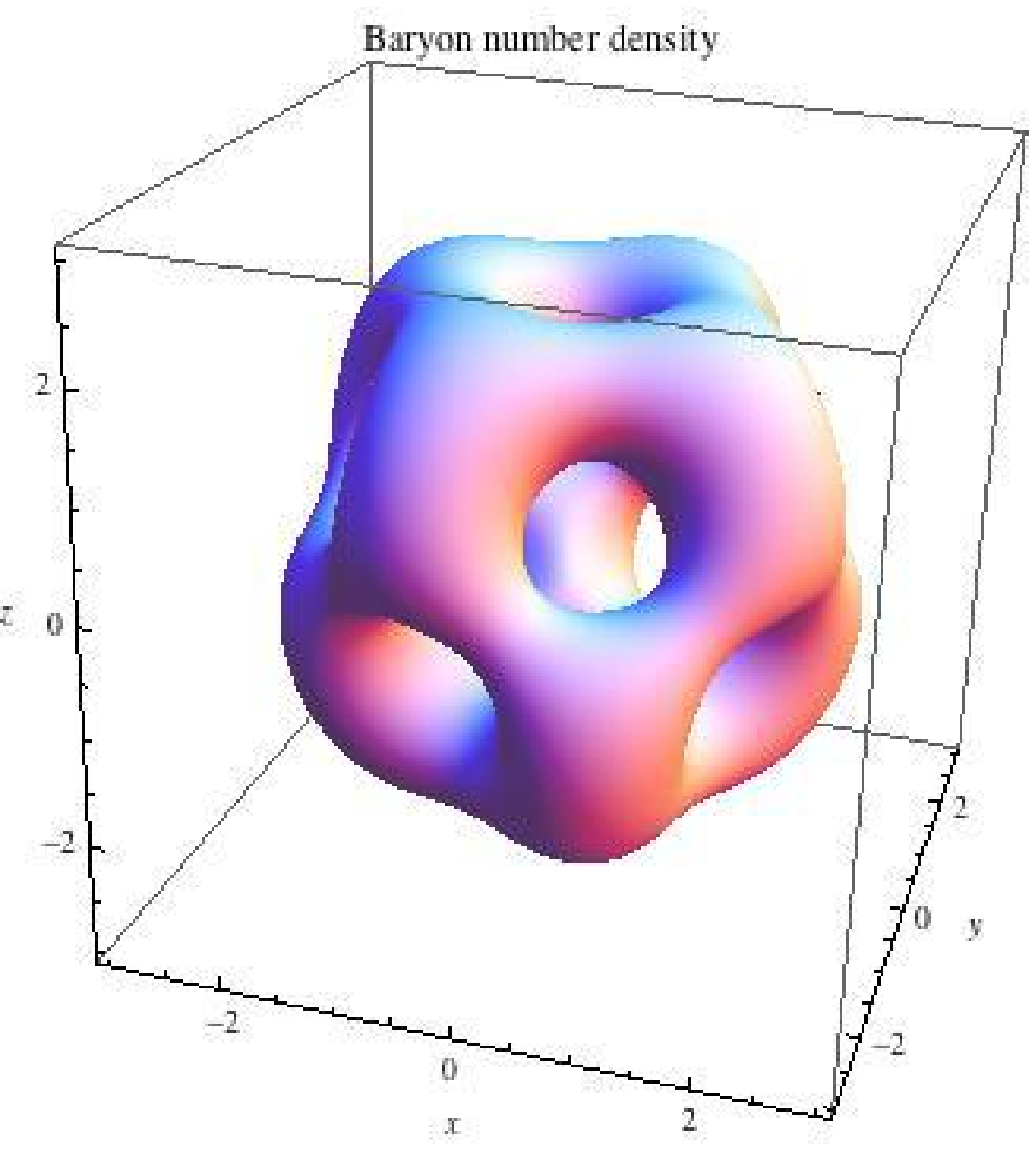} & 
\includegraphics[width=2.35cm]{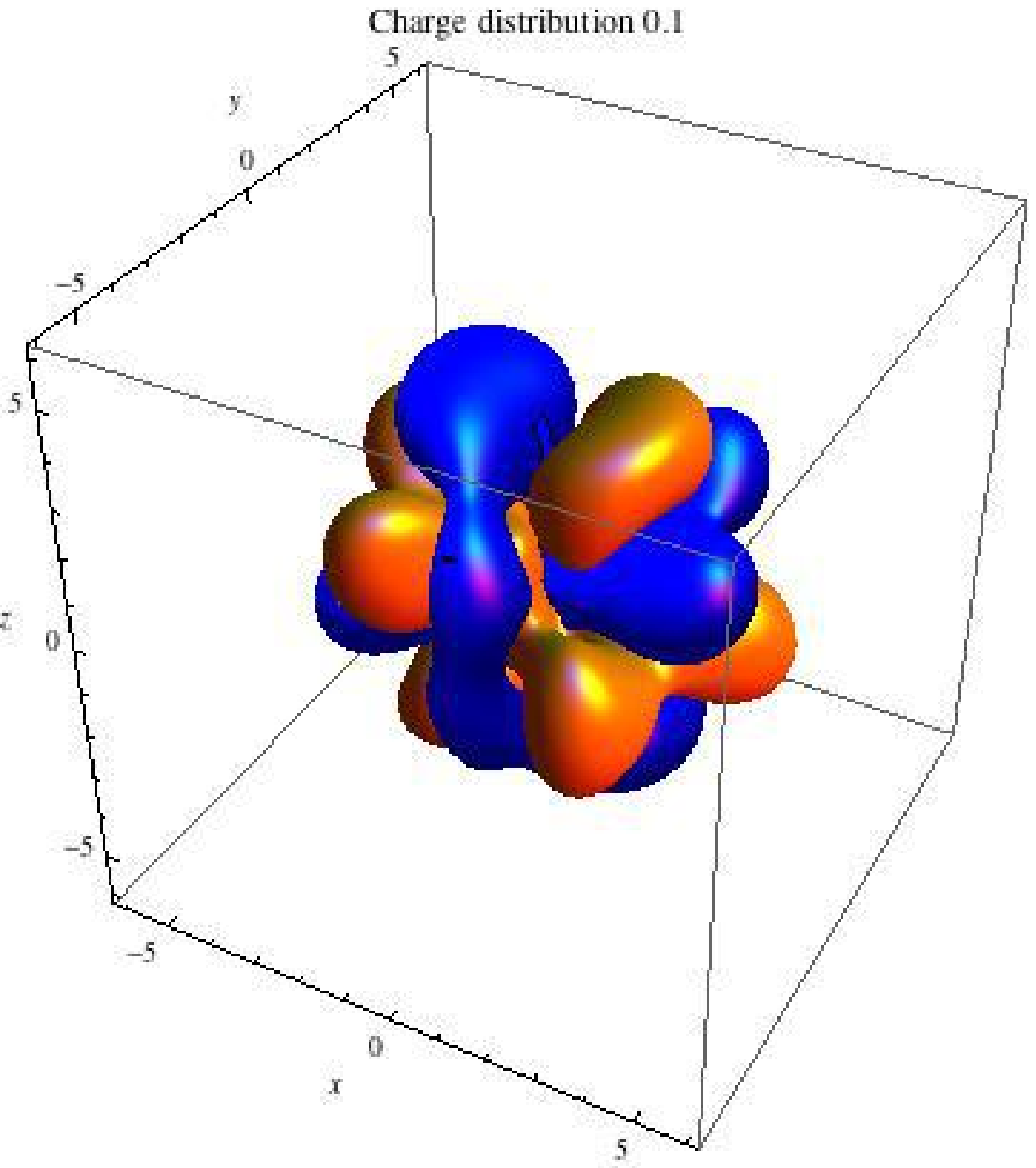} & 
\includegraphics[width=2.35cm]{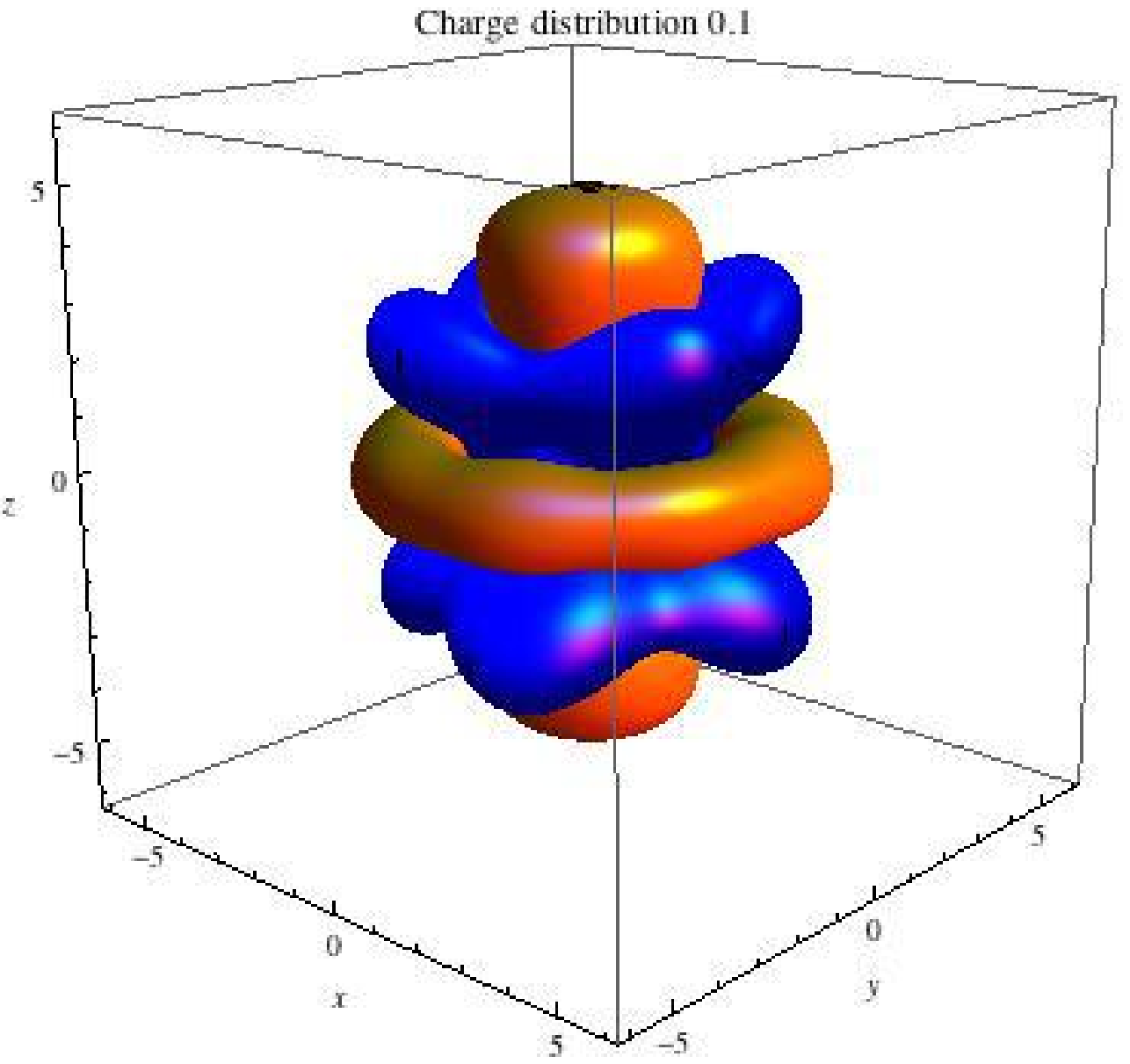}\\
\includegraphics[width=2.35cm]{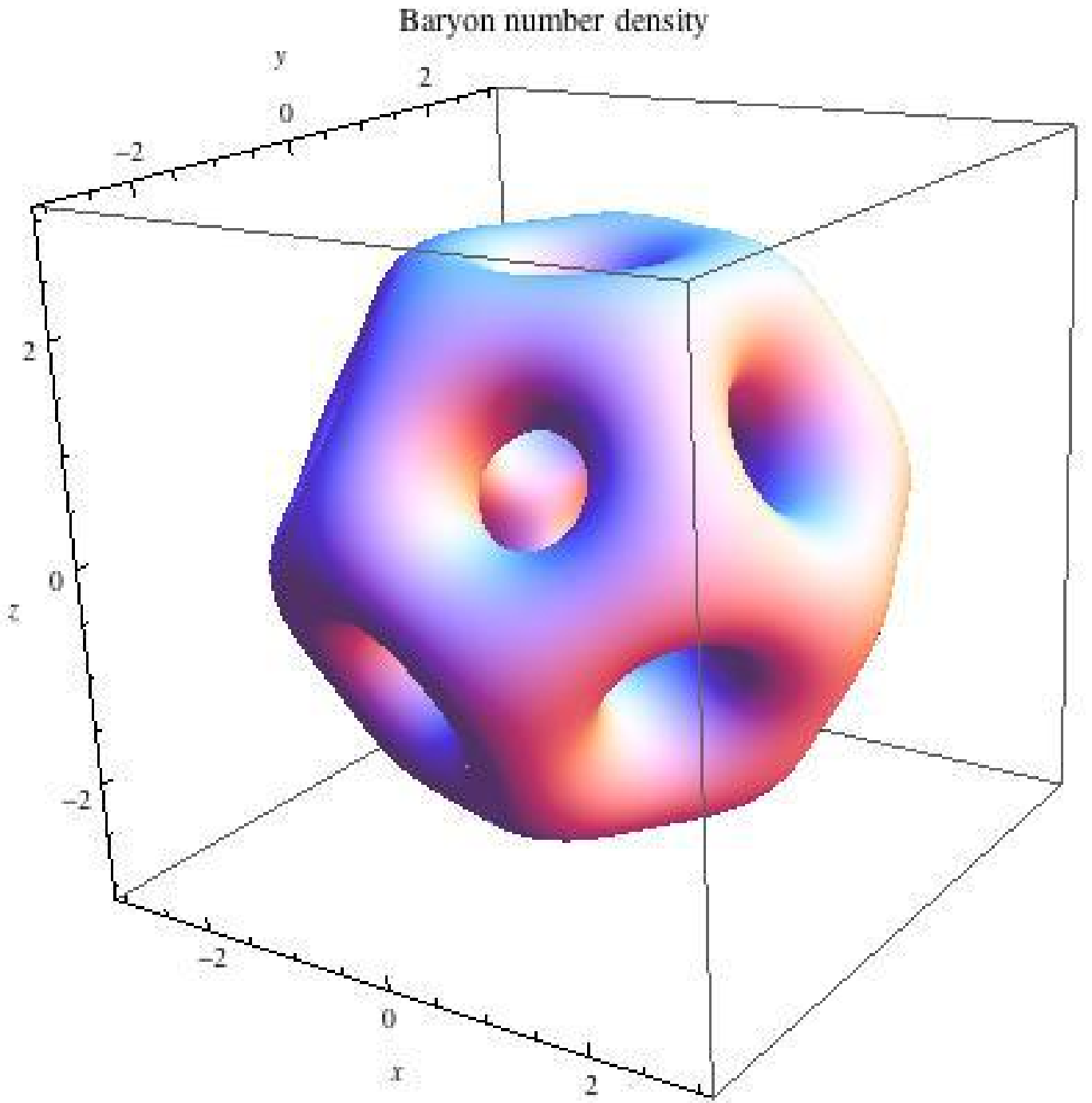} & 
\includegraphics[width=2.35cm]{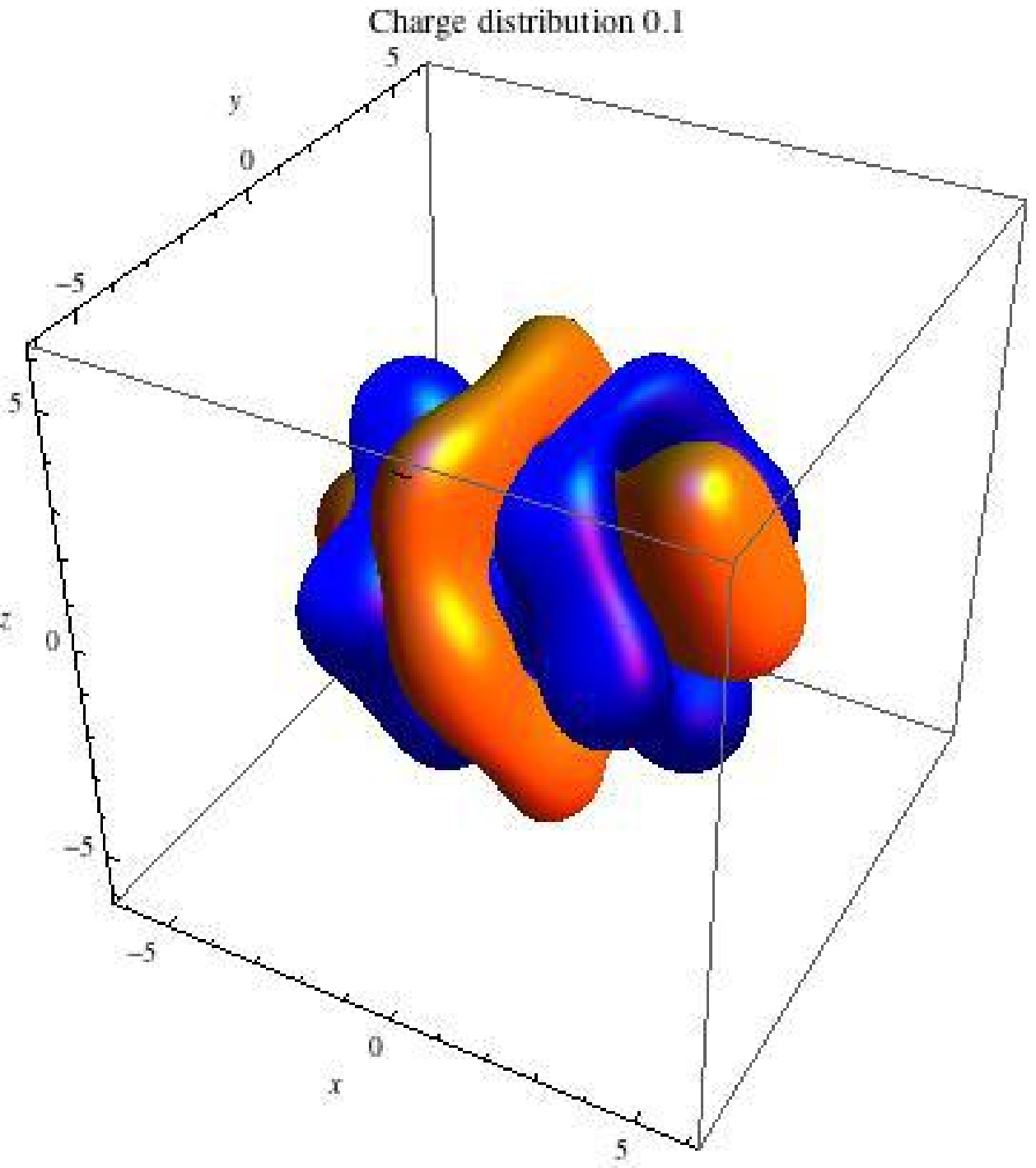} & 
\includegraphics[width=2.35cm]{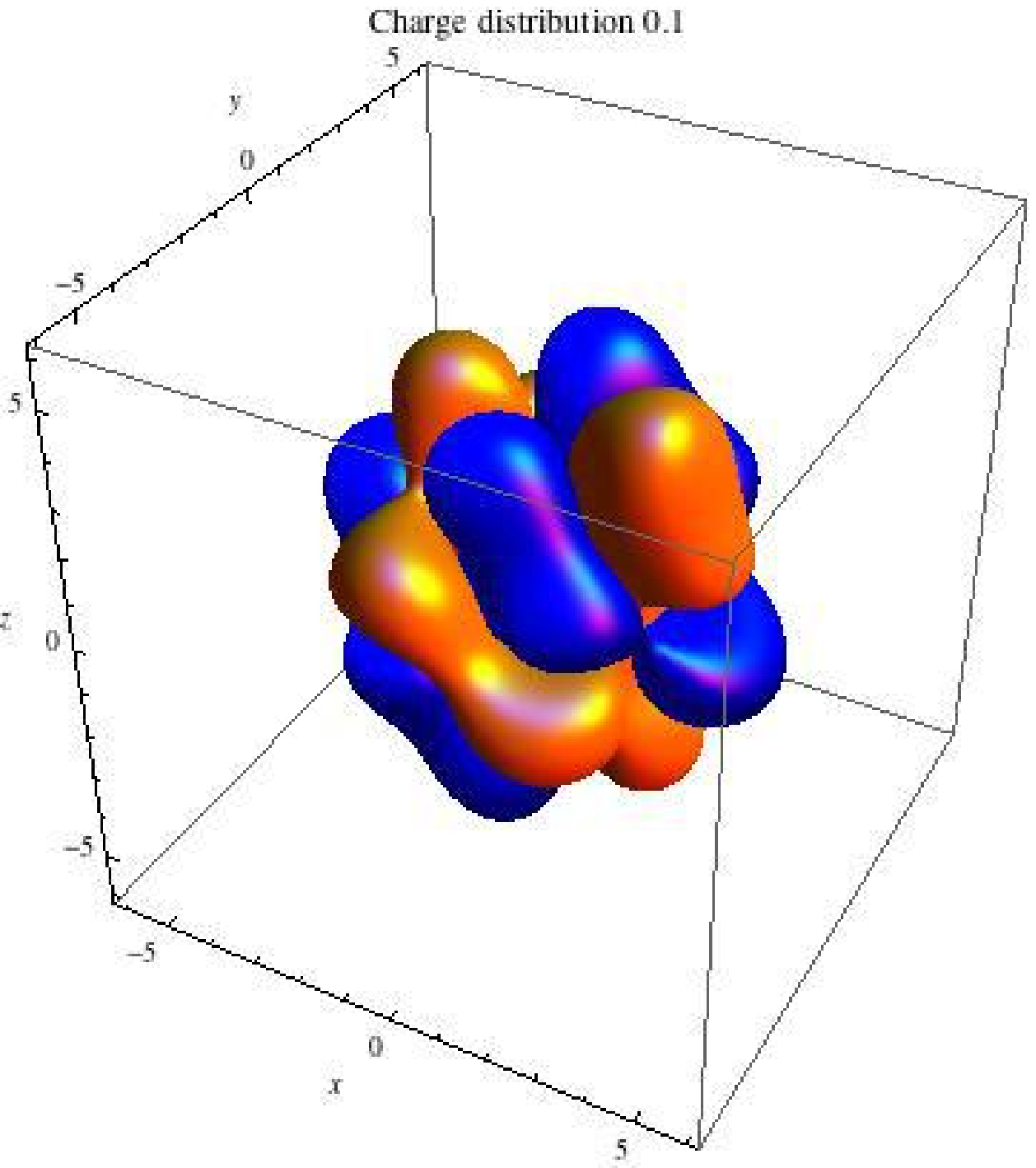}\\
\includegraphics[width=2.35cm]{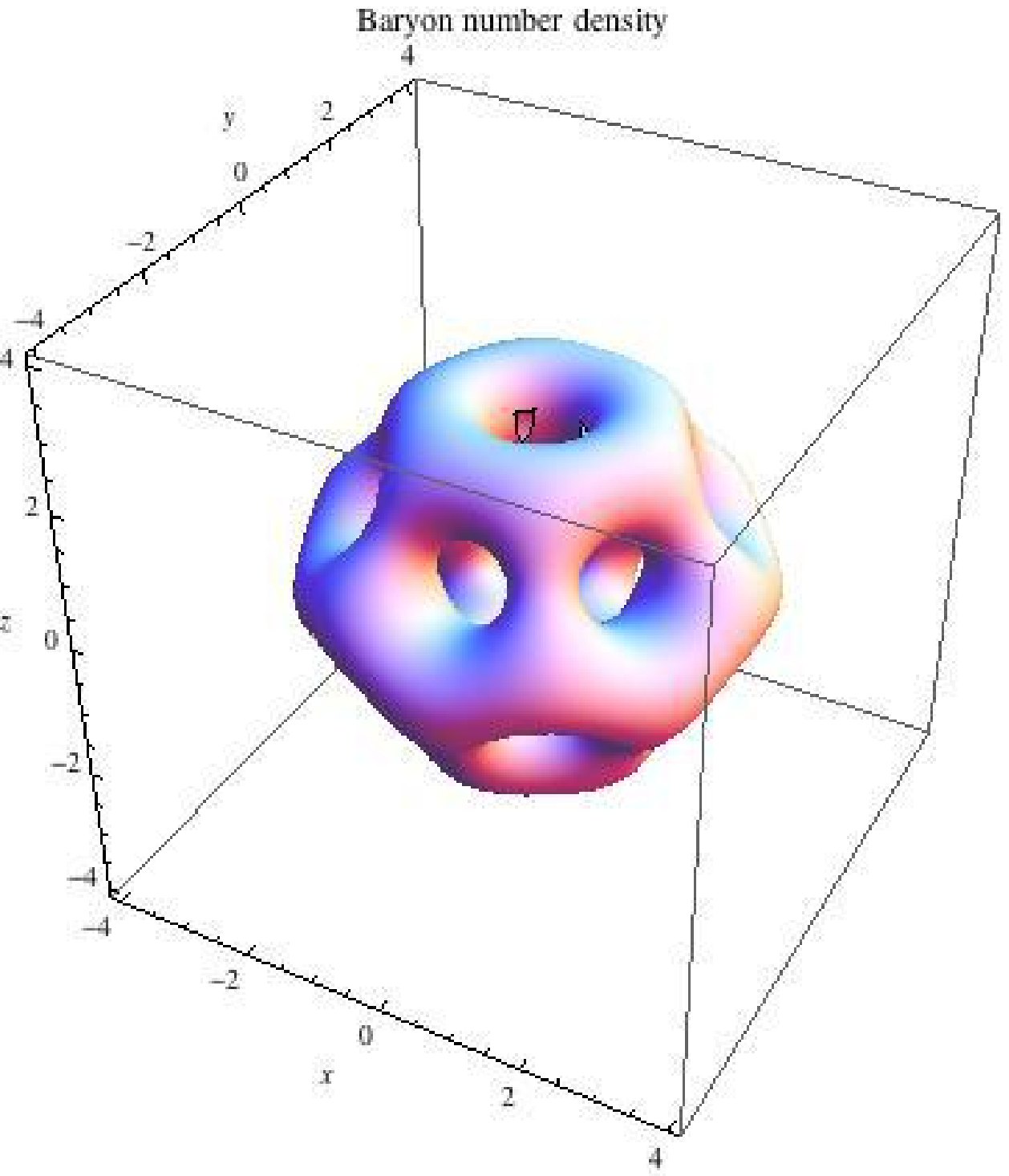} & 
\includegraphics[width=2.35cm]{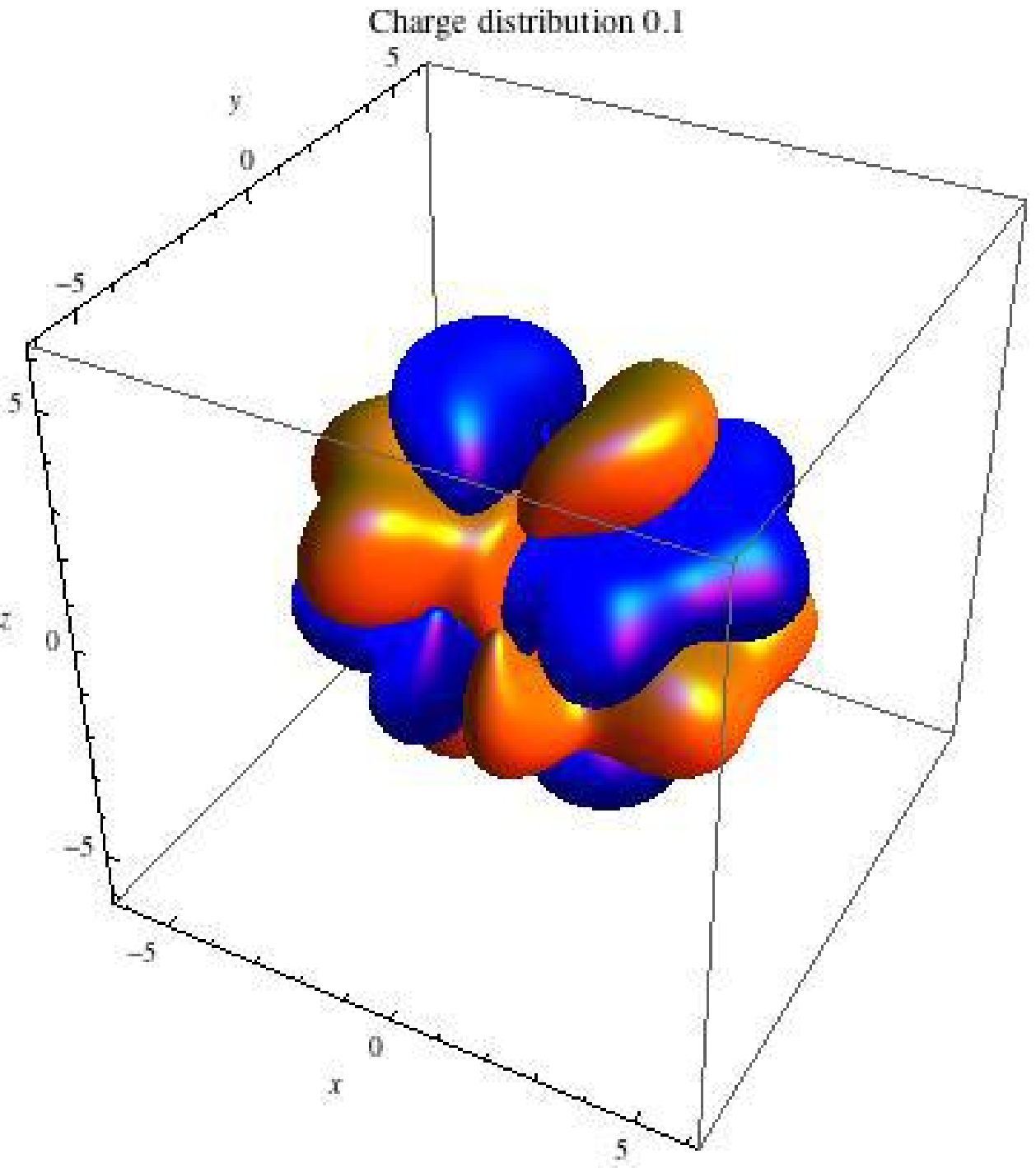} & 
\includegraphics[width=2.35cm]{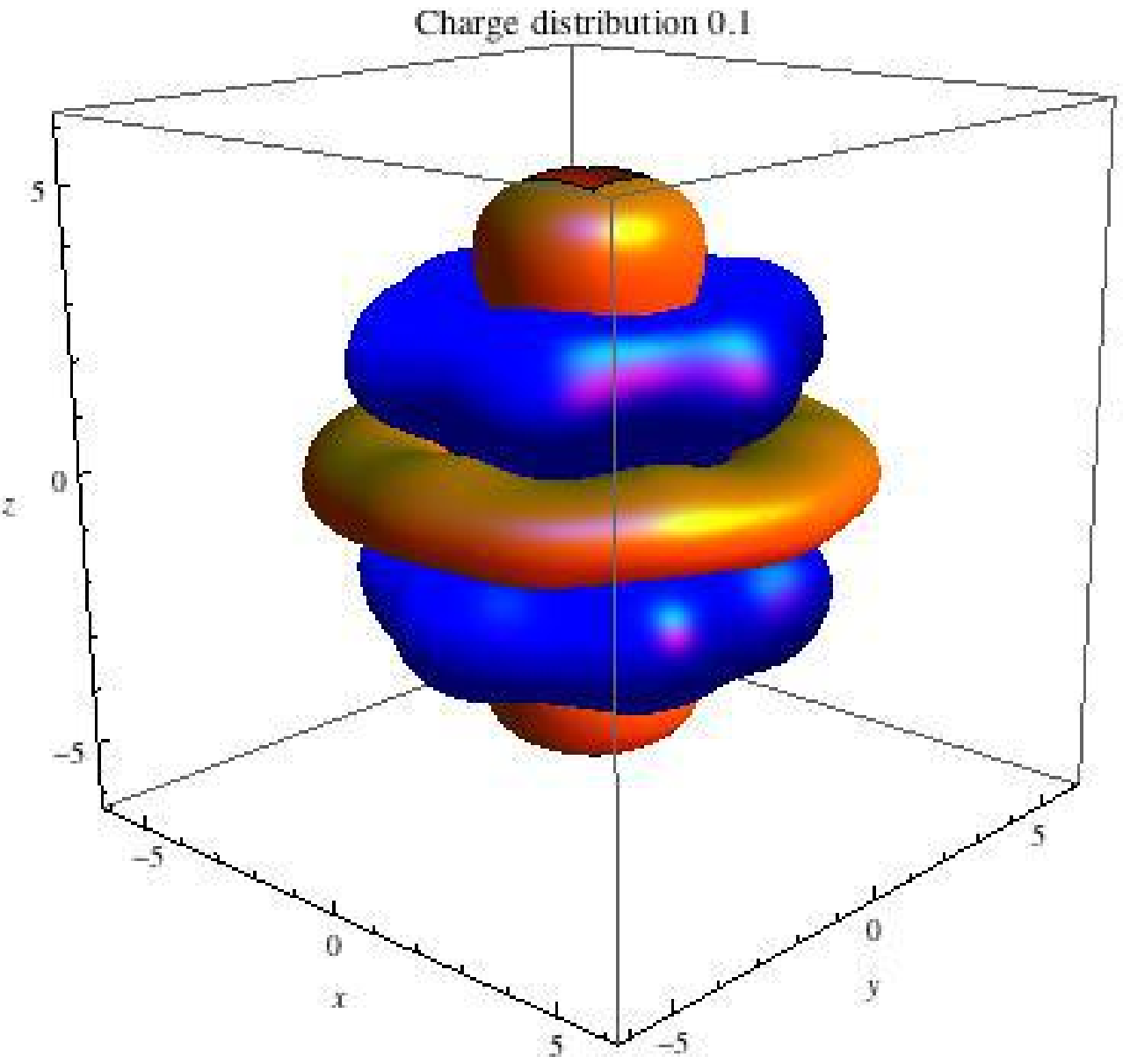}\\
\includegraphics[width=2.35cm]{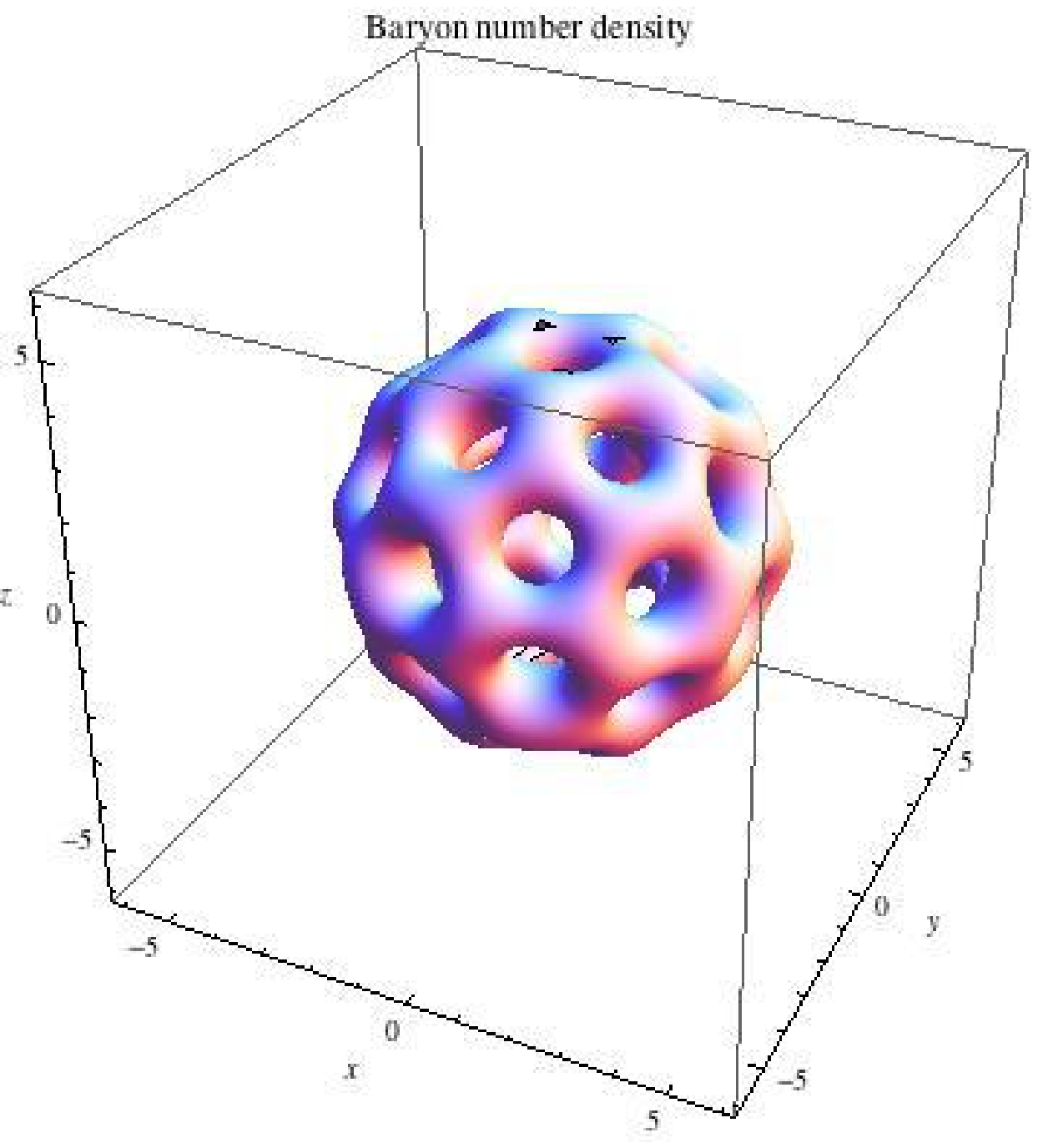} & 
\includegraphics[width=2.35cm]{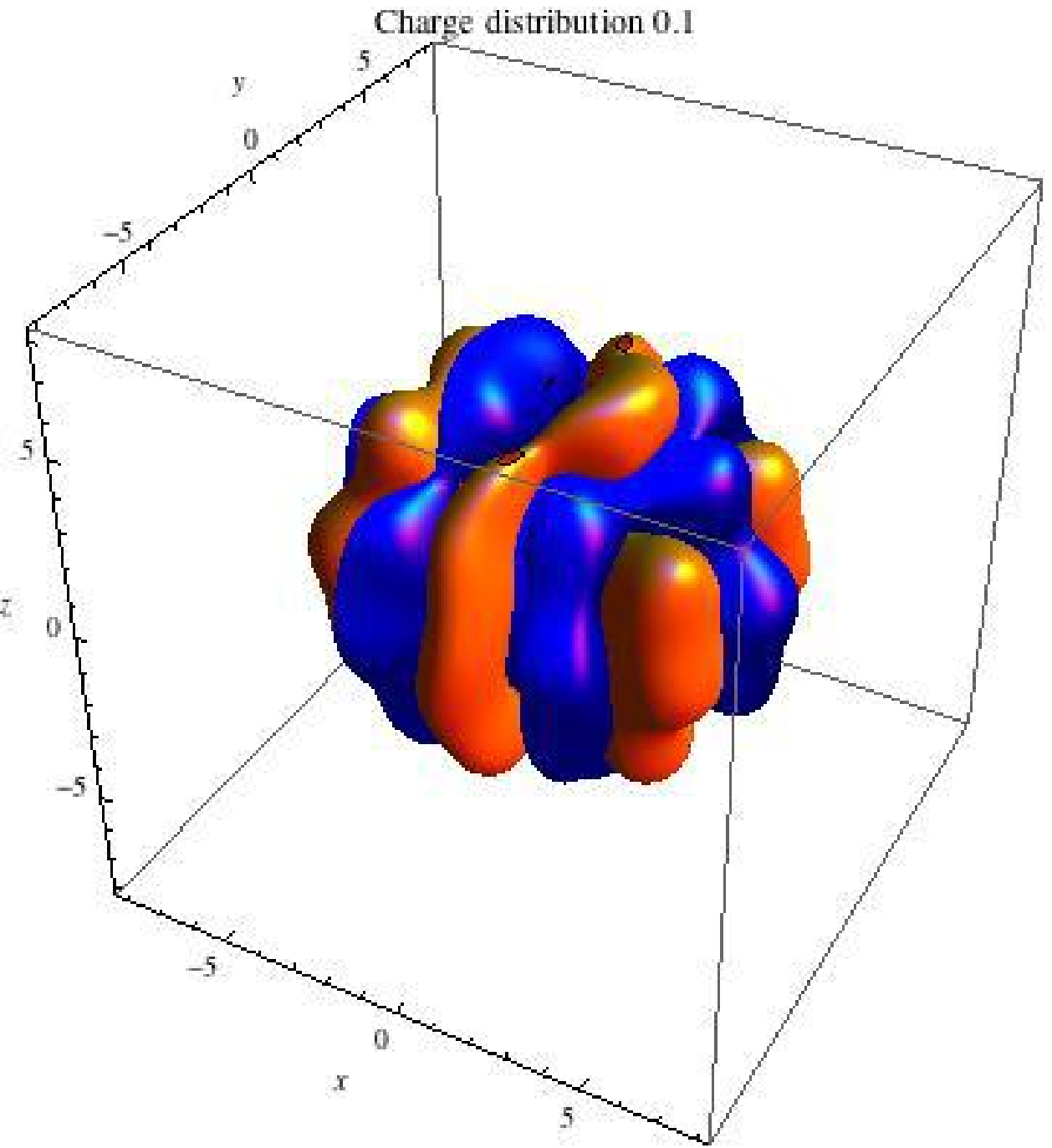} & 
\includegraphics[width=2.35cm]{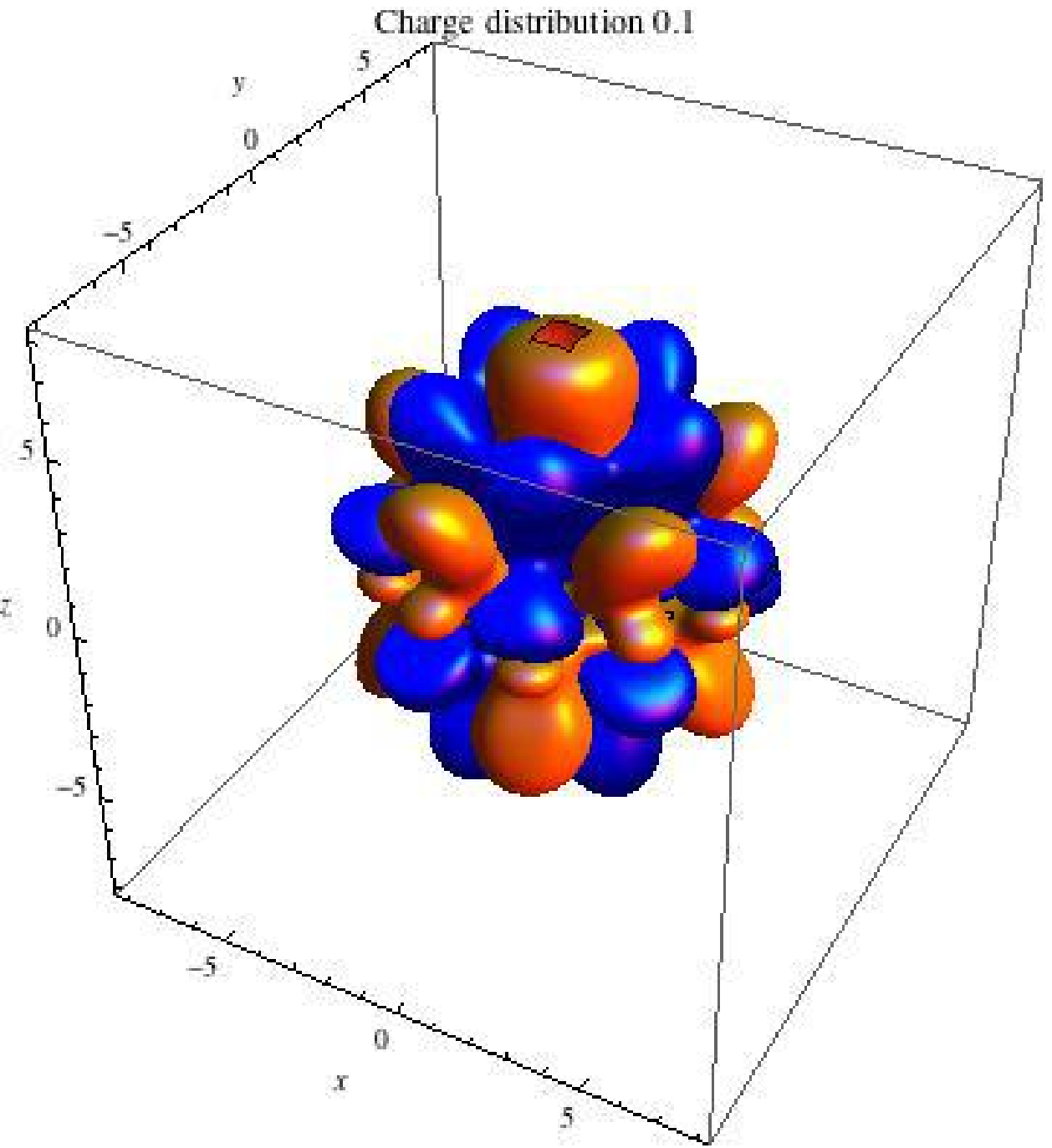}
\end{tabular}
\caption{Higher-charge Skyrmion solution. From left to right, the baryon number density,
$P_1$ and $P_3$, respectively.
$N_B=3,4,\cdots,8$ and 17 are shown from top to bottom.}
\label{fig:multi}
\vspace*{-1cm}
\end{center}
\end{figure}

\section{Conclusion and Discussion}
\label{sec:7}

We have evaluated the gauged WZW term for quantized Skyrmions under a background magnetic field 
in the expansion of 
the electromagnetic coupling constant. We have found that there is an anomaly-induced charge structure due
to the gauged WZW term. The detailed analysis of the total induced charge suggests 
that the pion cloud of the baryons can induce a net charge. The magnitude of the induced charge structure
is ${\cal O} (e^2 B)$, so it is quite small except for the case with strong magnetic field background.

We have calculated the anomaly-induced electric charge for any baryons which appear as 
quantum excitations of Skyrmions (Section \ref{sec:3}). The induced charge is non-vanishing when the
magnetic field is present along the axis of the quantization of the spin and the isospin of the baryon.
In Section \ref{sec:5}, we argued that this induced charge may not be cancelled by other possible electromagnetic
corrections to the Skyrmion, although a complete verification may need an explicit calculation of the
back-reaction of the Skyrmion solution in the magnetic field. We further examined explicitly
the anomaly-induced quadrupole moment (Section \ref{sec:4}) and also the cases with multi-baryons (Section \ref{sec:6}).

It is nontrivial that 
an additional electric charge of baryons is generated in magnetic fields. 
It may have an observable effect on physics related neutron stars and heavy ion collisions \cite{Eto:2011id}.

Finally we discuss possible origin of the anomaly-induced charge.
One may wonder if the constant magnetic field may be too artificial 
and it might be a reason for the anomaly-induced charge. In Appendix \ref{app:A}, we considered a magnetic field
generated by a circular electric current, and we found that the calculated induced charge is again nonzero.
It suggests that the induced charge is not an artifact of the everywhere-constant magnetic field.

Then what is the origin of the additional charge?
A good indication comes from the peculiar property of baryons.
As shown in Appendix \ref{app:B}, 
we found that the total induced charge is due to the multi-pion effect in the nonlinear sigma model. 
As the Skyrmion profile extends to the spatial infinity, the charge distribution also has a tail which
elongates to the spatial infinity. This would be the origin of the generation of the additional electric charge.
Obviously, if quarks are completely confined, the total charge of any baryon should be quantized
to be a half-integer.
However, in reality, any baryon is surrounded by a pion cloud, which means that quark-antiquark pair
can percolate out of the mean volume of the baryon. We can interpret our anomaly-induced charge
as an effective charge carried by the pion cloud surrounding the baryon.
To make sure our interpretation, it is important to calculate the complete effect of the magnetic field,
{\it i.e.} the back-reaction to the Skyrmion profile due to the magnetic field. 

The anomaly-induced charge may appear to violate the charge conservation. In general, 
any electromagnetic current should be conserved at on-shell when the total system is gauge-invariant,
and this applies surely to our case. However, we considered in our paper only a static situation, so
we have not considered the situation where one turns on the magnetic field gradually 
from zero to a nonzero value, in a time-dependent manner. 
To understand the origin of the additional charge concretely,
one needs to calculate the back-reaction and also the time-dependent magnetic fields. 
We leave it to our future work.

\vspace*{2cm}
{\noindent \it Acknowledgment.} ---
The authors would like to thank Aleksey Cherman, Kenji Fukushima, Deog-Ki Hong, Nicholas Manton, Makoto Oka, 
Masashi Wakamatsu, Nodoka Yamanaka, and Koichi Yazaki for useful comments and discussions. 
The work of M.~E. is supported in part by Grant-in Aid for Scientific Research (No. 23740226).
K.~H.\ is supported in part by the Japan 
Ministry of Education, Culture, Sports, Science and Technology.
H.I. is supported by Grant-in-Aid for Scientific Research on Innovative Areas
(No. 23105713).
T.~I.\ was supported by JSPS Research Fellowships for Young Scientists.

\vspace{2cm}
\appendix

\section{Anomaly induced charge in circular electric-current}
\label{app:A}
\def\simge{\mathrel{
       \rlap{\raise 0.511ex \hbox{$>$}}{\lower 0.511ex \hbox{$\sim$}}}}
\def\simle{\mathrel{
       \rlap{\raise 0.511ex \hbox{$<$}}{\lower 0.511ex \hbox{$\sim$}}}}
\newcommand{\br}{{\bf r}}
\newcommand{\bj}{{\bf j}}
\newcommand{\bA}{{\bf A}}
\newcommand{\bB}{{\bf B}}

In the above argument of the anomaly induced charge, 
 we have assumed a uniform external magnetic-field.
However the magnetic field should be always closed 
 unless the magnetic monopole appears.
In this appendix, we consider the anomaly induced charge
 in the external magnetic field generated by a circular electric-current,
 which is instructive for us because the magnetic field
 is closed with finite circular radius, whereas
  that becomes uniform 
  when a radius of the circular electric-current becomes infinity.
Here we suppose that an electric field is not induced by the electric current.
We will show that the anomalous charge is induced 
 in the circular electric-current even with the finite radius.

Let us suppose the circular electric-current density with a radius $a$
 on $xy$-plane as,
\begin{align}
\bj (\br) \equiv \frac{j_0a}{2\pi}
 \delta(z) \delta(\sqrt{x^2+y^2}-a)
 ( - \sin \zeta , \cos \zeta ,0)
,
\end{align}
where we assume that magnitude of the electric-current is proportional to the radius $a$.
The magnetic field generated by the electric-current density can be given by,
\begin{align}
\bB (\br) = \frac{\mu}{4\pi} \, {\rm rot} \int d \tilde \br \, 
 \frac{ \bj ( \tilde \br )}{| \tilde \br - \br |}
,
\end{align}
with $\mu$ being a magnetic permeability.
For simplicity, we omit the factor $\mu/4\pi$ in the following.
One can easily see that, in the large radius limit ($a \rightarrow \infty$), 
 the magnetic field becomes
\begin{align}
B_1 = B_2 = 0, \ \ B_3 = j_0.
\end{align}
This is the same situation with the uniform external magnetic-field
 to $z$-direction.

When the nucleon is located at the center of the circular electric-current,
the anomalous charge in the external magnetic field is 
 given by an integration of $\langle j_{\rm anm} \rangle^N_{I_3, S_3}$,
 shown in Eq.~\eqref{chargedensity}, over the whole space,
\begin{align}
Q_{\rm anm} &= \frac{ie^2N_c}{48\pi^2} \int d^3x \, B_i \langle P_i \rangle^N_{I_3,S_3}
.
\end{align}
Notice that the magnetic field is also functions of the coordinate variables.
Performing the integration over the whole angular-space,
 we can separate three components of the anomalous charge:
\begin{align*}
\rho_{xy}  (r) &= \int d \Omega_2 \, \hat x_1 \hat x_3 \tilde B_1 
                = \int d \Omega_2 \, \hat x_2 \hat x_3 \tilde B_2 ,\\
\rho_{z,1} (r) &= \int d \Omega_2 \, \tilde B_3 ,\ \ \ \ 
\rho_{z,2} (r) = \int d \Omega_2 \, \hat x_3^2 \tilde B_3
,
\end{align*}
where $\tilde B_i \equiv B_i / j_0 $.
Then the anomalous charge can be rewritten as
\begin{align*}
Q_{\rm anm} =& \frac{4eN_c}{27\pi} \, (I_3S_3) \, \frac{ej_0}{(e_sF_\pi)^2}
\left( c_{xy} + c_{z,1} + c_{z,2} \right)
,
\end{align*}
with the numerical coefficients:
\begin{align*}
c_{xy}  &= \frac{3}{4\pi} \int_0^\infty dr 
\left[ 2r^2 f^\prime - r \sin(2f) \right] \rho_{xy} (r) 
, \\
c_{z,1}  &= \frac{3}{8\pi} \int_0^\infty dr \, 
r \sin(2f) \, \rho_{z,1} (r) 
, \\
c_{z,2}  &=  \frac{3}{8\pi} \int_0^\infty dr 
\left[ 2r^2 f^\prime - r \sin(2f) \right] \rho_{z,2} (r) 
.
\end{align*}
Namely, we denote $c_{xy}$ ($c_{z,1}$ and $c_{z,2}$) as
 component(s) of the anomalous charge
 induced by $B_x$ and $B_y$ ($B_z$) generated by the circular electric-current.
With these definitions, one can also show that
 $c_{xy} + c_{z,1} + c_{z,2} = c_0$ at large radius limit
 ($a \rightarrow \infty$).

\begin{figure}[t]
  \includegraphics[width=85mm]{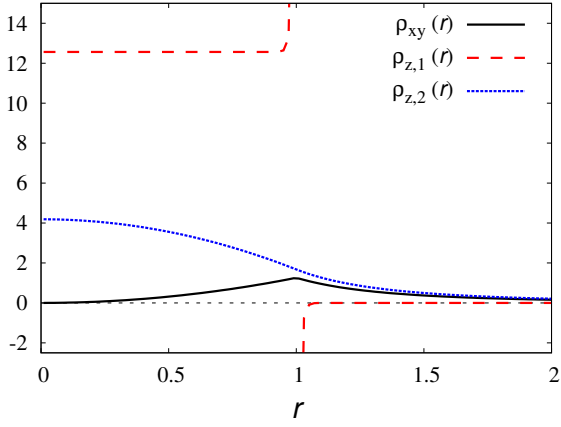}
  \caption{Results of $\rho_{xy}$, $\rho_{z,1}$ and $\rho_{z,2}$
  as a function of $r$ in the case of $a=1$.}
  \label{fig:rhox}
\end{figure}

\begin{figure}[t]
  \includegraphics[width=85mm]{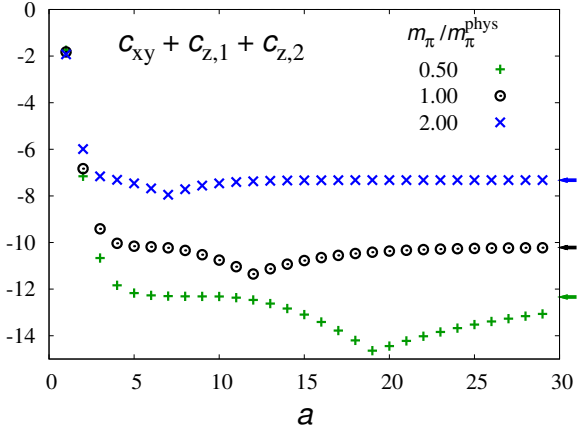}
  \caption{Magnitude of $c_{xy} + c_{z,1} + c_{z,2}$
  as a function of the radius $a$ of the circular electric-current 
  for $m_\pi/m_\pi^{\rm phys} = 0.5$, 1.0 and 2.0.
  The arrows on the right side denote magnitude of $c_0$.}
  \label{fig:inta}
\end{figure}

Fig.~\ref{fig:rhox} shows results of $\rho_{xy}$, $\rho_{z,1}$ and $\rho_{z,2}$
 as a function of $r$ in the case of $a=1$.
We find that $\rho_{xy}$ shows small but finite value with a peak at $r=a$,
 which implies that the anomalous charge is induced by not only $B_z$
  but also $B_x$ and $B_y$.
$\rho_{z,1}$ shows intrinsic behavior:
 it becomes constant at $r<a$, whereas it vanishes at $r>a$.
$\rho_{z,2}$ shows smooth behavior without any singularity at $r=a$.

Magnitude of the coefficients, $c_{xy} + c_{z,1} + c_{z,2}$,
 is shown in Fig.~\ref{fig:inta} as a function of the radius $a$
 for $m_\pi/m_\pi^{\rm phys} = 0.5$, 1.0 and 2.0.
The arrows on the right side denote magnitude of $c_0$.
The coefficients, $c_{xy} + c_{z,1} + c_{z,2}$,
 have finite value even with finite radius which implies 
  the anomaly induced charge
   by the closed magnetic field, and
  converge to $c_0$ at large radius.
It is also found that the coefficients shows minimum values
 at some radius, e.g. $a \sim 12$ for $m_\pi/m_\pi^{\rm phys}=1.0$.
This may be understood as follows:
 in the case of $a \simge m_\pi^{-1}$, the Skyrmion feels similar magnetic field 
  to the uniform one to $z$-direction, which induces similar anomalous charge,
  i.e. $c_{z,1} + c_{z,2} \sim c_0$.
Furthermore, since the anomalous charge is also induced by $B_x$ and $B_y$
 at finite radius discussed above, there is finite contribution, $|c_{xy}| > 0$.
This extra contribution gives larger induced charge than 
 that induced by the uniform magnetic field.

We also calculate the multipole moment due to the anomaly,
 and find that the results are similar to the case of the uniform magnetic field:
 the dipole moment vanishes, whereas
  the quadrupole moment $Q_{ij}$ shows finite values
   only for diagonal parts $(i=j)$.

\section{Multi-pion effect and comparison with point-particle picture}
\label{app:B}

Here we argue that the anomaly-induced electric charge is due to the pion cloud
which exists around any baryon. The pion ``cloud," which is the multi-pion 
effect, in the anomaly term is simply the terms with higher powers in the 
$\pi$ field. The anomaly term in the gauged WZW term  $S_{\rm WZW}$ can be expanded as
\begin{align}
S_{\rm WZW} \sim & \int \! d^4x \; A_0 B_3 P_3
\nonumber \\
\sim & \int \! d^4 x \; {\rm Tr}[\tau_3 U^\dagger \partial U] A_0 B_3
\nonumber \\
\sim & \int \! d^4x \; \left[
\partial \pi_0 + 
\pi \pi \partial\pi + \cdots
\right] A_0 B_3.
\label{expandpi}
\end{align}
The first term is responsible for the famous $\pi_0 \rightarrow 2 \gamma$ interaction,
while the remaining terms are the pion cloud. 

In the following, we shall see that, only with the first term, 
the anomaly-induced total charge $Q_{\rm anm}$ vanishes. So, our anomaly-induced
total charge is due to the pion cloud. 

For the Skyrme solution, we have $\pi_0 \sim f(r) \hat{x}_3$, so the total electric charge
induced by the first term in Eq.~\eqref{expandpi} is proportional to
\begin{align}
\int \! d^3 x \; \partial_3 \pi_0 
&= \int d^3x \partial_3 (f(r) \hat{x}_3)
\nonumber \\
& = 2\pi \int r^2 \sin\theta dr d\theta
\left[
\left( f' - \frac{f}{r}\right) \cos^2\theta + \frac{f}{r}
\right]
\nonumber \\
& =
\frac{4\pi}{3}
\int_0^\infty \! dr \; \left(
r^2 f' + 2 r f\right)
\nonumber \\
& = \frac{4\pi}{3}
\left[r^2 f\right]_{r=0}^{r=\infty} .
\end{align}
The last expression vanishes for nonzero pion mass, because $f(r)$ decays exponentially
at large $r$, and $f(0)$ is finite. So, the anomaly-induced total charge vanishes if one use
only the single-pion term in the anomaly term \eqref{expandpi}. 

It was discussed in \cite{Kharzeev:2011sq} that the anomaly-induced total charge of nucleon vanishes, by using
a generic argument without using the specific Skyrme model. The argument \cite{Kharzeev:2011sq}
uses only the single-pion term, so our result is consistent with it.

Before going to the multi-pion term, we note that, in the chiral limit where
the pion mass vanishes, the last expression is nonzero, since $f \sim r^{-2}$ at large $r$ (see \cite{Adkins1983}). 
So, in the chiral limit, contribution which comes from the single-pion term is nonzero. 
This is again consistent with the discussion in \cite{Kharzeev:2011sq} where the pion momentum is neglected 
compared to the pion mass to show the vanishing total charge. Note that 
this discussion on the chiral limit is suggestive but not so firm since various observables 
in the Skyrme model diverges in the chiral limit.

Now, let us evaluate the multi-pion term in Eq.~\eqref{expandpi}. The representative 3-pion term is evaluated 
as
\begin{align}
\int\! d^3x \; 
\pi\pi\partial\pi
\sim \int_0^\infty
r f(r)^3 dr,
\label{3pi}
\end{align}
which is nonzero for any pion mass. Therefore, we conclude that our anomaly-induced total charge
is due to the multi-pion effect.

The point-particle picture of \cite{Kharzeev:2011sq}
shows that 
the quadrupole moment is induced as a leading moment.
So
let us compare conclusion of the Skyrmion 
 with that of the point-particle picture.

The anomaly-induced quadrupole moment has been written in the point-particle
 picture as \cite{Kharzeev:2011sq}:
\begin{align}
Q^{ij}_{\rm pp} &=
-\frac{N_c \alpha}{6\pi} \frac{g_A}{(f_\pi m_\pi)^2}
 N^\dag \sigma^i \tau^3 N B^j
,
\end{align}
where $\alpha = e^2/4\pi$, and
  $g_A$ and $N$ are the axial coupling constant 
  and the nucleon wave function, respectively.
In the Skyrmion, the quadrupole moment due to the anomaly is given in Eq.~\eqref{eq:qud},
 where the pion-mass dependence of the coefficient becomes,
\begin{align}
c_2 \simeq \frac{A}{(m_\pi/e_sF_\pi)^2}
.
\end{align}
Using the formula of $g_A = - \pi D / 3 e_s^2$ in the Skyrme model, where
 $D$ is the numerical coefficients of $r$ integration 
 including the pion profile function \cite{Adkins1983},
  we can rewrite the quadrupole moment with familiar physical observables as,
\begin{align}
Q_{ij} = - \frac{8N_c \alpha}{45 \pi} \frac{A}{D} \, I_3S_3 \,  \frac{g_A}{(F_\pi m_\pi)^2}
 \tilde Q_{ijk} B_k
.
\end{align}
We have checked that the numerical coefficient $D$ does not show singular dependence
 on $m_\pi$.
This implies that the pion-mass dependence of the quadrupole moment is 
 qualitatively consistent between the point-particle picture and the Skyrme picture.

As a consequence of the comparison, 
we find no contradiction
 between the point-particle picture and the Skyrme picture.
For further understanding of the anomaly-induced charge,
 calculations of the multi-pion effect in the point-particle picture
  are required.


\end{document}